\documentclass{statsoc}

\usepackage{xr,bm,bbm}

\usepackage{amsfonts, amssymb, amsmath, rotating, latexsym,color}
\usepackage{amsmath,graphicx,mathtools}
\usepackage{amssymb,amsfonts}
\usepackage[T1]{fontenc}
\usepackage{optidef,accents}
\mathtoolsset{showonlyrefs}

\usepackage[linesnumbered,ruled,vlined]{algorithm2e}

\usepackage{nameref}

\usepackage[colorlinks,citecolor=blue,urlcolor=blue,linkcolor=blue,naturalnames=true]{hyperref}

\newcommand{\cln}{\color{black}}
\newcommand{\ubar}[1]{\underaccent{\bar}{#1}}
\usepackage{subfigure}
\usepackage[authoryear]{natbib}
\usepackage{enumerate,enumitem}
\newtheorem{mythm}{Theorem}[section]
\newtheorem{mydef}[mythm]{Definition}
\newtheorem{mylem}[mythm]{Lemma}

\newtheorem{proposition}[mythm]{Proposition}
\newtheorem{example}[mythm]{Example}
\newtheorem{myrem}[mythm]{Remark}

\newcommand{\Hh}{\mathcal{H}}
\newcommand{\Qc}{\mathcal{Q}}
\renewcommand{\hat}{\widehat}
\renewcommand{\tilde}{\widetilde}
\newcommand{\Rc}{\mathcal{R}}
\newcommand{\R}{\mathbb{R}}
\newcommand{\Ff}{\mathcal{G}}
\newcommand{\E}{\mathbb{E}}
\newcommand{\K}{\mathbb{K}}
\newcommand{\re}{\mathbb{R}}
\newcommand{\lle}{\le}

\definecolor{lgray}{gray}{0.70}
\newcommand{\rd}{\mathbb{R}^d}
\newcommand{\rn}{\mathbb{R}^n}
\newcommand{\cv}{\mathrm{Cv}}

\newcommand{\bx}{\boldsymbol{\xi}}
\newcommand{\bt}{\boldsymbol{\theta}}

\newcommand{\Kk}{\mathcal{K}}

\renewcommand{\tilde}{\widetilde}
\renewcommand{\hat}{\widehat}
\newcommand{\p}{\mathbb{P}}
\newcommand{\e}{\mathbb{E}}

\DeclareMathOperator*{\argmin1}{arg\,min}

\newcommand{\blind}{1}

\title[Estimation of Quasiconvex Functions] {Least Squares Estimation of a Quasiconvex Regression Function}
\author{Somabha Mukherjee$\dagger$  and Rohit K. Patra\footnote{\footnotesize{SM and RKP contributed equally to this work. RKP's work was partially supported by NSF grant DMS-2210662. SM's work was supported by the National University of Singapore start-up grant WBS A0008523-00-00 and the FoS Tier 1 grant WBS A-8001449-00-00.}}}
\address{Department of Statistics and Data Science, National University of Singapore, Singapore\\
	Department of Statistics, University of Florida, United States of America}
\email{rkumarpatra@gmail.com}

\author{Andrew L. Johnson}
\address{Industrial and Systems Engineering, Texas A\&M University, United States of America}
\author[Mukherjee and Patra {et al.}]{Hiroshi Morita}
\address{Graduate School of Information Science and Technology, Osaka University, Japan}
\begin{document}
	
	\begin{abstract} 
		We develop a new approach for the estimation of a multivariate function based on the economic axioms of quasiconvexity (and monotonicity). {\cln On the computational side, we prove the existence of the quasiconvex constrained least squares estimator (LSE) and provide a characterization of the function space to compute the LSE via a mixed integer quadratic programme. On the theoretical side, we provide finite sample risk bounds for the LSE via a sharp oracle inequality. Our results allow for errors to depend on the covariates and to have only two finite moments.  	
			We illustrate the superior performance of the LSE against some competing estimators via simulation. Finally, we use the LSE to estimate the production function for the Japanese plywood industry and the cost function for hospitals across the US.}
	\end{abstract}
	
	\maketitle
	\section{Introduction}\label{sec:intro}
	
	Production analysis has been an indispensable tool for economists, managers, and engineers in evaluating a firm's performance. Reliable estimates of production functions are of great importance because they can assist in accurate decision making. In this context, regression models enable us to identify relationships among resources and products. 
	
	Consider a production process that uses $d$ different resources to produce a single product or output, $Y\in \R$. The resources consumed are called the inputs, and we denote their quantity by  $\boldsymbol X\in \R^d $. We consider the following regression model
	\begin{equation}\label{mainmodel}
		Y=\varphi(\boldsymbol X)+\varepsilon,
	\end{equation}
	where the random variable $\varepsilon$ satisfies $\mathbb{E}(\varepsilon|\boldsymbol X) = 0$ and $\mathbb{E}(\varepsilon^2|\boldsymbol X) < \infty$ for almost every  $\boldsymbol X$.  
	Given $n$ i.i.d.~observations $\{(\boldsymbol X_j , Y_j )\}_{j=1}^n$ from the regression model~\eqref{mainmodel}, the goal of the paper is to estimate the {unknown} production function $\varphi: \R^d \to \R$,  subject to some basic shape constraints imposed by economic axioms.
	
	Production functions are linked to cost functions through a dual relationship,
	so axioms that hold for production functions imply similar axioms for cost functions \citep{shephard1953,diewert1982duality}. We will thus frame our discussion of axiomatic properties primarily in terms of production functions, recognizing that through duality, similar axioms are required for cost functions.  
	Microeconomic theory often implies qualitative assumptions on production functions, and  the most prominent of those assumptions is the monotonicity axiom~\cite[page 6]{varian1992microeconomic}, 
	which says that an increase in input resources should lead to no less output. 
	This argument is common and reasonable for establishments facing competition, see e.g., \citet[Pages 10--11]{beattie1985economics} and \cite{chambers1988applied}. 
	Formally,  the monotonicity axiom implies that 
	\begin{equation}\label{eq:monotone_axiom}
		\text{if  }\boldsymbol X_1  \lle \boldsymbol X_2, \text{ then } {\varphi(\boldsymbol X_1) \le \varphi(\boldsymbol X_2)},
	\end{equation}
	where for two vectors $\boldsymbol a := (a_1,\ldots, a_d), \boldsymbol b:= (b_1, \ldots, b_d) \in \R^d$, we say $\boldsymbol a \lle \boldsymbol b$ if $a_i \le b_i$ for all $i \in \{1, \ldots, d\}$.

	For a given output level $y,$ define the \textit{input requirement set}\footnote{We will use \textit{input set} and  \textit{input requirement set} interchangeably in this paper.} $V(y) \subset \R^d$ as the set of all input vectors $\boldsymbol x$ that produce {at least} $y$ units of output, i.e. 
	$V(y):=  \{ \boldsymbol x: \varphi(\boldsymbol x)\ge y\}.$
	Another prominent assumption about the production function is that the input requirement set $V(y)$ is convex for every $y\in \R$, i.e.
	\begin{equation}\label{eq:convex_axiom}
		\text{if  }\boldsymbol x_1, \boldsymbol x_2\in V (y), \text{  then  }\lambda \boldsymbol x_1 + (1 -\lambda )\boldsymbol x_2 \in V (y) \text{ for all }\lambda \in [0,1].
	\end{equation} 
	The economic motivation for this assumption is based on the fact that for most production technologies  there are optimal proportions  in which inputs should be used  and that deviations from the optimal proportion by decreasing the level of one input, such as capital, will require more than a proportional increase in another input, such as labor~\citep{johnson2018shape}. Furthermore \citet[Page 82]{varian1992microeconomic} argues that even if the production technology does not justify convexity, if the prices for inputs are positive, then operating in a nonconvex region of the input requirement set would be economically inefficient and should be avoided. 
	
	
	Estimates of production and cost functions are widely used in policy decisions. Thus estimation of these functions has received wide attention and a variety of estimators have been proposed; see e.g., the monographs \cite{tirole1988theory} and \cite{jorgenson2000econometrics}. 
	Nonparametric smoothing methods (such as the Nadaraya-Watson or smoothing/regression splines estimators) avoid the potential for functional misspecification and flexibly capture the nuances of the data. However, they are often difficult to interpret economically, require choice of \textcolor{black}{tuning} parameters whose values are hard to justify, and do not satisfy the basic axioms~\eqref{eq:monotone_axiom} and~\eqref{eq:convex_axiom}.  While parametric estimators (such as Cobb-Douglas~\citep[Page 4]{varian1992microeconomic} and translog estimators~\citep{berndt1973translog}) will satisfy the above economic axioms, they are likely to be misspecified because there is rarely a contextual motivation for the parametric specification selected.

	In between these two extremes lie many shape constrained estimators.  Semiparametric shape constrained models such as the monotone or convex single index models \citep{2017arXiv170800145K,balabdaoui2016least,score} \textcolor{black}{model the observation $Y$ as a univariate monotone or convex transform of a linear transformation of the covariates, rather than a multivariate shape-constrained transform of the entire set of covariates, which may not be realistic for many practical applications. Moreover, these shape-constrained single-index models are not guaranteed to satisfy assumptions~\eqref{eq:monotone_axiom} and~\eqref{eq:convex_axiom}, and consequently, cannot be applied to our framework.}
	\textcolor{black}{Several nonparametric methods for estimating multivariate monotone functions \cite{wu2015penalized,chernozhukov2009improving,ChatterjeeAOS1324,han,DengZhang} involving constrained/penalized nonparametric least squares, rearrangement, and block estimators have been developed in the last few years}.
	These estimators satisfy~\eqref{eq:monotone_axiom} but do not have convex input requirement sets.  \textcolor{black}{The monotonic and concave estimators in ~\cite{seijosen,kuosmanen2008representation,lim2012consistency,blanchet2019multivariate} and the recently proposed $S$-shape estimator~\citep{yagi2017SShape} will satisfy~\eqref{eq:monotone_axiom} and~\eqref{eq:convex_axiom}.
	}  However, these estimators are based on further restrictive  and unjustified assumptions about the production function, \textcolor{black}{which are not necessary in our framework}. \textcolor{black}{Monotonicity and convex input requirement sets arise naturally in many real life examples, and existing estimators can be unsatisfactory; see Section~\ref{sec:real_data} for more details on the Japanese production data. In most such examples the existing shape constrained estimators do not adequately incorporate the known shape of the nonparametric function (e.g., the  monotonic estimators) or impose additional stronger conditions (e.g., the monotonic and concave or $S$-shape estimators) This motivates us to propose an estimator that satisfies two most basic assumptions about production functions~\eqref{eq:monotone_axiom} and~\eqref{eq:convex_axiom} without enforcing \textit{any additional structure}.}

	Quasiconcave functions are defined as functions for which all upper level sets are convex. Thus a function satisfies both~\eqref{eq:monotone_axiom} and~\eqref{eq:convex_axiom} \textit{if and only if} it is quasiconcave and increasing. Note that, there is a very natural correspondence between quasiconcave, increasing functions and quasiconvex, decreasing functions, namely, if $f$ is a quasiconcave, increasing function, then $-f$ is quasiconvex and decreasing. In this paper, we focus on the estimation of quasiconvex and decreasing functions, and propose a least squares estimator that is guaranteed to be quasiconvex and decreasing. To be specific, given observations $\{(\boldsymbol X_i , Y_i) \in \rd\times \re\}_{i=1}^n$ from the regression model in~\eqref{mainmodel}, we study the following least squares estimator (LSE):
	
	\begin{equation}\label{eq:orig_def}
		\widehat{\varphi}_n \in \argmin1_{\psi \in \mathcal{C}} {\sum_{k=1}^n {(Y_k - \psi(\boldsymbol X_k))}^2 },
	\end{equation} 
	where 
	\begin{equation}\label{eq:C_def}
		\mathcal{C} := \big\{\psi : \mathbb{R}^d\to \mathbb{R}\, \big|\, \psi ~\textrm{is quasiconvex and decreasing}\big\}.
	\end{equation}
	
	An advantage of the above LSE is that it is \textit{tuning parameter free} and thus avoids fitting issues related to tuning parameter selection for other nonparametric estimators.

	{\cln Sections~\ref{sec:intro}--\ref{sec:qmest} focus on the quasiconvex and decreasing LSE. If one aims to find the quasiconcave and increasing LSE then she  needs to solve the problem \eqref{eq:orig_def} with $\{(\boldsymbol X_i, -Y_i)\}_{i=1}^n$ in place of $\{(\boldsymbol X_i, Y_i)\}_{i=1}^n$. The final estimator is then simply the negative of the above LSE. The development of the estimator \textit{without} the additional monotonicity assumption is almost identical and is described in Section~\ref{sec:qonly}.}
	
	
	\subsection{Our contributions} 
	\label{sub:contribution}
	\textcolor{black}{In this paper, we characterize the least-squares constraint space for multivariate, decreasing, and quasiconvex functions, and use this characterization to develop a mixed-integer quadratic optimization (MIQO) algorithm for computing the LSE, which is implemented in the \texttt{R} package~\if1\blind{~\texttt{QuasiLSE}~\citep{QuasiLSE}}. We also proposed a sample-splitting based algorithm to reduce the computational cost of the LSE. To the best of our knowledge, this is the first work studying the quasiconvex and decreasing LSE~\eqref{eq:orig_def}. We also provide finite-sample risk bound (via a sharp oracle inequality) for the LSE under a very general heteroscedastic model. Moreover, we show that the quasiconvex LSE is minimax rate optimal when $d\ge 4$. Finally, the performance of the LSE is illustrated through simulations and analysis of two real datasets, namely the Japanese plywood production data and the US hospital cost data.}

	To the best of our knowledge, the only other  estimator in the nonparametric regression framework that satisfies~\eqref{eq:monotone_axiom} and~\eqref{eq:convex_axiom} without any additional assumptions is proposed in~\cite{chen2018shapeenforcing}.~\cite{chen2018shapeenforcing} propose a functional operator that can modify any existing estimator and enforce the shape constraint of quasiconvexity and monotonicity. Their procedure is very general and they show that ``shape-enforced point estimates are closer to the target function than the original point estimates.'' However, their approach is \textit{ex post} and the performance of the shape enforced estimator is directly related to the initial estimator (such as the kernel or splines based estimators), the performance of which, in turn, will often depend on the smoothness assumption on the true regression function and tuning parameters. Thus the improvement due to the operator is only relative to the performance of the initial estimator. Furthermore, the estimator in~\cite{chen2018shapeenforcing} does not have a clear interpretation as a minimizer of any loss function. {\cln However, it is worth noting that there are settings under which the~\cite{chen2018shapeenforcing} estimator will perform better than the LSE as well as settings under which the opposite is true; see Section~\ref{sec:simulation_study} for further discussion.}

	\subsection{Organization} 
	\label{sub:organization} Our exposition is organized as follows. In Section \ref{notationanddef}, we introduce some preliminary notations and definitions that will be used throughout the paper. In Sections \ref{sec:qmest} and~\ref{sec:qonly}, we establish existence and almost sure uniqueness of the LSE  and provide an algorithm to compute the LSE for the quasiconvex and monotone LSE and the quasiconvex only LSE, respectively. In Section \ref{sec:consist}, we provide a finite sample risk bound for the  quasiconvex (and monotone) LSE. In Section \ref{sec:simulation_study}, we compare the performance of our quasiconvex and increasing LSE with that of the Nadaraya-Watson estimator, the estimator due to \cite{chen2018shapeenforcing}, and other existing shape constrained estimators through simulations. In Section \ref{sec:real_data}, we apply our techniques to a real production dataset. The paper ends with Section~\ref{sec:conc}, where we give a brief discussion and provide some exciting future directions. 
	
	All the sections, lemmas, definition, and remarks in the supplementary file have the prefix ``S.'' In
	Section~\ref{sec:analysis_of_cost_data} of the supplementary, we describe the cost data on US hospitals and show that a quasiconcave and increasing regression leads to valuable insights.  The proofs of the results in the main paper can be found in Sections~\ref{sec:techResults}--\ref{sec:necLemmas} of the supplement.
	
	\section{Notations and definitions}\label{notationanddef}
\begin{figure}
		\begin{center}
			\includegraphics[width=2.5in]{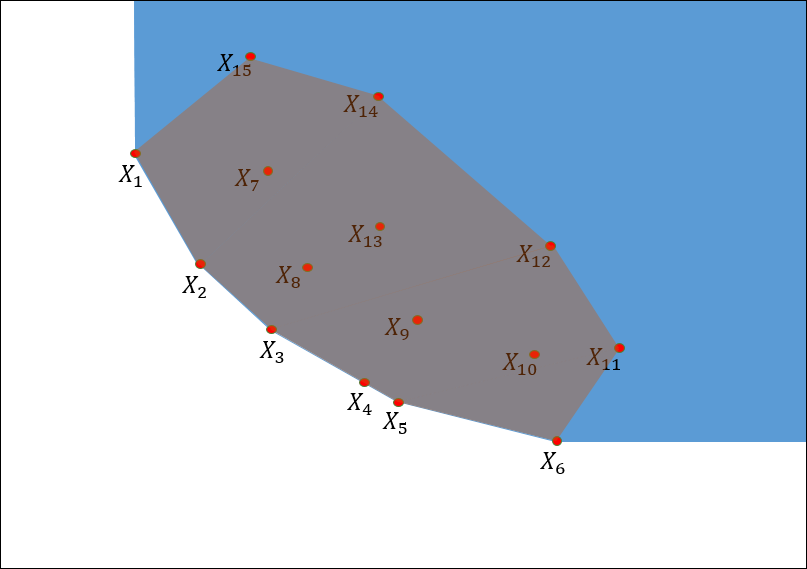}
		\end{center}
		\caption{The set $\mathrm{Cv}(\{\boldsymbol X_i: 1\leq i\leq 15\})$ in $\mathbb{R}^2$ is shaded gray. The set $\mathrm{Cv}^\dagger(\{\boldsymbol X_i: 1\leq i\leq 15\})$ in $\mathbb{R}^2$ is the union of areas shaded blue and gray.\label{chull}}
	\end{figure}

	In this section, we introduce some notations and definitions that will be used throughout the rest of the paper. We use bold letters to denote vectors, matrices, and tensors. The $d$-dimensional vector with all entries equal to zero will be denoted by $\mathbf{0}_d$. For any positive integer $m$, we will denote the set $\{1,2,\ldots,m\}$ by $[m]$. For a function $\psi: \rd \to \re$ and $\alpha \in \re$, the $\alpha$-lower level set of $\psi$ is defined as: 
	$$S_\alpha(\psi):= \psi^{-1}((-\infty,\alpha]) = \{\boldsymbol x \in \rd : \psi(\boldsymbol x) \leq \alpha\}~.$$ For $\boldsymbol X \in \rd$ and a set $A \subseteq \rd$, the upper orthants of $\boldsymbol X$ and $A$ are defined as
	\begin{equation}\label{eq:upper_orthant}
		\boldsymbol X^\dagger := \{\boldsymbol Y \in \rd: \boldsymbol X \lle \boldsymbol Y\}\qquad\text{and}\qquad A^\dagger:= \cup_{\boldsymbol X \in A} \boldsymbol X^\dagger,   
	\end{equation}
	where $\boldsymbol X \le \boldsymbol Y$ denotes that $X_i \le Y_i$ for all $i \in [d]$; see Figure \ref{chull} for an illustration.\footnote{Two crucial properties of the set $A^\dagger$, which we will use later, are proved in Lemmas \ref{dagconv} and \ref{dagclosed} in Section~\ref{sec:techResults} of the supplement.} The convex hull of a set $A \subseteq \mathbb{R}^d$ is denoted by $\mathrm{Cv}(A)$, and is defined as the intersection of all convex subsets $C$ of $\mathbb{R}^d$ such that $A \subseteq C$. For notational convenience, we will use $\cv^\dagger(A)$ to denote  the upper  orthant of $\cv(A)$. Throughout the paper, $\|\cdot\|$ will stand for the Euclidean norm of a vector.  
	
	Below, we define the central objects of importance in this paper, namely quasiconvex and decreasing functions with multivariate entries.
	
	\begin{mydef}
		A function $\psi: \rd \to \re$ is said to be quasiconvex, if \[\psi\left(\lambda \boldsymbol X + (1-\lambda)\boldsymbol Y\right) \leq \max \left\{\psi(\boldsymbol X) , \psi(\boldsymbol Y)\right\} \text{ for all } \boldsymbol X , \boldsymbol Y \in \rd\text{ and }\lambda \in [0,1],\]
		and decreasing, if $\psi(\boldsymbol X) \geq \psi(\boldsymbol Y)$ for all $\boldsymbol X \lle \boldsymbol Y \in \rd$.
	\end{mydef}
	The following alternative definition will turn out to be more useful in many of our proofs.
	\begin{mydef}
		A function $\psi: \rd \to \re$ is quasiconvex if $S_\alpha(\psi)$ is a convex set for all $\alpha \in \re$, and is quasiconvex and decreasing if $~\cv^\dagger\left(S_\alpha(\psi)\right) = S_\alpha(\psi)$ for all $\alpha \in \re$.
	\end{mydef}  
	
	\section{The quasiconvex-decreasing regression problem}\label{sec:qmest}
	The goal of this section is to estimate the {unknown} function $\varphi$ under the assumption that it is quasiconvex and decreasing function. In~\eqref{eq:orig_def}, we proposed  the tuning parameter free least squares estimator $\widehat{\varphi}_n$ based on the data $\{(\boldsymbol X_i , Y_i)\}_{i=1}^n$. The first observation is that the seemingly  infinite  dimensional  optimization  problem~\eqref{eq:orig_def} can be reduced  to  a finite  dimensional  optimization problem by observing that the loss function in~\eqref{eq:orig_def} depends on $\psi$ only through its values at $\boldsymbol X_1,\ldots, \boldsymbol X_n$.  Letting $\widehat{\boldsymbol \theta}=(\widehat\varphi_n(\boldsymbol X_1),\ldots,\widehat\varphi_n(\boldsymbol X_n)),$ we have:
	\begin{equation}\label{opt}
		\widehat{\boldsymbol \theta}\in \argmin1_{\boldsymbol \theta \in \Qc}  \hspace {.2 cm}\sum_{k=1}^n {(Y_k - \theta_k)^2 }
	\end{equation}
	where
	\begin{equation}\label{ca}
		\Qc := \big\{(\psi(\boldsymbol X_1), \ldots ,\psi(\boldsymbol X_n)) \in \R^n: \psi \in \mathcal{C}\big\},
	\end{equation}
	for $\mathcal{C}$ defined in~\eqref{eq:C_def}.
	Some immediate and natural questions arise: (i) does $\widehat{\boldsymbol \theta}$ exist?; (ii) is $\widehat{\boldsymbol \theta}$ unique?; and (iii) how can we compute $\widehat{\boldsymbol \theta}$?  We answer the questions  (i) and (ii) in the affirmative in Section~\ref{sec:primal} and provide a way to compute $\widehat{\boldsymbol \theta}$ in Section~\ref{sec:dual}.

	Observe that, $\widehat{\boldsymbol\theta}$ is only the first step in estimating $\varphi.$ There are indeed many  quasiconvex and decreasing functions satisfying $\widehat{\varphi}_n(\boldsymbol X_i) = \widehat\theta_i$ for all $i =1,\ldots,n$.  Any such function can act as a least squares estimator.\footnote{This type of behavior is not uncommon in nonparametric maximum likelihood or least squares problem, e.g., see~\cite{npmle} for an example where the NPMLE exits but is not unique and see~\cite{seijosen} for an example in the regression setting; also see~\cite{zheng2017fitting}.}
	In this paper, however,  we use a simple piecewise constant version defined on the whole of $\mathbb{R}^d$.   The function can be computed from $\widehat{\bt}$ in an inductive way. We describe the process now. First arrange the elements of $\widehat{\bt}$ in an increasing order $\widehat\theta_{(1)} \leq \widehat\theta_{(2)} \leq \ldots \leq \widehat\theta_{(n)}$, and suppose that $\boldsymbol{X}_{(i)}$ is the data point corresponding to the estimate $\widehat\theta_{(i)}$. Set $\widehat{\varphi}_n(\boldsymbol X) = \widehat\theta_{(1)}$ for all $\boldsymbol X \in  \boldsymbol X_{(1)}^\dagger.$ Now, assume inductively, that $\widehat{\varphi}_n$ has been defined on $\cv^\dagger(\{\boldsymbol X_{(1)},\ldots,\boldsymbol X_{(m-1)}\})$ for some $1 < m \leq n$. For all $\boldsymbol X \in \cv^\dagger(\{\boldsymbol X_{(1)},\ldots,\boldsymbol X_{(m)}\}) \setminus \cv^\dagger(\{\boldsymbol X_{(1)},\ldots,\boldsymbol X_{(m-1)}\})$, we define 
	\begin{equation}\label{eq:thetatophi}
		\widehat{\varphi}_n (\boldsymbol X) = \widehat\theta_{(m)}.
	\end{equation} 
	This completes the definition of $\widehat{\varphi}_n$ on $\cv^\dagger(\{\boldsymbol X_{(1)},\ldots,\boldsymbol X_{(n)}\}) = \cv^\dagger(\{\boldsymbol X_1,\ldots,\boldsymbol X_n\}).$ Finally, we define $\widehat{\varphi}_n(\boldsymbol X) = \widehat\theta_{(n)}$ for all $\boldsymbol X \notin\cv^\dagger(\{\boldsymbol X_1,\ldots,\boldsymbol X_n\})$. The rather delicate issue of defining the ordered entries $\widehat\theta_{(i)}$ in case of the presence of ties in the entries of $\widehat{\bt}$, is addressed rigorously in the proof of Lemma \ref{lem:primal}.  
	{\cln 
		\begin{myrem}[Interpolation]\label{rem:thetatophi}
			We believe that it might be possible to find a piecewise linear or even smooth interpolation of $\widehat\theta_{(m)}$  that satisfies the quasiconvexity (and monotonicity) constraint. However, we couldn't formulate such a procedure. The main difficulty with this approach is that, the boundary smoothing (at boundaries of the convex upper hulls) must be carried out in such a way, that the smoothed function is still quasiconvex and monotone. In this sense, the piecewise constant interpolation is the only \textit{practical} option for us. From a theoretical perspective, we will show in Section~\ref{sec:consist} that our main theoretical results hold for \textit{any} interpolation of $\widehat\theta_{(m)}$, and that the asymptotic behavior of the estimator does not depend on the interpolation technique used. 
	\end{myrem}}
	
	%
	
	\subsection{Primary characterization, existence, and uniqueness}\label{sec:primal}
	In this section, we provide a characterization of the constraint space $\Qc$. This primary characterization will help us prove the existence of the LSE. A secondary characterization of $\Qc$ (given in Section \ref{sec:dual}) will be crucial for the computation of the LSE.

	Let $\mathcal{X} := \{\boldsymbol X_1,\ldots, \boldsymbol X_n\}$ and let $\mathcal{L}(\mathcal{X})$ be defined as:
	\begin{equation}\label{eq:L_def}
		\mathcal{L}(\mathcal{X})\coloneqq\big\{(i,S): i \in [n],\, S \subseteq [n], \text{ and }\boldsymbol X_i \in \cv^\dagger\big(\{\boldsymbol X_j: j \in S\}\big) \big\}.
	\end{equation}
	\begin{mylem}[Primary characterization]\label{lem:primal} 
		\begin{equation}\label{eq:primalQc}
			\mathcal{Q} = \Big\{\boldsymbol z \in \mathbb{R}^n: z_i \leq \max_{j \in S} z_j~\textrm{for all}~(i,S) \in \mathcal{L}(\mathcal{X})\Big\}.
		\end{equation}
	\end{mylem}
	The above characterization of $\Qc$ (proved in Section~\ref{primalproof} of the supplement) will play a key role in proving the existence and uniqueness of $\widehat{\boldsymbol \theta}$; see Theorem~\ref{thm:exun} below. Furthermore, it will later help us develop a method for its computation (see Section~\ref{sec:dual}). A crucial difference between other shape constraints such as monotonicity \citep{brunk,zhang} and  convexity \citep{seijosen,kuosmanen2008representation}, and quasiconvexity, is that  the set $\Qc$ is not convex.\footnote{\cln It is easy to see this via the following simple example. Let $A$ and $B$ be two convex sets on $\R^d$ such that $A\cup B$ is not convex. Then observe that both $h_1(\boldsymbol x) := \mathbf{1}(\boldsymbol x\in {A^c})$ and  $h_2(\boldsymbol x) := \mathbf{1}(\boldsymbol x\in {B^c})$ are quasiconvex but $(h_1+h_2)/2$ is \textit{not} quasiconvex.\label{foo:nonconvex}} Consequently, a minimizer for \eqref{opt} may not be unique.  However, in the result below (proved in Section~\ref{sec:proof_of_theorem_thm:exun} of the supplement) we show that $\widehat{\boldsymbol\theta}$ is unique \textit{almost surely} if $Y$ has a density with respect to the Lebesgue measure on $\R.$
	\begin{figure}
		\begin{center}
			\includegraphics[width=3in]{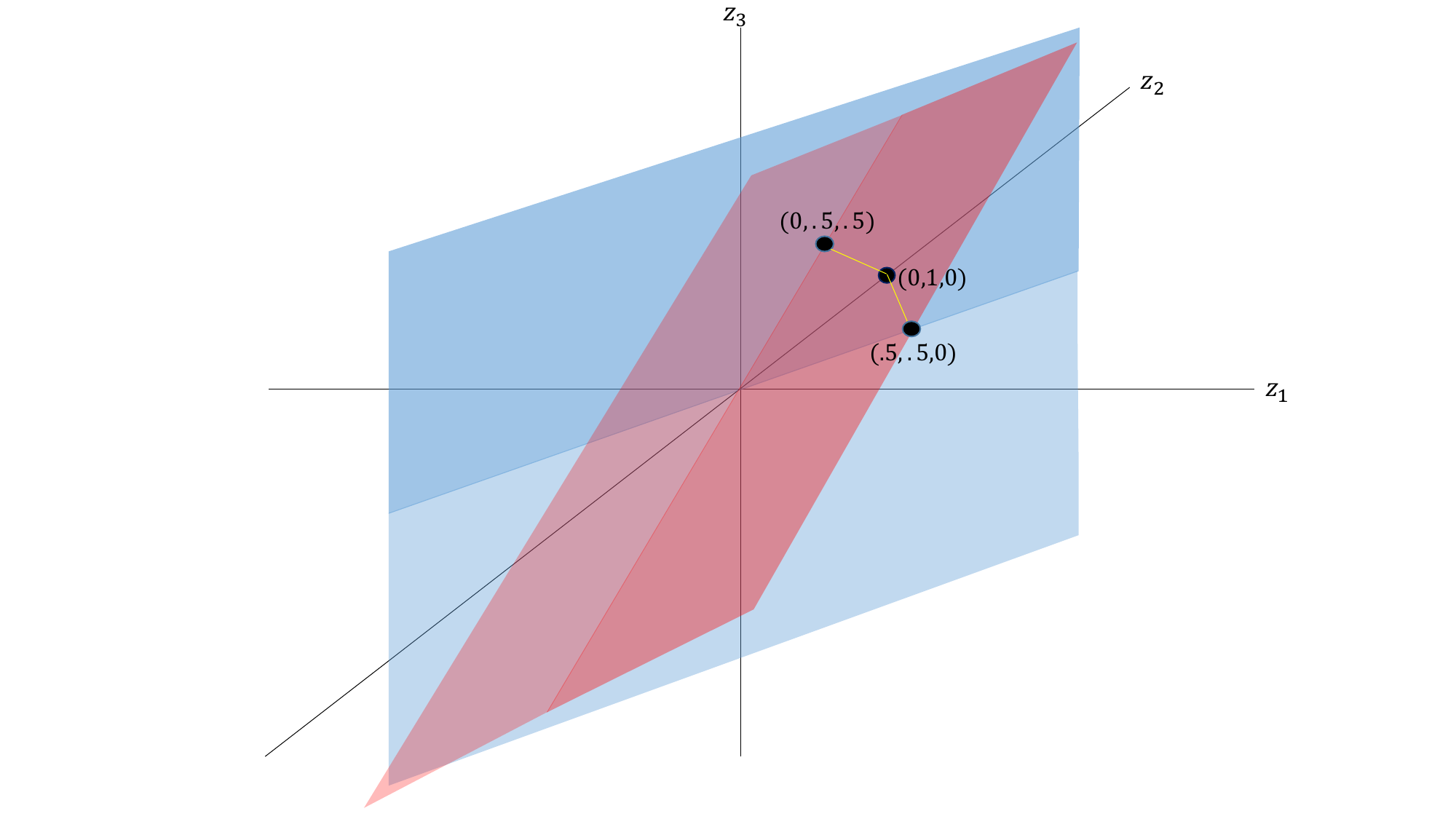}
		\end{center}
		\caption{Figure for Example~\ref{nonunique}. Both the points $(0,.5,.5)$ and $(.5,.5,0)$ are projections of $(0,1,0)$ on $\Qc$. The blue plane has equation $z_1 = z_2$ and the red plane has equation $z_2 = z_3$.\label{nonunproj}}
	\end{figure}
	\begin{mythm}[Existence and uniqueness]\label{thm:exun}
		The optimization problem \eqref{opt} has a minimizer $\widehat{\bt}$ in $\Qc$. Moreover, if $\,Y\,$has a density with respect to the Lebesgue measure on $\mathbb{R}$, then $\widehat{\bt}$ is unique  with probability $1$.
	\end{mythm} 
	
	\begin{example}[Non-uniqueness of minimizer]\label{nonunique}
		Since the constraint space $\Qc$ is not convex, there are points lying outside $\Qc$ that have two different projections on $\Qc$. Consequently a minimizer of~\eqref{opt} may not be unique. For example, take $n=3$, $d=2$, $\boldsymbol X_1= (1,0), \boldsymbol X_2 = (0.75,0.75)$, and $\boldsymbol X_3 = (0,1)$. It follows from Lemma \ref{lem:primal} that $\Qc = \{\boldsymbol z \in \mathbb{R}^3: z_2 \leq z_1\vee z_3\}$. One can easily check that both the points $(0.5,0.5,0)$ and $(0,0.5,0.5)$ are projections of the point $\boldsymbol u := (0,1,0)$ on $\Qc$ (see Figure \ref{nonunproj}). However as shown in the second part of the proof of Theorem~\ref{thm:exun}, this happens only when $u$ is in a set of Lebesgue measure zero. This example is interesting from another aspect too. Since $\boldsymbol u \notin \Qc$, no function $f: \mathbb{R}^2 \to \mathbb{R}$ passing through $(\boldsymbol X_1,u_1), (\boldsymbol X_2,u_2)$ and $(\boldsymbol X_3,u_3)$ (i.e. $f(\boldsymbol X_i) = u_i$ for $i = 1,2,3$), is both quasiconvex and decreasing. However, one can construct functions $f_1:\R^2 \to \R$ and $f_2:\R^2 \to \R$ passing through $(\boldsymbol X_1,u_1), (\boldsymbol X_2,u_2)$ and $(\boldsymbol X_3,u_3)$, such that $f_1$ is quasiconvex and $f_2$ is decreasing. This shows that the constraint space $\Qc$ for the ``quasiconvex and decreasing" regression problem is not equal to, but a proper subset of the intersection of the constraint spaces for the quasiconvex regression and the decreasing regression problems.
		
		{\cln The proof of Theorem \ref{thm:exun} reveals that as long as the error $\varepsilon$ has a density with respect to the Lebesgue measure on $\mathbb{R}$, the LSE over any set $K \subseteq \mathbb{R}^n$ (not only $\mathcal{Q}$) is unique with probability $1$. However, this is not true if $\varepsilon$ does not have a continuous distribution. As an example, consider the setup in the previous paragraph, and assume that the distribution of $\varepsilon$ assigns positive mass to the points $r-\varphi(\boldsymbol X_1)$ and $r-\varphi(\boldsymbol X_3)$ for some real number $r$, where $\varphi$ is the true function. Then, the random vector $\boldsymbol Y := (Y_1,Y_2,Y_3)$ lies on the line $z_1=z_3=r$ with positive probability, and hence, as long as the support of $\varepsilon$ is unbounded above (to make sure that $\varepsilon_2$ can take arbitrarily large values, so that $Y_2 > r$ with positive probability), $\boldsymbol Y$ has two different projections on the set $\mathcal{Q}$ with positive probability.}
	\end{example}
	
	\begin{algorithm}\label{alg1} 
		\KwData{$\boldsymbol z, \boldsymbol X_1,\ldots, \boldsymbol X_n$}
		\KwResult{out = 1 denotes $\boldsymbol z \in \Qc$, out = 0 denotes $\boldsymbol z \notin \Qc$ }
		$i=1$\;
		$\mathrm{out} = 1$\;
		\While{$i \leq n$ and $\mathrm{out} = 1$}{
			set $S = \{j \in [n]: z_j < z_i\}$\;
			\eIf{$\boldsymbol X_i \in \cv^\dagger(\{\boldsymbol X_j: j \in S \}) $}{
				$\mathrm{out} = 0$\;
			}{
				$i=i+1$\;
			}
		}
		\caption{Checking whether a given point $\boldsymbol z$ belongs to $\Qc$.}
	\end{algorithm}
	We now use the characterization of $\Qc$ in Lemma~\ref{lem:primal} to construct an algorithm to  check if a given point in $\R^n$ is in the feasible region. Algorithm~\ref{alg1} below determines whether a set of $n$ real values are realizations of a quasiconvex and decreasing function on the data points. It may seem at first that, in order to apply Lemma \ref{lem:primal} for this purpose, we need to go through each of the $n$ data points $X_1,\ldots,X_n$ and for each of the data points go through the $2^n$ subsets $S$ of $[n]$ and pull out all cases such that $\boldsymbol X_i \in \cv^\dagger\left(\{\boldsymbol X_j: j \in S\}\right)$ to check whether $z_i \leq \max_{j \in S} z_j$ in each of these cases. In the following algorithm, we show that this is not the case. In fact, we need to check \textit{only} $n$ subsets of $[n]$; see step 4 of Algorithm~\ref{alg1}. 

	A short proof of the validity of Algorithm~\ref{alg1} is given in Section~\ref{sec:proof:alg1} of the supplement. The \textit{if} statements in Algorithm \ref{alg1} involves checking the condition whether a given point $\boldsymbol p \in \mathbb{R}^n$ belongs to the upper orthant of the convex hull of some other points $\boldsymbol p_1,\ldots, \boldsymbol p_m \in \mathbb{R}^n$. This can be done efficiently by checking whether the following linear program (LP) has a feasible solution:
	
	\begin{mini}|l|
		{\boldsymbol{\lambda}, \boldsymbol v}{0}{}{} 
		\label{chullcheck}
		\addConstraint{\lambda_1,\ldots,\lambda_m}{\geq 0,\hspace{0.2cm}}{\boldsymbol v }{\in \mathbf{0}_d^\dagger,\hspace{0.2cm}}{\sum_{i=1}^m \lambda_i}{=1,\hspace{0.2cm}}{\sum_{i=1}^m \lambda_i \boldsymbol p_i}{= \boldsymbol p- \boldsymbol v.}
	\end{mini}
	
	\noindent where $\boldsymbol{\lambda} := (\lambda_1,\ldots,\lambda_m)$. Thus, Algorithm \ref{alg1} has a complexity that is linear in the sample size $n$, modulo performing the $O(n)$-many linear programs \eqref{chullcheck}, and hence, is computationally efficient. One can alternatively use built-in software functions to check whether a multivariate point belongs to the convex hull of others, which will likely make the process even more efficient. {\cln See \cite{chazelle1993optimal} for a deterministic algorithm for computing the convex hull of $n$ points in $\mathbb{R}^d$ which has computational complexity $O(n \log n + n^{\lfloor d/2\rfloor})$. }
	

	\subsection{Secondary characterization and computation of the LSE}\label{sec:dual}
	Although Lemma \ref{lem:primal} can be used to (efficiently) check if a vector is a feasible solution for the program in \eqref{opt}, this  characterization of $\Qc$ is not computationally amenable to be used as a constraint in the quadratic program in~\eqref{opt}. With this purpose in mind, we give a secondary characterization of $\Qc$. In this section, we will reduce \eqref{opt} to a mixed-integer quadratic optimization (MIQO) problem. 
	\begin{mylem}[Secondary characterization]\label{dual}
		A vector $\boldsymbol z \in \Qc$ if and only if there exist vectors $\bx_1,\ldots,\bx_n \in \mathbf{0}_d^\dagger$ such that 
		\[ \bx_j^\top(\boldsymbol X_i - \boldsymbol X_j) > 0 \text{ for every }i, j \text{ such that } z_i < z_j.\]
	\end{mylem}
	By Lemma \ref{dual}, $\boldsymbol z \in \Qc$ if and only if the following LP (with variables $\bx_1,\ldots,\bx_n$) has a feasible solution:
	\begin{equation}\label{lpdual1}
		\min~0\quad\textrm{subject to}\;\; \bx_1,\ldots,\bx_n \in \mathbf{0}_d^\dagger \text{ and }\bx_j^\top(\boldsymbol X_i - \boldsymbol X_j) >0,~\text{whenever}~z_i < z_j.
	\end{equation}
	Thus, Lemma \ref{dual} enables us to rewrite the  quadratic optimization problem in~\eqref{opt}:
	
	
	\begin{mini}|l|
		{\boldsymbol z, \boldsymbol{\Xi}}{\sum_{k=1}^n (Y_k - z_k)^2}{\label{lpdual}}{} 
		\addConstraint{\bx_1,\ldots,\bx_n}{\in \mathbf{0}_d^\dagger}{}
		\addConstraint{\bx_j^\top(\boldsymbol X_i - \boldsymbol X_j)}{> 0,\quad}{\text{for every} ~(i, j) \text{ such that } z_i < z_j,}
	\end{mini}

	\noindent where $\boldsymbol z = (z_1,\ldots,z_n)$ and $\boldsymbol{\Xi}:= (\bx_1,\ldots,\bx_n)$. We would  like to emphasize that, the set of constraints also depends on $z_1,\ldots, z_n.$ The optimization problem \eqref{lpdual} cannot be solved in its exact form because of the presence of implication constraints that include the variables of optimization (i.e., $z_1,\ldots, z_n$). However, the implication constraint
	\begin{equation}\label{eq:implocat_const}
		z_i < z_j \implies \bx_j^\top(\boldsymbol X_i - \boldsymbol X_j) > 0,
	\end{equation} in~\eqref{lpdual}, can  easily be framed as the following logical constraint
	\begin{equation}\label{eq:or_const}
		z_j - z_i \leq 0 \quad \texttt{or}\quad \bx_j^\top(\boldsymbol X_i - \boldsymbol X_j) > 0.
	\end{equation} Now note that, the \texttt{or} constraint in~\eqref{eq:or_const} can be converted into a standard constraint  by introducing binary variables $u_{ij}$. To elaborate, let us consider the following logical constraints:
	\begin{align}\label{eq:bigM}
		\begin{split}
			z_j - z_i &\leq M u_{ij},\\
			\bx_j^\top(\boldsymbol X_i - \boldsymbol X_j)&> M(u_{ij}-1),
		\end{split}
	\end{align}
	where $u_{ij} \in \{0,1\}$ and $M $ is an arbitrarily large number. If $u_{ij}=0$, then the first constraint in~\eqref{eq:bigM} reads $z_j - z_i \leq 0$ and the second constraint becomes essentially unconstrained, since $M$ is large. On the other hand, if $u_{ij}=1$, then the first constraint in \eqref{eq:bigM} becomes essentially unconstrained, while the second constraint reads $\bx_j^\top(\boldsymbol X_i - \boldsymbol X_j) > 0$. The above discussion is formalized in Lemma \ref{miqeqv} below and  proved in Section~\ref{sec:proof:miqeqv} of the supplement. 
	
	\begin{mylem}\label{miqeqv}
		Let \begin{align}\label{eq:R_M_def}
			\begin{split}
				\Rc_M := \Big\{\big(\boldsymbol z^\top,\bx_1^\top,&\ldots,\bx_n^\top,( (u_{ij}))_{i\neq j}\big) \in \mathbb{R}^n \times [0,\infty)^{nd}\times \{0,1\}^{n^2-n}:\\
				& z_j - z_i \leq Mu_{ij},~\bx_j^\top (\boldsymbol X_j - \boldsymbol X_i) > M(u_{ij}-1)~\forall ~i\neq j \in [n] \Big\}
			\end{split}
		\end{align} and let $\Pi_n$ denote the projection function onto the first $n$ coordinates of a vector. Then, $\Pi_n(\Rc_M) \uparrow \Qc$ as $M \rightarrow \infty$. In fact, there exists $M_0 \geq 1$ such that $\Pi_n(\Rc_M) = \Qc$ for all $M > M_0$. Finally, the minimizer of~\eqref{miqop} matches $\widehat{\boldsymbol\theta}$  (the minimizer of~\eqref{opt}) for large enough $M.$
	\end{mylem}
	
	Thus the optimization problem~\eqref{opt} and~\eqref{lpdual}  can be framed as the following mixed-integer quadratic program:
	\begin{mini}|l|
		{\boldsymbol z, \boldsymbol{\Xi},\mathbf{u}}{\sum_{k=1}^n (Y_k - z_k)^2}{}{} 
		\label{miqop}
		\addConstraint{z_j - z_i}{\leq M u_{ij},}{\forall~ i\neq j \in [n]}
		\addConstraint{\bx_j^\top(\boldsymbol X_i - \boldsymbol X_j)}{> M(u_{ij}-1),\quad}{\forall~ i\neq j \in [n]}
		\addConstraint{u_{i,j}}{\in \{0,1\},\quad}{\forall~ i\neq j \in [n]}
		\addConstraint{\bx_1,\ldots,\bx_n}{\in \mathbf{0}_d^\dagger,}{}
	\end{mini}
	\noindent where $\boldsymbol z := (z_1,\ldots,z_n)$, $\boldsymbol{\Xi} := (\bx_1,\ldots,\bx_n)$ and $\mathbf{u} := ((u_{ij}))_{1\leq i \neq j \leq n}$. {\cln The above MIQO is implemented in the \texttt{R} package \texttt{QuasiLSE}~\citep{QuasiLSE}; with a slight computational modification to account for the strict inequality in the second constraint above.}
	
	It is  important to note that there is a strict inequality in the constraint $\bx_j^\top(\boldsymbol X_j - \boldsymbol X_i) > M(u_{ij}-1)$ in \eqref{miqop}. It would be incorrect to use `$\geq$' instead of `$>$', since in that case, $z_i = Y_i$, $\bx_i = \mathbf{0}_d$ for all $i$, and $u_{ij}=1$ for all $i\neq j$ would be a feasible solution of \eqref{miqop}, which makes the optimal objective $0$. However, in case a closed constraint formulation is necessary, one can take a very small positive quantity $\epsilon$, and work with the slightly stricter (but closed) constraints $\bx_j^\top(\boldsymbol X_j - \boldsymbol X_i) \geq M(u_{ij}-1) + \epsilon$. As long as $\epsilon > 0$, smaller the value of $\epsilon$ one takes, closer are the optimum objective values of the new and the original problems. This is what we do in our implementation of the above MIQO in the R package \texttt{QuasiLSE}~\citep{QuasiLSE}.
	
	
	\subsection{A note on the quasiconvex and increasing LSE}\label{rem:QuasiInc_existence}
	Suppose now that $\varphi$ is known to be quasiconvex and \textit{increasing}. All the above discussions and results will go through  with only minor modifications. Let $$\Qc':= \{(\psi(\boldsymbol X_1),\ldots,\psi(\boldsymbol X_n)) \in \mathbb{R}^n : \psi~\textrm{is quasiconvex and increasing}\}.$$ 
	In this case, we define the set $\mathcal{L}'(\mathcal{X})$ as:
	\begin{equation}\label{eq:L_defprime}
		\mathcal{L}'(\mathcal{X})\coloneqq\big\{(i,S): i \in [n],\, S \subseteq [n], \text{ and }\boldsymbol X_i \in \cv_\dagger\big(\{\boldsymbol X_j: j \in S\}\big) \big\},
	\end{equation}
	where for any $\boldsymbol X\in \R^d$ and $A\subset \R^d$, $\boldsymbol X_\dagger$ and $A_\dagger$ denote their lower orthants and  are defined as
	\begin{equation}\label{eq:lower_orthant}
		\boldsymbol X_\dagger := \{\boldsymbol Y \in \rd: \boldsymbol Y \lle \boldsymbol X\}\qquad\text{and}\qquad A_\dagger:= \cup_{\boldsymbol X \in A} \boldsymbol X_\dagger,   
	\end{equation}respectively.
	The primary characterization of the set $\mathcal{Q}'$ becomes
	\begin{equation}\label{eq:primalQcpr}
		\mathcal{Q}' = \Big\{\boldsymbol z \in \mathbb{R}^n: z_i \leq \max_{j \in S} z_j~\textrm{for all}~(i,S) \in \mathcal{L}'(\mathcal{X})\Big\}.
	\end{equation}
	Needless to say that the only change in Algorithm \ref{alg1} for checking whether a given point $\boldsymbol z \in \Qc'$, would be to replace the upper orthants of the convex hulls by their lower orthants. For the secondary characterization of $\Qc'$, the only change in the statement of Lemma \ref{dual} would be $\bx_1,\ldots,\bx_n \lle \mathbf{0}_d$.
	
	{\cln
		\section{The quasiconvex regression problem}\label{sec:qonly}
		It is natural to ask what happens if the function $\varphi$ in \eqref{mainmodel} is assumed to be quasiconvex only (not necessarily decreasing or increasing). The LSE in this scenario is:
		\begin{equation}\label{qconvonly}
			\widetilde{\bt} \in \argmin1_{\boldsymbol z \in \widetilde{\Qc}}  \hspace {.2 cm}\sum_{k=1}^n {(Y_k - z_k)^2 }
		\end{equation}
		where $\widetilde{\Qc} := \{(\psi(\boldsymbol X_1), \ldots ,\psi(\boldsymbol X_n))\, |\, \psi:\mathbb{R}^d\to \mathbb{R}~\textrm{is quasiconvex}\}.$ 
		\begin{mythm}[Existence and uniqueness]\label{thm:exun11}
			The optimization problem \eqref{qconvonly} has a minimizer $\widetilde{\bt}$ in $\widetilde{\Qc}$. Moreover, if $\,Y\,$has a density with respect to the Lebesgue measure on $\mathbb{R}$, then $\widetilde{\bt}$ is unique  with probability $1$.
		\end{mythm}

		It turns out that the primary and secondary characterizations of the space $\widetilde{\Qc}$ are very similar to those of $\Qc$. If we define $\widetilde{\mathcal{L}}(\mathcal{X})$ as the set of all tuples $(i,S)$ with $i \in [n]$ and $S\subseteq [n]$, such that $\boldsymbol X_i \in \cv(\{\boldsymbol X_j : j \in S\})$, then we have the following primary characterization of $\widetilde{\Qc}$:
		\begin{mylem}[Primary characterization]\label{primalqconly}
			\[\widetilde{\Qc} = \left\{\boldsymbol z \in \mathbb{R}^n: z_i \leq \max_{j \in S} z_j~\textrm{for all}~(i,S) \in \widetilde{\mathcal{L}}(\mathcal{X})\right\}.\]
		\end{mylem}
		For the secondary characterization of $\widetilde{\Qc}$, all that we need to do, is drop the nonnegativity assumptions on the vectors $\bx_1,\ldots,\bx_n$ from the statement of Lemma \ref{dual}. Formally, we have
		\begin{mylem}[Secondary characterization]\label{dualqconly}
			$\boldsymbol z \in \widetilde{\Qc}$ if and only if there exist vectors $\bx_1,\ldots,\bx_n \in \mathbb{R}^d$ such that 
			\[ \bx_j^\top(\boldsymbol X_i - \boldsymbol X_j) > 0 \text{ for every }i, j \text{ such that } z_i < z_j.\]
		\end{mylem} 
		The proofs of Theorem \ref{thm:exun11}, Lemma \ref{primalqconly}, and Lemma \ref{dualqconly} are identical to the proofs of Theorem \ref{thm:exun}, Lemma \ref{lem:primal}, and Lemma \ref{dual},  respectively, so we skip them.
		Further, the optimization problem~\eqref{qconvonly} can also be framed as the mixed-integer quadratic program similar to~\eqref{miqop}, the \textit{only} change being that now  $\bx_1,\ldots,\bx_n$ are unconstrained. The code to compute $\tilde{\boldsymbol \theta}$ is made available in the \texttt{R} package \texttt{QuasiLSE} \citep{QuasiLSE}.}
	
	\section{Asymptotic properties of the  LSE}\label{sec:consist}
	The LSEs obtained from problem \eqref{opt} or~\eqref{qconvonly} are almost surely unique  but is not consistent without any restriction on the design $\mathcal{X}$. Probably the simplest example is to take $\mathcal{X} \subset \{\boldsymbol x \in \mathbb{R}^d:  \boldsymbol x \lle \mathbf{0}_d~\textrm{and}~\|\boldsymbol x\| = 1\}$ and assume that all elements of $\mathcal{X}$ are distinct. In this case, one can verify that $(i,S) \notin \mathcal{L}(\mathcal{X})$ if $i \notin S$, and hence, by the primary characterization of $\Qc$  in Lemma \ref{lem:primal}, $\Qc = \mathbb{R}^n$. The problem \eqref{opt} is thus unconstrained,  the minimum is attained at $\widehat{\boldsymbol \theta} = \boldsymbol Y$, and the estimator is not consistent. The above example shows the need to impose additional structure on the design points in order to have consistency.  The optimization problem~\eqref{qconvonly} has a similar property.

	{\cln We will now provide risk upper bounds for the two LSEs under the standard nonparametric regression setup described in~\eqref{mainmodel}. We stress that we do not assume independence between $\varepsilon$ and $\boldsymbol X$. Let $P_{\boldsymbol X}$ denote the distribution of $\boldsymbol X$ and let $P$ denote the joint distribution of $\boldsymbol X$ and $Y$.
		Let $\mathcal{H}_{d,\Gamma}$ be any arbitrary subset of the set of all quasiconvex functions on $\mathbb{R}^d$ bounded by $\Gamma$. For example, $\mathcal{H}_{d,\Gamma}$ may denote the set of all quasiconvex and decreasing functions on $\mathbb{R}^d$ bounded by $\Gamma$ or it may denote the set of all quasiconvex functions on $\mathbb{R}^d$ bounded by $\Gamma$ (with out any additional monotonicity assumption). The least squares estimate (LSE) of $\varphi$ in the class $\mathcal{H}_{d,\Gamma}$ is defined as:
		$$\widehat{\varphi} := \argmin1_{g \in \mathcal{H}_{d,\Gamma}} \sum_{i=1}^n \left(Y_i - g(\boldsymbol X_i)\right)^2,$$ 
		where $\hat\varphi$ is piecewise constant function defined as in~\eqref{eq:thetatophi}.
		The following result, proved in Section \ref{sub:proof_of_theorem_mainthm} of the supplement, provides an upper bound on 
		$$\mathcal{R}_{L_2^2(P)}(\widehat{\varphi},\varphi) :=  \int_{\mathbb{R}^d}\left(\widehat{\varphi}(\boldsymbol x) - \varphi(\boldsymbol x)\right)^2~\mathrm{d} P_{\boldsymbol X}(\boldsymbol x),$$  the $L_2^2(P)$ risk of the LSE $\widehat{\varphi}$ in estimating $\varphi.$
		
		\begin{mythm}\label{mainthm}
			Assume that $d \ge 2$ and  $(\varepsilon_1, \boldsymbol X_1),\ldots,(\varepsilon_n, \boldsymbol X_n)$ are i.i.d. Suppose that $\varepsilon$ has a continuous density with respect to the Lebesgue measure on $\R.$ Let $f$ denote the density of $\boldsymbol X$ with respect to the Lebesgue measure on $\R^d$, and suppose that
			\begin{equation}\label{eq:Tail_cond_main}
				f(\boldsymbol x) \le C_f (1+ \|\boldsymbol x\|)^{-r} \quad \text{ for some }\quad r> 
				(d^2+ 1)/(d-1).
			\end{equation} 
			Further, suppose $\|\E(\varepsilon|\boldsymbol X)\|_{\infty}$, $\mathrm{var}(\varepsilon)$, and $\|\varphi\|_{\infty}$ are finite. Then
			\begin{equation}\label{eq:Main_result}
				\E \mathcal{R}_{L_2^2(P)}(\widehat{\varphi},\varphi) \le  \mathcal{R}_{L_2^2(P)}({\varphi},\mathcal{H}_{d,\Gamma})+  C_d C_f \Gamma (\Gamma+ C_\varepsilon+C_\varphi) \times \begin{cases}
					n^{-1/2}  &\text{ when } d = 2,\\
					n^{-1/2} \log n  &\text{ when } d = 3,\\
					n^{-2/(d+1)} &\text{ when } d\ge 4,
				\end{cases}
			\end{equation}
			where \[\mathcal{R}_{L_2^2(P)}({\varphi},\mathcal{H}_{d,\Gamma}) = \inf_{g \in \mathcal{H}_{d,\Gamma}} \mathbb{E}\left({\varphi}(\boldsymbol X) - g(\boldsymbol X)\right)^2,\] 
			and $C_\varepsilon$, $C_\varphi$, $C_f$, and $C_d$ depend only on $\|\E(\varepsilon|X)\|_{\infty} +\mathrm{var}(\varepsilon)$, $\|\varphi\|_{\infty}$, $f,$ and  $d,$ respectively.
		\end{mythm}
		
		\textcolor{black}{The risk bound in~\eqref{eq:Main_result} is finite sample. A bound of this type is often called an oracle inequality and describes the ``bias-variance'' or the ``approximation-estimation'' trade-off for the shape constrained LSE when estimating $\varphi$. If the model is well specified, i.e., $\varphi\in \mathcal{H}_{d,\Gamma},$ then $\mathcal{R}_{L_2^2(P)}({\varphi},\mathcal{H}_{d,\Gamma}) =0$. Also, the ``bias" ("approximation'') in~\eqref{eq:Main_result} is zero and the ``variance'' (``estimation'') term determines the estimation error of $\widehat\varphi.$  The leading constant for the ``bias'' term in~\eqref{eq:Main_result} is 1. Such oracle inequalities have been called ``exact'' or ``sharp'' in the literature;~\cite{lecue2012general,bellec}. Sharp oracle inequalities are more ``valuable'' from the statistical point of view as they can be used to provide both prediction and estimation risk bounds~\citep[Chapter 3.4]{lecue2012general}. Also, note that although the risk bound is in expectation, using standard concentration inequalities it can be extended to a high probability bound for $\mathcal{R}_{L_2^2(P)}(\widehat{\varphi},\varphi)$ as well.}
		
		\textcolor{black}{Theorem \ref{mainthm} holds for any function that lies in $\mathcal{H}_{d,\Gamma}$ and interpolates the points $\{(\boldsymbol X_i, \hat{\varphi}(\boldsymbol X_i))_{i=1}^n\}$. We focus on the piecewise constant interpolation $\hat\varphi$ in this paper, as it is the only computable/practical interpolation of $\{(\boldsymbol X_i, \hat{\varphi}(\boldsymbol X_i))_{i=1}^n\}$ that is guaranteed to maintain quasiconvexity; see Remark~\ref{rem:thetatophi}.}

		\textcolor{black}{If $\mathcal{H}_{d,\Gamma}$ is the class of quasiconvex and isotonic functions bounded by $\Gamma$, and $\varphi\in\mathcal{H}_{d,\Gamma}$ (i.e., the model is well specified), then Theorem~\ref{mainthm} implies that incorporating the additional constraint of quasiconvexity in the LSE leads to a significantly faster rate of convergence. The quasiconvex and isotonic LSE converges at a $n^{-2/(d+1)}$ rate with respect to $L_2^2(P)$ risk, while the (only) isotonic LSE converges at only a $n^{-1/d}$ rate, a significantly slower rate;~\citet[Theorem 3.6]{han}. }
		
		\textcolor{black}{The theoretical results in this section are given in the context of bounded regression functions. The bound $(\Gamma)$  on functions in $\mathcal{H}_{d,\Gamma}$ can be thought of as a tuning parameter. However, in real-world applications  such a bound is often known; e.g., in the context of the Japanese plywood production data presented in Section \ref{sec:real_data}, there are natural upper-bounds on the maximum possible production value of a factory. We would also like to point out that the characterizing results in Section \ref{sec:qmest} can be easily modified to apply to the bounded LSE setting of this section, by simply adding an additional linear constraint $\|\boldsymbol z\|_\infty \le \Gamma$ in the MIQO \ref{lpdual}.}
		
		\textcolor{black}{Although the bound ($\Gamma$) on the regression function is known beforehand in many real examples, a natural question is what should one do when there is no known estimate of the bound $\Gamma$. In that case, we suggest minimizing the square error loss over $\mathcal{H}_{d,\infty}$. At first glance it might seem that in this scenario, the bound in~\eqref{eq:Main_result} leads to a  trivial upper bound, but that is not the case.  In Lemma~\ref{simplequasi} of the supplementary, we show that 
			\[ \argmin1_{g \in \mathcal{H}_{d,\infty}} \sum_{i=1}^n \left(Y_i - g(\boldsymbol X_i)\right)^2 \equiv  \argmin1_{g \in \mathcal{H}_{d,\max_{i\in [n]} |Y_i| }} \sum_{i=1}^n \left(Y_i - g(\boldsymbol X_i)\right)^2.\] Thus in case there is no known bound on $\Gamma$, we {can} find the LSE over $\mathcal{H}_{d,\infty}$ by fixing $\Gamma=\max_{i\in [n]} |Y_i|$. Moreover, if there exist finite $q\ge 2$ and $\K_q < \infty$ such that $\E(|\varepsilon|^q)\le \K_q^q$, then it is easy to see that $\Gamma\le C_{K_q} n^{1/q} + \|\varphi\|_{\infty}$ with high probability (w.h.p). Thus, we have
			\[ \argmin1_{g \in \mathcal{H}_{d,\infty}} \sum_{i=1}^n \left(Y_i - g(\boldsymbol X_i)\right)^2 \equiv  \argmin1_{g \in \mathcal{H}_{d, (C_{K_q} n^{1/q} + \|\varphi\|_{\infty}) }} \sum_{i=1}^n \left(Y_i - g(\boldsymbol X_i)\right)^2\qquad \text{w.h.p.}\]
			Hence, Theorem~\ref{mainthm} implies that
			\begin{equation}\label{eq:Main_result1}
				\E\mathcal{R}_{L_2^2(P)}(\widehat{\varphi},\varphi) \le  \mathcal{R}_{L_2^2(P)}({\varphi},\mathcal{H}_{d, C n^{1/q}})+  C n^{2/q} \begin{cases}
					n^{-1/2}  &\text{ when } d = 2,\\
					n^{-1/2} \log n  &\text{ when } d = 3,\\
					n^{-2/(d+1)} &\text{ when } d\ge 4,
				\end{cases}
			\end{equation}
			with high probability for sufficiently large $n$. If $\varepsilon$ is sub-Gaussian or sub-exponential, then $\Gamma \le  C \log n$ with high probability and hence $n^{1/q}$ in \eqref{eq:Main_result1} can be replaced by $\log n.$} 
		
		\begin{myrem}[Assumptions in Theorem~\ref{mainthm}]\label{Assump} The assumptions in Theorem~\ref{mainthm} are quite mild. \textcolor{black}{The covariates are not required  to be bounded; common continuous distributions such as sub-Gaussian or log-concave distributions satisfy~\eqref{eq:Tail_cond_main} for every $d\ge 2$. The only assumption on $\varphi$ (the true conditional mean) is that it is bounded, i.e., $\|\varphi\|_{\infty} < \infty$. For example, Theorem~\ref{mainthm} allows for mis-specification  and does not require $\varphi$  to be quasiconvex and/or monotone. The assumptions also allow for heteroscedastic errors, i.e., errors that can depend on the covariates arbitrarily. This is a significant improvement over the assumption of independence between $\varepsilon$ and $\boldsymbol X$  in most of the shape constrained literature. Theorem~\ref{mainthm} requires the errors to have only 2 finite moments as opposed to sub-Gaussianity of the error distributions required in most works; \cite{zhang}, \cite{mendelson2016upper}, \cite{han2017sharp,han2021set}, and \cite{kuchibhotla2019least} being a few notable exceptions.}
		\end{myrem}

		
		\subsection{Minimax Optimality} 
		\label{sub:MinimaxOptimality}
		
		In this section, we will show that the bound in Theorem \ref{mainthm} is tight \textcolor{black}{when $d\ge 4$}, and is achieved, for example, when the underlying distribution $P_{\boldsymbol X}$ is uniform on the $d$-dimensional Euclidean ball  $B_d(0, 1)$. We will do this by comparing the quasiconvex regression to that of the bounded convex  regression. 
		
		\cite{han2016multivariate} proved the following lower bound for the bounded convex regression problem when $P_{\boldsymbol X}$ is the uniform measure on $B_d := B_d(0, 1)$ and $d\ge 4:$
		\begin{equation}\label{convr}
			\inf_{\tilde{\varphi}} \sup_{\varphi\in \mathcal{C}_{d,\Gamma}} \mathcal{R}_{L_2^2(\mathrm{Unif}(B_d))}(\tilde{\varphi}, \varphi) = \Theta_{d, \Gamma}\left(n^{-{2}/{(d+1)}}\right),
		\end{equation}
		where $\mathcal{C}_{d,\Gamma}$ denotes the set of convex functions on $B_d$, bounded by $\Gamma$ and the infimum is over all estimators of $\varphi$. Let $\mathcal{G}_{d,\Gamma}$ denote the set of all quasiconvex functions on $B_d$, bounded by $\Gamma$.  Since $\mathcal{G}_{d,\Gamma} \supset \mathcal{C}_{d,\Gamma}$,  Theorem~\ref{mainthm} and \eqref{convr} implies:
		\begin{proposition}\label{prop1}
			Let $\mathcal{G}_{d,\Gamma}$ denote the set of all quasiconvex functions on $B_d$, bounded by $\Gamma$. Then for $d\ge 4$,
			$$\inf_{\widetilde{\varphi}} \sup_{\varphi\in \mathcal{G}_{d,\Gamma}}\mathcal{R}_{L_2^2(\mathrm{Unif}(B_d))}(\widetilde{\varphi},  \varphi) = \Omega_{d, \Gamma}\left(n^{-{2}/{(d+1)}}\right).$$ Consequently, for every $d\ge 4$,
			$$\inf_{\widetilde{\varphi}}\sup_P \sup_{ \varphi\in \mathcal{G}_{d,\Gamma}} \mathcal{R}_{L_2^2(P)}(\widetilde{\varphi},  \varphi) = \Theta_{d, \Gamma}\left(n^{-{2}/{(d+1)}}\right),$$ 
			where the supremum is over all distributions $P$ that satisfy the assumption of Theorem~\ref{mainthm} and the infimum is over all estimators of $\varphi.$
		\end{proposition}
		The above result is remarkable because it shows that the quasiconvex  and convex regression problems have the same minimax rate when $d\ge 4$. In this case, even though quasiconvexity is a significantly weaker assumption than convexity, the rate of recovery is surprisingly the same under both of these assumptions.} 

	
	\section{Simulation study} 
	\label{sec:simulation_study}
	{\cln In this section, we illustrate the finite sample performance of the quasiconvex and increasing LSE using synthetic data. The most widely used estimator in the nonparametric regression setting of~\eqref{mainmodel} is the Nadaraya-Watson estimator. However, the kernel estimator is not guaranteed to be either quasiconvex or increasing.~\cite{chen2018shapeenforcing} propose a functional operator that can enforce quasiconcavity and monotonicity \textit{ex post} on any estimator when the domain of the covariates is a rectangle; see~\citet[Remark 2]{chen2018shapeenforcing}.\footnote{There are no such domain restrictions for the LSE proposed here. The \texttt{ChenEtAl} estimator is however, not as computationally expensive as the proposed LSE. } Just as in~\cite{chen2018shapeenforcing}, we use the Nadaraya-Watson estimator as the initial estimator and compute the shape enforced estimator that is both quasiconvex and increasing.  In this section, we compare the performance of the quasiconvex and monotone LSE with the: (1) Nadaraya-Watson estimator (\texttt{NW}); (2) shape enforced version of the Nadaraya-Watson estimator (\texttt{ChenEtAl}); (3) bivariate convex LSE (\texttt{Cvx}); (4) bivariate monotonic LSE (\texttt{Iso}); and (5) the penalized isotonic regression spline estimator proposed in~\cite{meyer2013simple} (\texttt{IsoPen}). The Nadaraya-Watson estimator requires a choice for the bandwidth parameter; we use the cross-validated choice for its bandwidth~\cite[Page 66]{li2007nonparametric}. For the penalty parameter for~\texttt{IsoPen}, we use the default choice in the R package~\texttt{isotonic.pen};~\cite{isotonicpen}. Following the discussion after Theorem~\ref{mainthm}, for the LSE,  we fix $\Gamma= \max_{i\in [n]} |Y_i|$. The \texttt{R} code for computing~\texttt{ChenEtAl} was kindly provided to us  via private communication by~Scott Kostyshak.  In the following two subsections, we consider two simulation settings: (1) well-specified setting i.e., where the conditional mean function is quasiconvex and increasing; and (2) mis-specified setting i.e., where the conditional mean function is increasing but \textit{not} quasiconvex.

		\subsection{Well-specified setting}\label{sub:uds} 
		\label{sub:well_specified_setting}
		We now describe the well-specified  regression setup. As a first step, we have $n$ i.i.d.~observations from the model 
		\begin{equation}\label{eq:sim_model}
			Y = \psi(\boldsymbol X) + \varepsilon,\quad \text{where}\quad \psi(\boldsymbol x) := \lfloor \|\boldsymbol x\|_2^2 \rfloor,\quad \varepsilon\sim N(0,\sigma^2), \quad \boldsymbol X\sim \text{Uniform }[0,1]^d.
		\end{equation} Note that the function $\psi$ is both increasing and quasiconvex, but it is not continuous and not convex. We use $\psi$ as the basis for all the functions considered in this section. For the first modification, we introduce a ``smoothness'' parameter $\xi$ which can vary between $0$ and $1$, with $\xi = 1$ denoting a completely smooth function, and $\xi=0$ recovering the piecewise constant function $\psi$. To be precise, we define a smoothing function $s_\xi: [0,1]\mapsto \mathbb{R}$ as:
		$$s_\xi(t) := \frac{t-(1-\xi)}{\xi} \mathbf{1}\{t \ge 1-\xi\}~.$$ The next step is to modify the function $\psi(\boldsymbol x) := \lfloor \|\boldsymbol x\|_2^2\rfloor$ by the following ``smoothed" version:
		\begin{equation}\label{eq:Smooth_step_function}
			\psi_\xi(\boldsymbol x) :=  \lfloor \|\boldsymbol x\|_2^2\rfloor + s_\xi\left(\|\boldsymbol x\|_2^2 - \lfloor \|\boldsymbol x\|_2^2\rfloor\right).
		\end{equation}
		Note that $$\psi_0 (\boldsymbol x) = \psi(\boldsymbol x) = \lfloor \|\boldsymbol x\|_2^2\rfloor\quad \textrm{and}\quad \psi_1(\boldsymbol x) = \|\boldsymbol x\|_2^2~.$$ Once again, all the functions $\{\psi_\xi\}_{\xi \in [0,1]}$ are increasing and quasiconvex. But only $\psi_1$ is convex. 
		\begin{figure}
			\begin{center}
				\includegraphics[width=.85\textwidth]{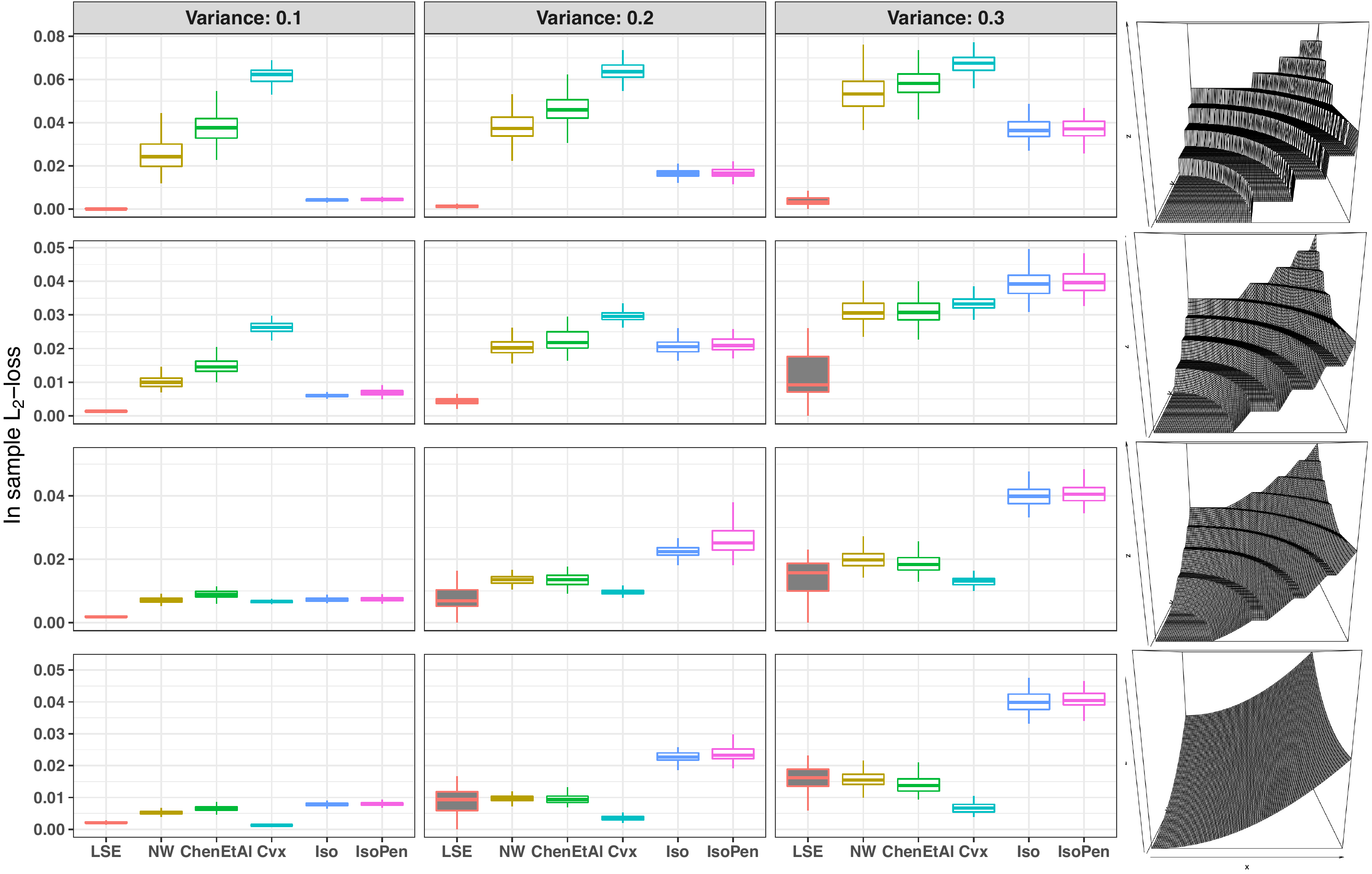}
			\end{center}
			\caption{Box plots comparing the performance of our quasiconvex and monotone estimator with other competing estimators in~\eqref{eq:sim_model} when $n=400$ and $d=2$. The smoothness parameter $\xi$ increases along $0.01, 0.34, 0.67, 1$ as one goes down the rows  and error variance increases along $0.1, 0.2, 0.3$ as one goes from left to right. The rightmost column plots  the true conditional mean $\psi_\xi$ as $\xi$ increases. The box plots summarize  results over 100 replications.  \label{fig:2dim}}
		\end{figure}

		In Figure \ref{fig:2dim}, we show box plots for $\sum_{i=1}^n (\tilde{\varphi}(\boldsymbol X_i)-\varphi(\boldsymbol X_i))^2$ (in sample $L_2$-loss) comparing the performance of our quasiconvex and monotone LSE  with the other four competing estimators when $d=2$ in~\eqref{eq:sim_model} with~\eqref{eq:Smooth_step_function}.  As we go from left to right, the noise variance  increases from $0.1$ to $0.3$ in increments of $0.1$. The smoothness parameter $\xi$ increases from $0.01$ to $1$ in increments of $0.33$ as we go from top to bottom in Figure~\ref{fig:2dim}. The sample size in each case is taken to be $400$, and the box plots are created over $100$ replications. In each of the settings, the proposed LSE performs significantly better than  monotonicity (only) based estimators (\texttt{Iso} and \texttt{IsoPen}). When the true conditional mean is convex, \texttt{Cvx} has the best performance (unsurprisingly). When the true conditional mean function is piecewise constant \textit{or} the noise variance is low, the LSE has much better performance when compared to shape enforced estimator~\texttt{ChenEtAl}. However when the true conditional mean function is smooth (bottom row) \textit{and} the noise variance is high then both the LSE and the shape enforced estimator~\texttt{ChenEtAl} have comparable performance. A similar relationship between the shape enforced operator based on rearrangement and isotonic LSE is observed in the case of univariate monotone regression \citep[Section 2.4]{chernozhukov2009improving}.

		\begin{figure}
			\begin{center}
				\includegraphics[width=.7\textwidth]{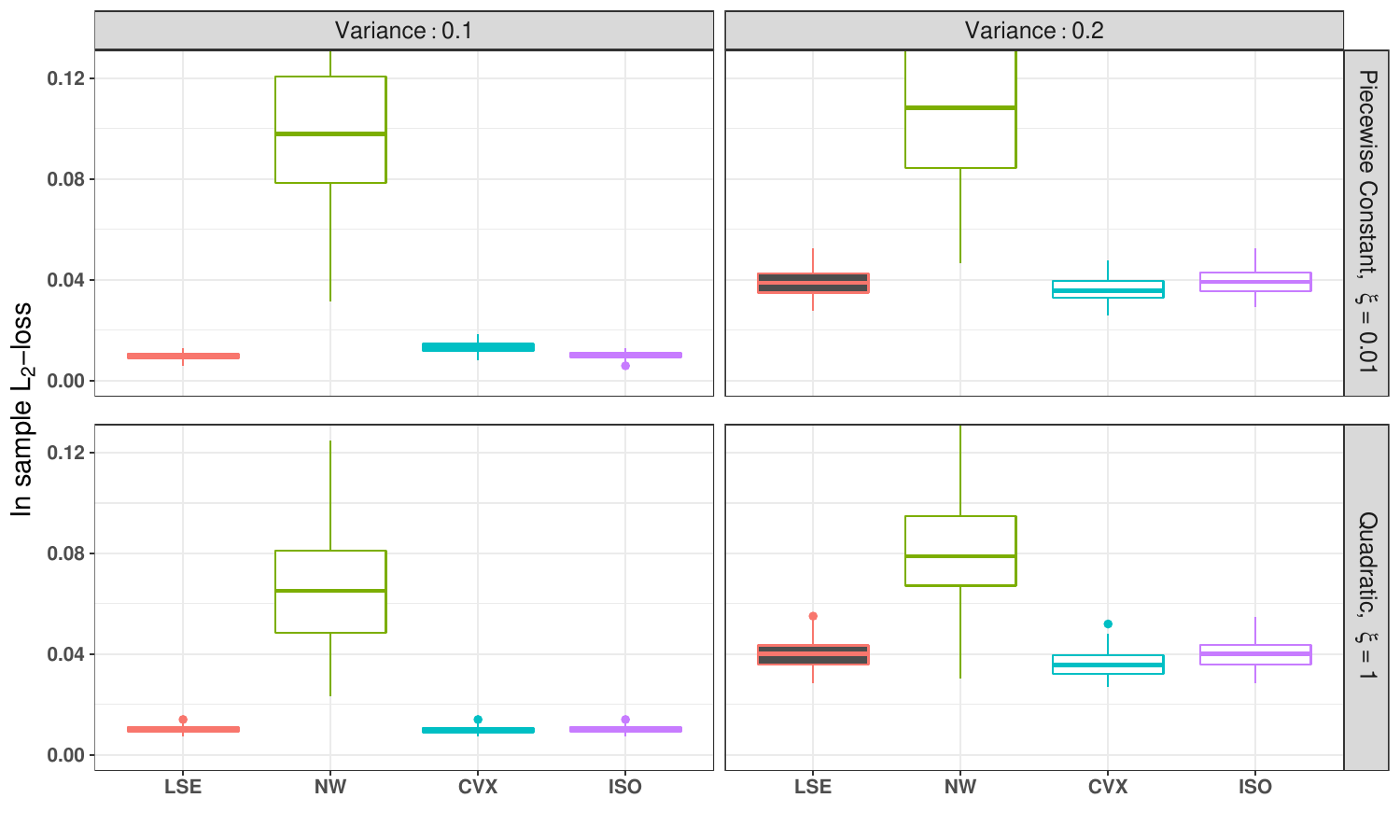}
			\end{center}
			\caption{Box plots comparing the in sample $L_2(P)$-loss of our proposed LSE with the 3 competing estimators when the true conditional mean function is as defined in~\eqref{eq:Smooth_step_function}, with $n=100$, and  $d=4$. The box plots summarize results over 100 replications. \label{fig:d4}}
		\end{figure}
		
		Figure \ref{fig:d4} deals with~\eqref{eq:sim_model} when $d=4$. It  compares the performance of our quasiconvex and monotone LSE with that of the convexity constrained LSE.\footnote{Figure~\ref{fig:d4} does not include~\texttt{ChenEtAl} and~\texttt{IsoPen} as we could not find any implementation for them when $d\ge 3.$} The plot provides numerical justification for  the optimality of  the quasiconvex LSE established in Section~\ref{sub:MinimaxOptimality}.  The sample size is taken to be $100$ and the error variances are allowed to be $0.1$ and $0.2$.
		In all of  the cases in Figure~\ref{fig:d4}, the proposed LSE performs well and its average $L_2$ error is close to that of the convex LSE. This is especially remarkable when $\xi=1$, as  then the true conditional mean function is convex and convex LSE is minimax optimal in this setting. This reaffirms the remarkable behavior of the quasiconvex LSE that it performs as well as the convex LSE when $d\ge 4$, even when the true conditional mean function is convex. 
		
		
		
		\subsection{Misspecified setting} 
		\label{sub:misspecified_setting}
		
		We also consider regression setup where the true mean is not quasiconvex. We do this by perturbing the functions $\psi_\xi$ (defined in~\eqref{eq:Smooth_step_function}) slightly, so that the resulting true conditional mean function is \textit{not} quasiconvex.  To be specific, in Figure~\ref{fig:mis-spec}, we consider the following perturbed version of $\psi_\xi$:
		\begin{equation}\label{misp}
			{\psi}^\dagger_\xi(\boldsymbol x) = 
			\begin{cases}
				\lfloor \|\boldsymbol x\|_2^2 \rfloor + 1 &\quad\text{if}~ \boldsymbol x\ge r(\boldsymbol x),\\
				\psi_\xi(\boldsymbol x) &\quad\text{otherwise},
			\end{cases}
		\end{equation}
		where 
		\[   
		r(\boldsymbol x) := 
		\begin{cases}
			\sqrt{\lfloor \|\boldsymbol x\|_2^2 \rfloor/2} &\quad\text{if} ~\|\boldsymbol x\|_2^2 \ge 1,\\
			\frac{1}{2\sqrt{2}}\left(\sqrt{\lceil \|\boldsymbol x\|_2^2 \rceil} + \sqrt{\lfloor \|\boldsymbol x\|_2^2\rfloor}\right) &\quad\text{otherwise} \\ 
		\end{cases}
		\]
		The perturbed function ${\psi}^\dagger_\xi$ introduces small ``bumps'' in each step of the piecewise constant function $\psi_\varepsilon$ in such a way, that the function is no longer quasiconvex (it continues to be monotone); see the rightmost panel in Figure~\ref{fig:mis-spec}. In each of the $100$ replications the sample size is set to be $400$.  As expected, the two monotonicity based estimators outperform all the other estimators in this setting. The proposed quasiconvex and monotone LSE performs reasonably well when compared to the shape enforced estimator \texttt{ChenEtAl} and the convex LSE. The Nadaraya-Watson estimator performs better than both (but worse than the monotonicity based estimators), since it does not assume any shape constraint, and hence is not affected by misspecification from quasiconvexity. 
		

	}

	
	\begin{figure}
		\begin{center}
			\includegraphics[width=.8\textwidth]{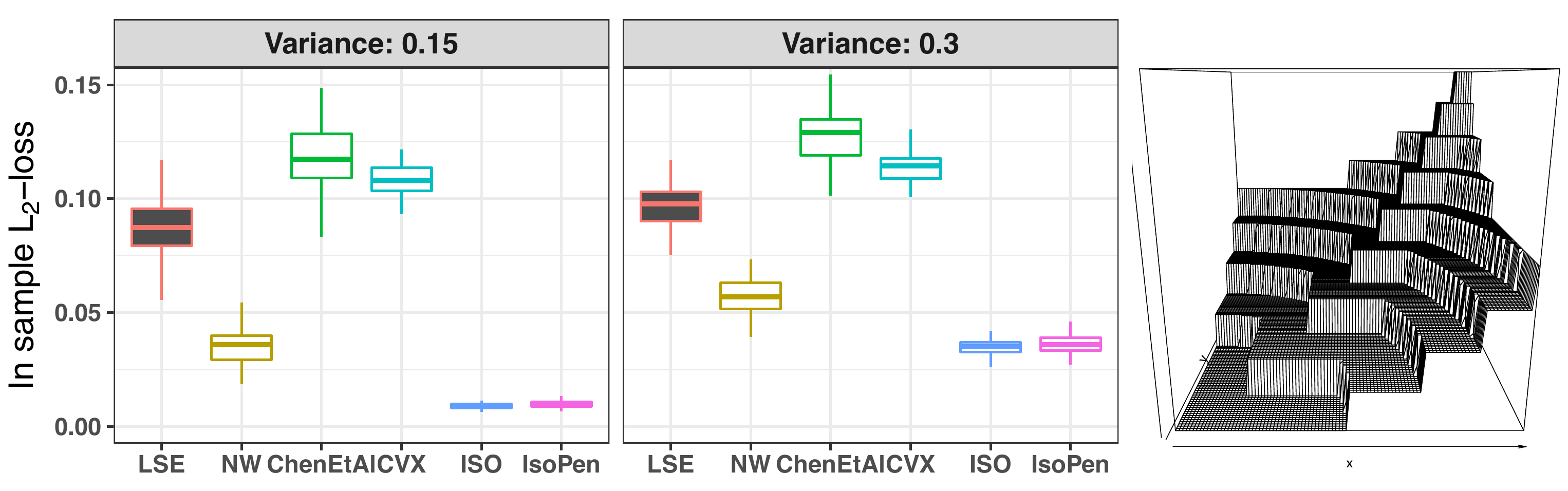}
		\end{center}
		\caption{Box plots comparing the performance of the proposed LSE with other competing estimators in the mis-specified setting \eqref{misp}. The sample size is set at $400$ and the number of replications is taken to be 100. \label{fig:mis-spec}}
	\end{figure}

	\begin{myrem}\label{rem:Tilting}
		{\cln  Another possible competitor may be tilting based estimators which are viable and important estimators when enforcing various shape constraints. However, currently tilting based estimators can only enforce monotonicity and convex  shapes.  \cite{du2013nonparametric} discuss the enforcement of quasiconcavity only in passing and without any technical details. The codes to compute tilting estimator under monotonicity or convexity were kindly provided to us by Jeffery Racine. However, we did not include them in our simulation due to various technical problems faced by the  R package~\texttt{quadprog}. }
	\end{myrem}

	\subsection{Numerical Studies under General Covariate Distributions}
\begin{figure}
		\begin{center}
			\includegraphics[width=\textwidth]{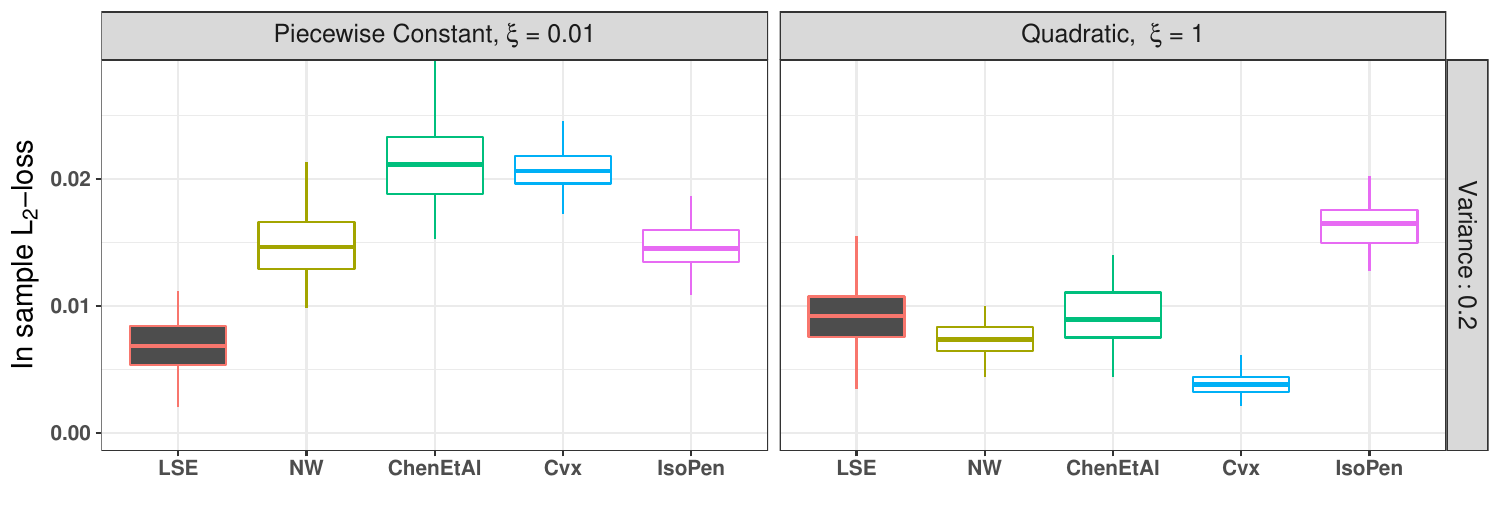}
		\end{center}
		\caption{\textcolor{black}{Box plots comparing the in sample $L_2(P)$-loss of our proposed LSE with the 4 competing estimators when the true conditional mean function is as defined in~\eqref{eq:Smooth_step_function} and covariate distribution as defined in~\eqref{eq:nonuniform_design}, with $n=400$, and  $d=2$. The box plots summarize results over 100 replications. \label{fig:d2_nonunif}}}
	\end{figure}

	\textcolor{black}{In Section \ref{sub:uds}, we assumed that the covariates are distributed Uniformly. To better understand the behavior of the LSE under a more complex covariate distribution, we consider:
		\begin{equation}\label{eq:nonuniform_design}
			\boldsymbol X :=(\psi \cos \eta, \psi \sin \eta)\text{, where}~\psi \sim \text{Unif}\,[0,2.5]\text{ and }\eta \sim \text{Unif}\,[0.05,\pi/2 - 0.05]
		\end{equation}
		This above distribution was used in \cite{yagi2017SShape} and~\cite{olesen2014maintaining} to better replicate real data distribution observed in practice. The conditional mean functions considered is the same as the ones in Section \ref{sub:uds}, and we consider the same 4 estimators as in Section \ref{sub:uds} and compare their performance. The results are summarized in Figure~\ref{fig:d2_nonunif}.} 
	\subsection{Approximate LSE for Large Sample Sizes via Sample Splitting and Minkowski Averaging}\label{fastalg2}
	\textcolor{black}{While the MIQO formulation~\eqref{miqop} allows one to compute the LSE for the first time, due to the number of constraints in the computation of the LSE, the memory requirement for the proposed MIQO can make it prohibitive when sample sizes are large ($\gg 500$). In this section, we propose a sample splitting based method to enable approximate computation of the quasiconvex LSE. The sample splitting procedure allows for parallelization of the computation allowing for arbitrarily large sample size. The first step is to split the sample into $K$ splits, one then computes the quasiconvex LSE by applying the MIQO algorithm in Section \ref{sec:dual} on each of these splits, to compute estimators $\widehat{\varphi}^{(1)},\ldots,\widehat{\varphi}^{(K)}$. The final estimator is then obtained by an aggregation of the above $K$ estimators. In case of the standard non-parametric regression, one can aggregate the estimators by taking a simple pointwise average of the regression function estimates. However, in our case, we need to aggregate the estimators in a way that the resulting estimator is also quasiconvex. Simple averaging doesn't preserve quasiconvexity as the sum of two quasiconvex functions is not necessarily quasiconvex~\cite{volle1998duality}. We propose to aggregate the $K$ estimators via the following modified version of \textit{infimal convolution} \citep{volle1998duality} (or level averaging~\cite{traore1996level}). We define the aggregate function as:
		\begin{equation}\label{eq:infimal_conv}
			x \mapsto \widehat{\varphi}(\boldsymbol{x}) := \inf\big\{ \vee_{i=1}^K \widehat{\varphi}^{(i)}(\boldsymbol{x}_i) : K^{-1}\sum_{i=1}^K \boldsymbol{x}_i = \boldsymbol{x}\big\}.
		\end{equation}
		In Lemma \ref{infconv7} (Section \ref{sec:techResults} of the supplement),  we show that for every  $\alpha \in \R$,
		$$\widehat{\varphi}^{-1}((-\infty,\alpha]) = \frac{1}{K}\sum_{i=1}^K \widehat{\varphi}^{(i) ~-1}((-\infty,\alpha]),$$
		where for sets $A$ and $B$ and $\alpha \in [0,1]$, we define their \textit{Minkowski average} as: $\alpha A + (1-\alpha) B := \{\alpha a + (1-\alpha) b: a\in A, b\in B\}.$ 
		The quasiconvexity of $\widehat{\varphi}$ follows immediately from the fact that Minkowski sums (and averages) of convex sets are convex~\citep{krein1940regularly}. Finally, one can easily show that the $\widehat{\varphi}$ is consistent for $\varphi$ under assumptions discussed in Section~\ref{sec:consist}.}
	
\begin{figure}[ht] 
		\begin{minipage}[b]{0.5\linewidth}
			\centering
			\includegraphics[width=\linewidth]{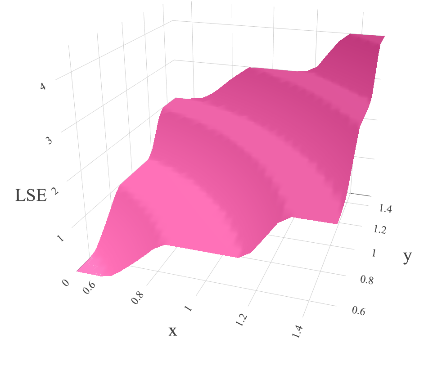} 
		\end{minipage}
		\begin{minipage}[b]{0.5\linewidth}
			\centering
			\includegraphics[width=\linewidth]{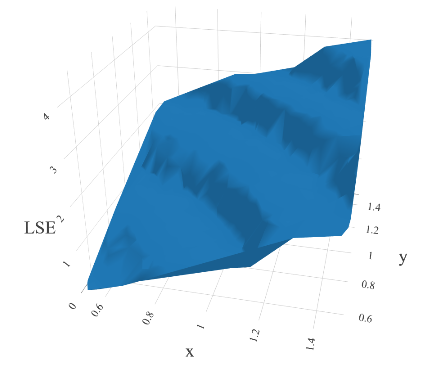} 
		\end{minipage} 
		\caption{The left panel shows the true function surface and the right panel shows the Minkowski-aggregated LSE surface.}\label{Minplott1}
	\end{figure}

\begin{figure}
		\begin{center}
			\includegraphics[width=3in]{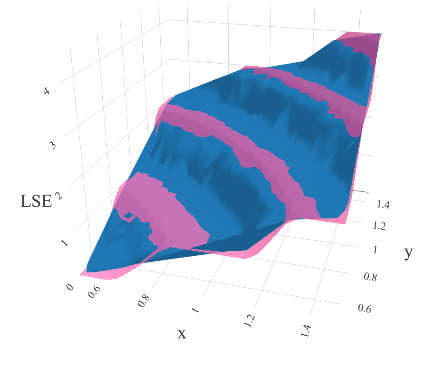}
		\end{center}
		\caption{Simultaneous plot showing the true function surface and the Minkowski-aggregated LSE surface. The pink surface denotes the true function, the blue surface denotes the LSE estimate obtained by aggregation through Minkowski averaging.}\label{Minplott2}
	\end{figure}

	\textcolor{black}{We have added this new estimator to our existing R package \texttt{QuasiLSE}~\citep{QuasiLSE}. We now provide a small simulation to show that the above aggregation works well in practice for larger sample sizes.  We implement the new algorithm on a data consisting of $2,000$ bi-variate samples simulated from $\mathrm{Uniform} [0.5,1.5]^2$. The regression functions are exactly same as those in Section \ref{sub:uds}, and we fix the error variance at $0.2$, whereas the smoothness parameter $\xi$ is fixed at $0.34$. The entire sample is split into $5$ equal parts, followed by applying the MIQO algorithm \eqref{miqop} on each of these splits, and combining the resulting estimators by Minkowski averaging to get the final estimator, plotted in Fig \ref{Minplott1} and \ref{Minplott2}. We see that the Minknowski-aggregated LSE approximates the true function surface very well.}

	\section{Analysis of the Japanese plywood production data}\label{sec:real_data}
	
\begin{figure}
		\centering
		\subfigure{\includegraphics[width=.3\textwidth]{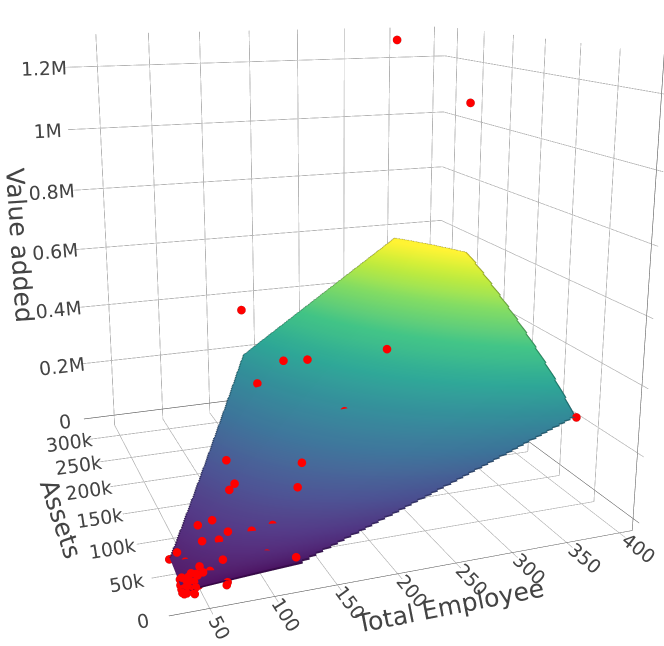}
			\label{fig:plywood_cobb}}
		\quad
		\subfigure{\includegraphics[width=.3\textwidth]{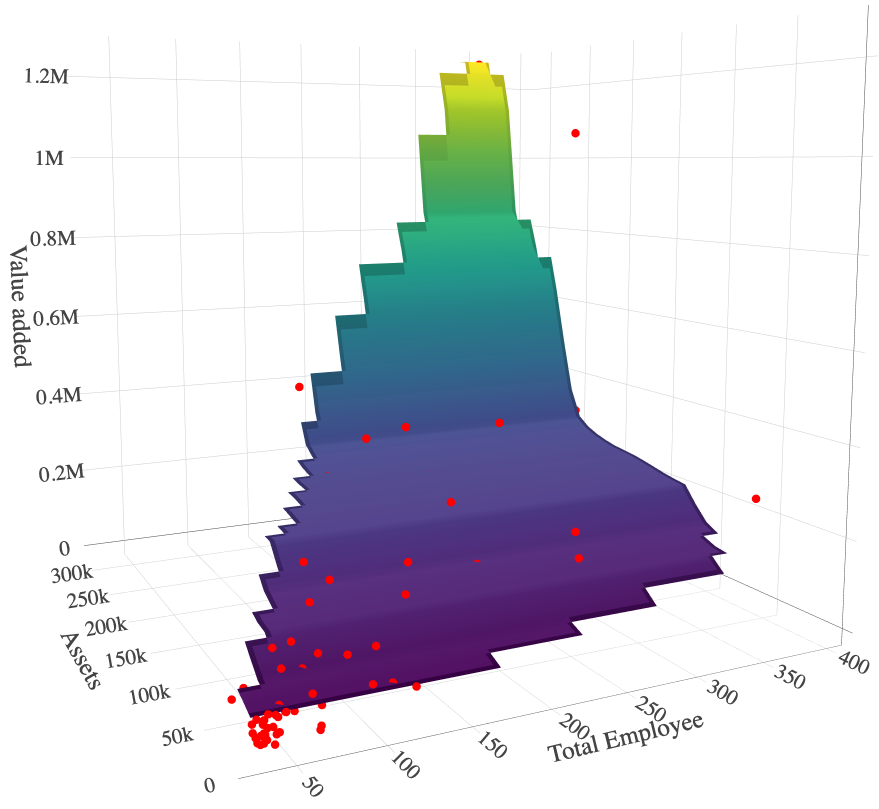}
			\label{fig:plywood_shape}}
		\subfigure{\includegraphics[width=.3\textwidth]{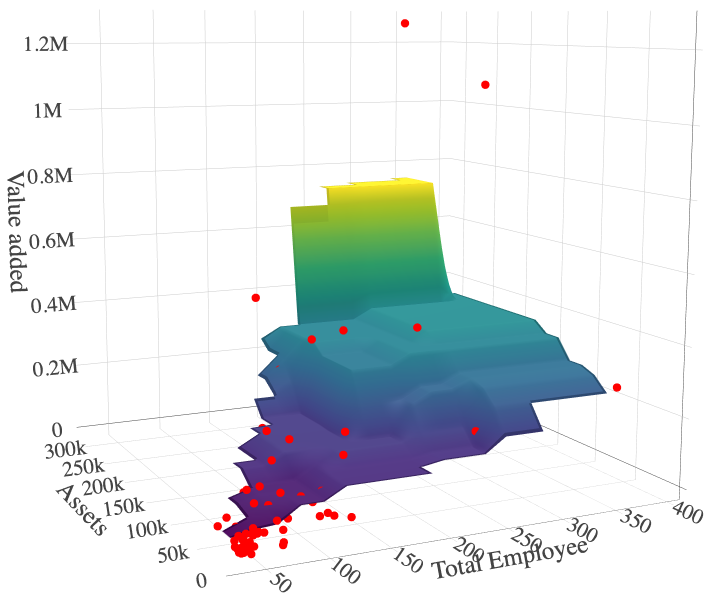}}
		\caption[]{Production surface estimates for the Japanese plywood industry for the year 2007; Left panel:  least squares Cobb-Douglas estimate; \textcolor{black}{Middle panel: shape enforced version (\texttt{ChenEtAl}) of the Nadaraya-Watson estimate} with bandwidth chosen through least squares cross validation~\cite[Page 69]{li2007nonparametric} using the \texttt{np} package in \texttt{R}~\citep{np}; \textcolor{black}{Right panel: the penalized isotonic regression spline estimator proposed in~\cite{meyer2013simple}}.}
		\label{fig:plywood_intro}
	\end{figure}

	\textcolor{black}{\cite{foster2008reallocation} studied the production surface in the US plywood industry. Their goal was to predict the value added by a company based on two input variables: Total Employees and Assets. In this section, we consider the production data of $78$ Japanese mid to large plywood factories for the year 2007. To provide a preliminary study of the production surface, in~Figure~\ref{fig:plywood_intro},  we plot the least squares  Cobb-Douglas \footnote{The least squares Cobb-Douglas estimator is the least squares estimator for the linear regression model between log of the inputs and log  of the output; see~Definition~\ref{def:cobb} and Remark~\ref{rem:cobb-Douglas} for more details on the Cobb-Douglas production function.} and  the shape enforced (quasiconcave and increasing function) version of the Nadaraya-Watson estimator for the data. 
		The least squares Cobb-Douglas estimate satisfies the economic assumptions of monotonicity and convex input requirement set. 
		Furthermore, this parametric estimator suggests that output for the factories in the data
		increases by more than the proportional change in inputs.\footnote{This property is called increasing returns to scale; see Definition~\ref{def:CRS} and Remark~\ref{rem:cobb-Douglas} in Section~\ref{sec:economic_background_and_terminologies} of the supplement for more details.} This, however, is inconsistent with the common understanding of microeconomic theory, as the production data contains a mixture of young and mature factories \citep{list2007internal,haltiwanger2016high}.
		``Young'' factories generally exhibit increasing returns to scale, while ``mature'' factories exhibit  decreasing returns to scale; see~\cite{arrow1971economic} and 
		as there is a mix of young and mature factories in the data, other shape constrained estimators such as concave  or $S$-shape estimators will impose additional unjustified structure on the estimator.} 
	
	We now elaborate on the Japanese production data introduced above, and apply the developed methodology to estimate the production and cost functions. 
	The Japanese plywood data is part of a larger dataset  collected by the Japanese Ministry of Economy, Trade, and Industry. The dataset contains production data for various Japanese industries. Japanese industry data is considered to be of high quality for the following reasons: (1) Japan has a large and developed manufacturing industry; (2) Japanese economy was stable during the data collection period; (3) The work practices of the Japanese census are known to be set at very high standards. The above factors result in a high-quality dataset compared to many other countries \citep{censusofmanufacture_2007}. In this paper, we study the 2007 cross-sectional dataset. \cite{foster2008reallocation} argue that plywood production data is particularly suitable for production function estimation using cross establishment data, as plywood establishments produce physically homogeneous products. As discussed above, the least squares Cobb-Douglas estimator fails to properly fit the data.  The data contains both young and mature establishments as measured by the establishment date. Young and mature establishments are likely to have different returns to scale.\footnote{See Definition~\ref{def:CRS} in Section~\ref{sec:economic_background_and_terminologies} of the supplement for a definition.} However, the Cobb-Douglas estimator can only have either increasing or decreasing returns to scale. 
	Furthermore, as all the establishments operate on a narrow cone of input ratios, the model assumption of $S$-shape is also too restrictive for this data. 
	\begin{figure}
		\centering
		\subfigure{\includegraphics[width=.3\textwidth]{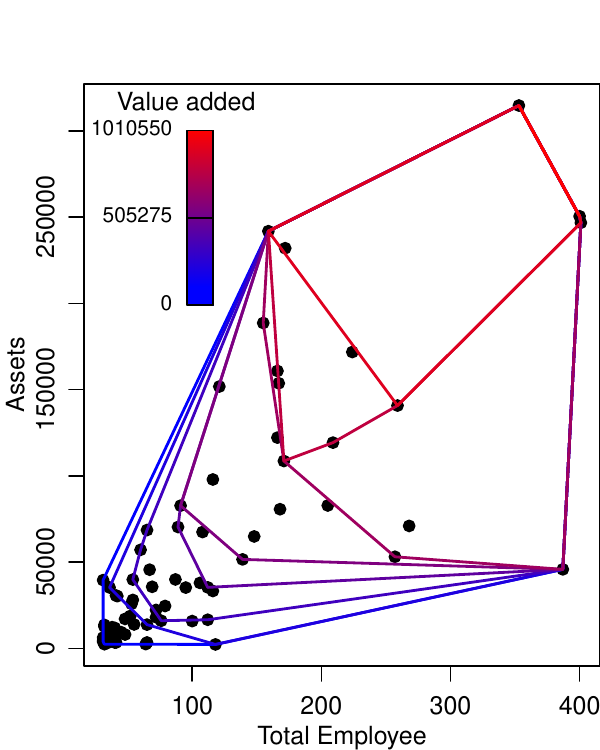}}
		\quad
		\subfigure{\includegraphics[width=.3\textwidth]{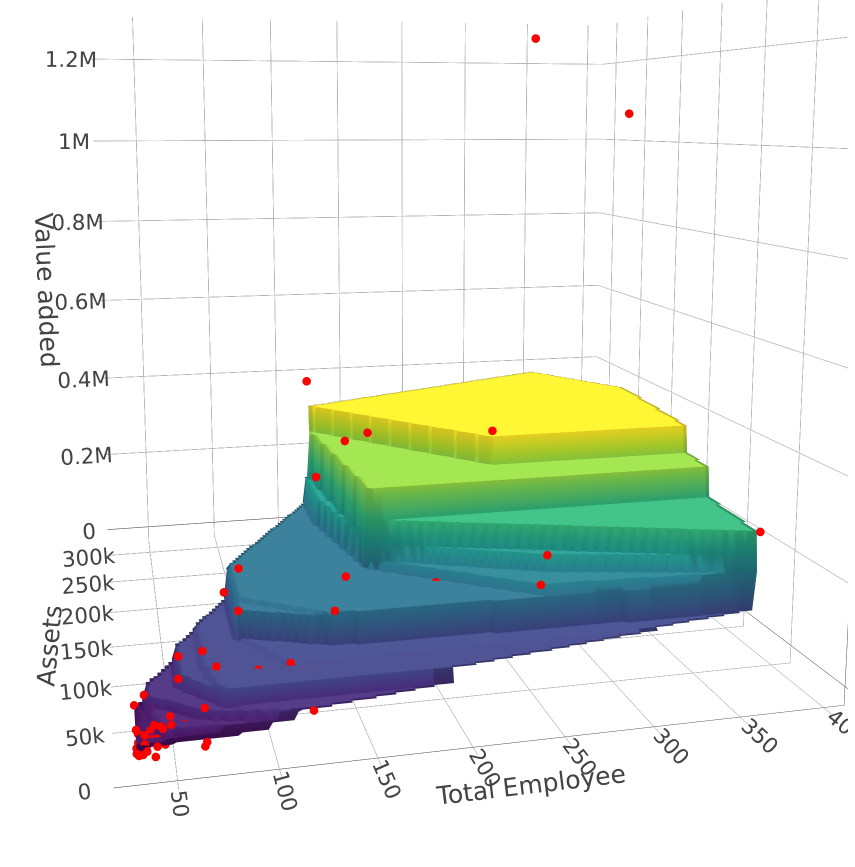}}
		\caption{Plot for the Japanese plywood production data from 2007: Left panel -- contour plot for the \textit{upper} level sets for the estimate of the production function; Right Panel --  plot of the estimated production function on the convex hull of the observed input variables.}
		\label{fig:Plywood}
	\end{figure}
	The left panel of Figure~\ref{fig:Plywood} shows the input requirement sets for $\widehat\varphi_n$ and the right panel shows the surface plot of the production function. Notice that as value-added increases, the establishments become more capital intensive.  This illustrates the typical pattern of capital deepening as production expands \citep{kumar2002technological}. We also observe that establishments are operating at different scales of production throughout the domain of the production function. Our proposed estimator captures the characteristics of the data as flexibly as possible while maintaining the fundamental axioms of monotonicity and quasiconcavity. To further understand the predictive performance of the various estimators discussed in Section~\ref{sec:simulation_study}, we estimate the out-of-sample prediction error by randomly and {\color{black}repeatedly partitioning the data ($100$ times) into 80\%/20\% training/test splits. The average test error of the competing estimators relative to the LSE is: 1.07 (\texttt{NW}),  1.09 (\texttt{ChenEtAl}), and 1.11 (\texttt{Iso}); the LSE has a relative error of 1}  and a lower number is better. 
	
	\begin{myrem}[Additional real data example]\label{rem:Additonal examples}
		{\cln In Section~\ref{sec:analysis_of_cost_data} of the supplementary file, we analyze the data of the cost function for hospitals across  the US using the 2007 Annual Survey Database from the American Hospital Association studied in~\cite{MR3993460}. We show that just as in the case of the Japanese production data, existing estimators either overfit or do not adequately incorporate the
			known shape of the nonparametric function when estimating the cost function.}
	\end{myrem}

	\section{Future work}\label{sec:conc}
	
	Several interesting future directions of work follow. The \textcolor{black}{optimal rates of convergence are} not known for $d\le 3$. We plan to study this in the near future. Even though the MIQO developed in Section~\ref{sec:dual} is new, the \texttt{R} package \texttt{QuasiLSE} \citep{QuasiLSE} uses CPLEX/gurobi (two off-the-shelf programs) to compute the minimizer. The memory requirement for the proposed MIQO can make it prohibitive when sample sizes are large ($\gg 500$). However, there have been recent developments (see e.g.,~\cite{dedieu2020learning}) that provide approximate solutions to mixed-integer programs. We are currently working towards developing an approximate algorithm that will be computationally less expensive. 
	\section{Acknowledgement} 
	Somabha Mukherjee (SM) and Rohit Kumar Patra (RKP) contributed 
	equally to this work. Hiroshi Morita provided the Japanese plywood production data used in the paper.   RKP is the senior statistics author and a bulk of the work was done when SM	was a PhD student at the University of Pennsylvania. We thank Arun K. Kuchibhotla and Bodhisattva Sen for their helpful discussions throughout the preparation of the manuscript. We also  thank the Joint Editor, the Associate Editor, and the two anonymous referees for their careful reading and constructive comments that led to a much improved paper. 
	\bibliographystyle{apalike}
	\bibliography{References}

	\appendix

	\newpage
	\begin{center}
		\textbf{\large Supplement to ``Least Squares Estimation of a Quasiconvex Regression Function''}
	\end{center}

\noindent\textbf{Summary.} ~In Section~\ref{sec:analysis_of_cost_data}, we estimate the cost variation across hospitals in the US to illustrate the usefulness of the proposed estimator. Section~\ref{sec:economic_background_and_terminologies} reviews some basic concepts and definitions from economics, with the aim of providing a background behind the assumptions for the shapes of the production and cost functions. In Section~\ref{sec:techResults}, we prove some technical results, that are used throughout the paper. The proofs of Lemma \ref{lem:primal}, Theorem \ref{thm:exun}, Lemma \ref{dual}, and Lemma \ref{miqeqv} are given in Sections \ref{primalproof},  \ref{sec:proof_of_theorem_thm:exun},  \ref{dualproof}, and  \ref{sec:proof:miqeqv}, respectively. The validity of Algorithm \ref{alg1} is established in Section \ref{sec:proof:alg1}.
		
	\section{Analysis of Hospital cost data} 
	\label{sec:analysis_of_cost_data}

	In this section, we analyze the cost variation across  hospitals in the US. The analyzed  data is from the American Hospital Association's Annual Survey Database for 2007. The  reported cost includes payroll, employee benefits, infrastructure depreciation, interest, supply, and other expenses. For every patient at each of the hospitals, all procedures received are recorded via the International Classification of Diseases, Ninth Revision, Clinical Modification (ICD9-CM) codes~\citep{zuckerman1994measuring}. Following~\cite{pope2013returns} and~\cite{MR3993460}, we map the codes into four categories of procedures, specifically ``Minor Diagnostic,'' ``Minor Therapeutic,'' ``Major Diagnostic,'' and ``Major Therapeutic''. 
	Finally, we add up the number of procedures in each of the  categories (for every hospital) to construct the hospital specific output variables.  
	\begin{figure}
		\centering
		\subfigure{\includegraphics[width=.45\textwidth]{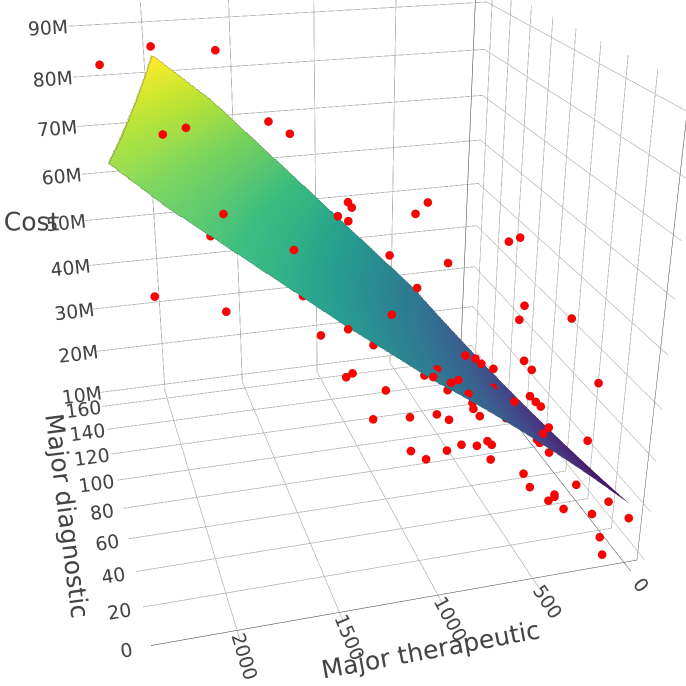}
			\label{fig:Major_quad}}
		\quad
		\subfigure{\includegraphics[width=.45\textwidth]{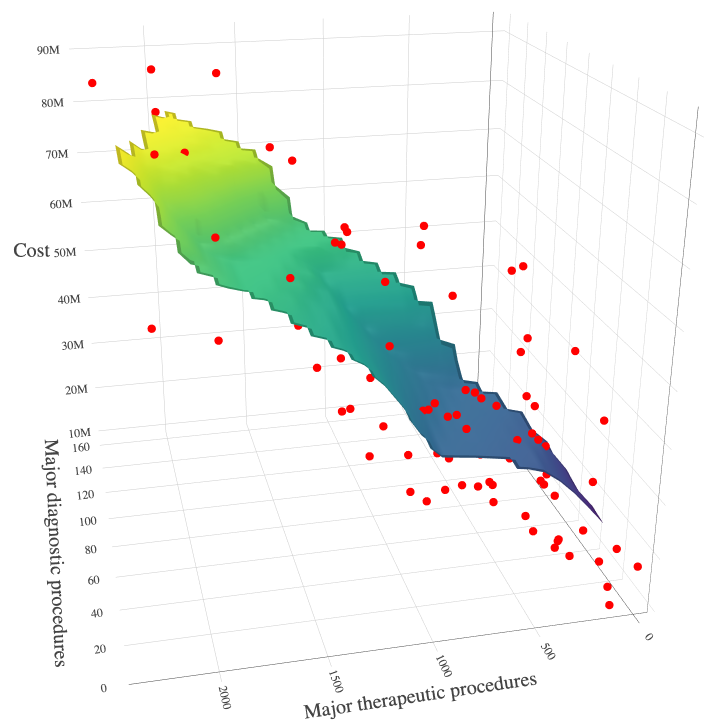}
			\label{fig:Major_Nw}}
		\caption[]{Cost function estimates for the Hospital data on the convex hull of the data for the year 2007. The number of Minor therapeutic and diagnostic procedures are  held constant around their median values. Left panel: fit based on a quadratic regression model (without interaction); Right panel:  Nadaraya-Watson estimator \textcolor{black}{corrected by \texttt{ChenEtAl}} with bandwidth chosen by least squares cross validation using the \texttt{np} package in R.}
		\label{fig:Major_intro}
	\end{figure}
	
\begin{figure}
		\centering
		\subfigure{\includegraphics[width=.4\textwidth]{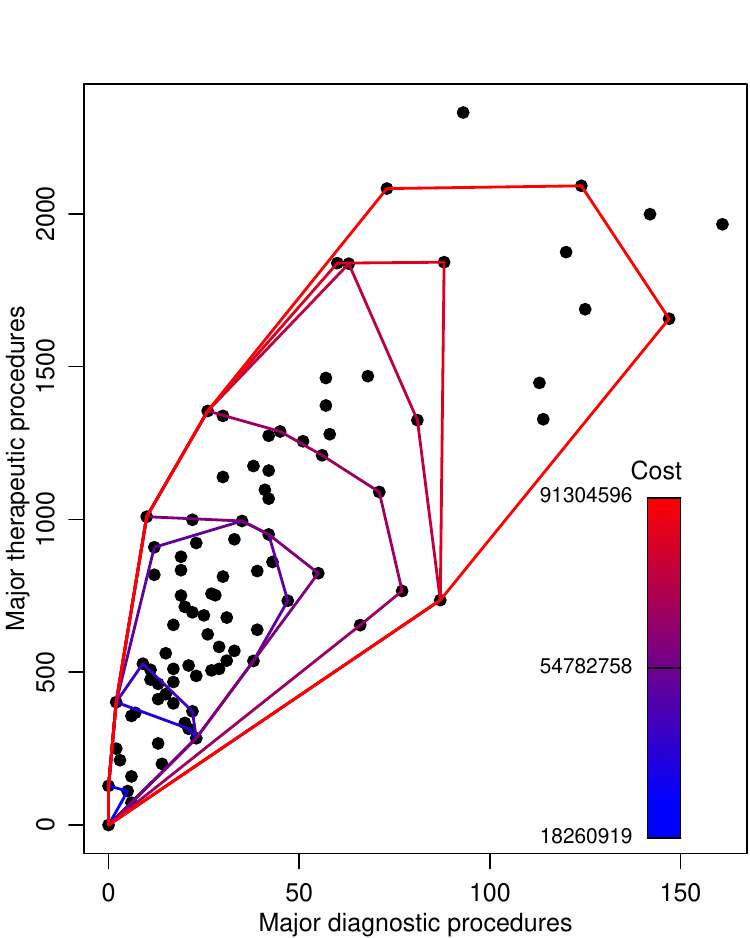}
			\label{fig:hospitalsub1}}
		\quad
		\subfigure{\includegraphics[width=.5\textwidth]{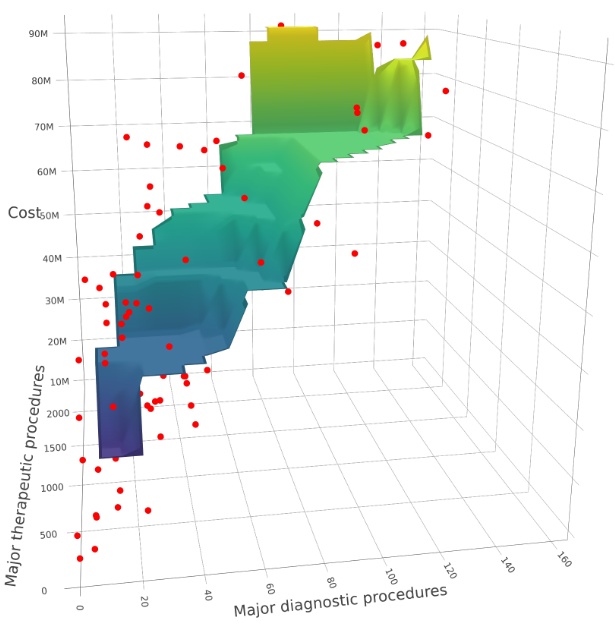}
			\label{fig:hospitalsub2}}
		\caption{Estimated cost functions when regressing the operating cost of  hospitals across the US on the number of  major diagnostic and therapeutic procedures, while keeping the number of minor procedures fixed around their median. Left panel: contour plot for the \textit{lower} level sets for the estimated cost function; Right panel: estimated cost function on the convex hull of the output variables.}
		\label{fig:HospitalMajor}
	\end{figure}

\begin{figure}
		\centering
		\subfigure{\includegraphics[width=.4\textwidth]{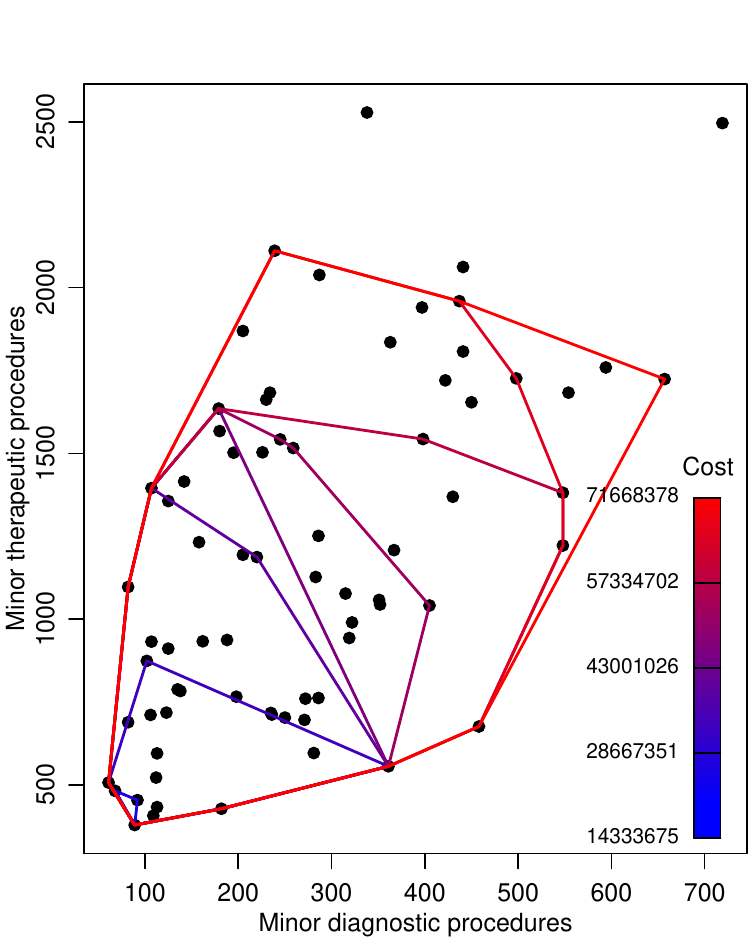}
			\label{fig:hospitalsub3}}
		\quad
		\subfigure{\includegraphics[width=.4\textwidth]{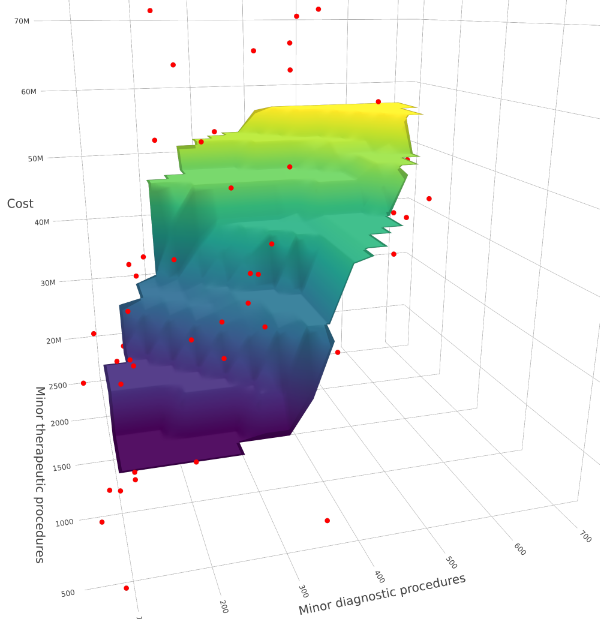}
			\label{fig:hospitalsub4}}
		\caption{Estimated cost functions when regressing the operating cost of  hospitals across the US on the number of  minor diagnostic and therapeutic procedures, while keeping the number of major procedures fixed around their median. Left panel: contour plot for the \textit{lower} level sets of the estimated cost function; Right panel: estimated cost function on the convex hull of the output variables.}
		\label{fig:HospitalMinor}
	\end{figure}

	After some preliminary cleaning up of the data, there are 523 hospitals in our dataset.~\cite{MR3993460} conclude that the above four regressors are statistically significant for predicting the cost of the hospitals. However, to keep the results interpretable and be able to plot the cost function, we fix two of the four variables around their median and estimate the two dimensional cost function assuming the two constrained variables to be constant, i.e., we demonstrate the cost function estimator on a ``slice'' of the data. We consider two different slices of the data. In the first slice, we consider hospitals for which the number of both Minor Therapeutic and Diagnostic procedures are between their respective second and third quartiles; see Figures~\ref{fig:Major_intro} and~\ref{fig:HospitalMajor}. The second and third quartiles are chosen so  that we have a reasonable amount of the hospitals in the data slice. The estimates in Figures~\ref{fig:Major_intro} and~\ref{fig:HospitalMajor} are based on data from $92$ hospitals. The second slice reverses the role of major  and minor procedures and Figure~\ref{fig:HospitalMinor} is based on $73$ hospitals.

	In Figure~\ref{fig:Major_intro}, we plot the estimated cost functions using: (1) the fit based on a quadratic model without the interaction term (left panel); and (2) the Nadaraya-Watson estimator with cross-validated choice of the tuning parameter (right panel).
	A quadratic model (without the cross-product terms) is often used in productivity and efficiency analysis of healthcare data; see e.g.,~\cite{fare2010functional},~\cite{MR3993460}, and~\cite{ferrier2018directional}. However, in Figure~\ref{fig:Major_intro}, we see that the quadratic cost estimate shows very little substitutability between major therapeutic and major diagnostic procedures in contrast to the nonparametric estimator. On the other hand, the  Nadaraya-Watson estimator overfits the data and does not maintain the monotonic structure implied by the standard axioms of the cost function, \cite{shephard1970}. In Figures~\ref{fig:HospitalMajor} and~\ref{fig:HospitalMinor}, we fit a quasiconvex and increasing function to the two slices of the data. The quasiconvex and  increasing LSE is able to estimate a function that characterizes the trade-off between the two outputs for any given cost level, while still maintaining the monotonic structure, implying increasing costs for increasing production, consistent with the basic axioms of production. {To further understand the predictive performance of the various estimators discussed in Section~\ref{sec:simulation_study}, we estimate the out-of-sample prediction error by randomly and repeatedly partitioning $100$ times the data into 80\%/20\% training/test splits. The average test error of the competing estimators relative to the LSE for predicting Major therapeutic and diagnostic procedures is: 0.97 (\texttt{NW}),  0.94 (\texttt{ChenEtAl}), and 3.45 (\texttt{Iso}); and  predicting Major therapeutic and diagnostic procedures is: 1.01 (\texttt{NW}),  0.98 (\texttt{ChenEtAl}), and 2.70 (\texttt{Iso}), the LSE has a relative error of 1 and a lower number is better.}
	

	\section{Economic background and terminologies} 
	\label{sec:economic_background_and_terminologies}
	In this section, we review some key concepts and definitions from economics. The goal is to provide a basic background behind the assumptions for the characteristics and shapes of the production and cost functions.
	\begin{mydef}[Return to scale]\label{def:CRS}
		A production function $f$ is said to exhibit constant returns to scale, if $f(\lambda \bm{x}) = \lambda f(\mathbf{x})$ for all inputs $\bm{x}$ and all $\lambda > 0$, increasing returns to scale, if $f(\lambda \bm{x}) > \lambda f(\bm{x})$ for all inputs $\mathbf{x}$ and all $\lambda > 1$, and decreasing returns to scale, if $f(\lambda \bm{x}) < \lambda f(\bm{x})$ for all inputs $\bm{x}$ and all $\lambda > 1$ (see \citep[Section 18.10]{varianint}).
	\end{mydef}    

	\begin{mydef}[Cobb-Douglas production function]\label{def:cobb}
		The Cobb-Douglas production function is defined as $F(X_1,X_2) = A X_1^\alpha X_2^\beta$, where $\alpha, \beta > 0$ are the output elasticities of the inputs $X_1$ and $X_2$, respectively (see \cite[Section 18.3]{varianint}). 
	\end{mydef}  
	
	\begin{myrem}\label{rem:cobb-Douglas}
		The Cobb-Douglas production function exhibits constant returns to scale if and only if $\alpha + \beta = 1$, increasing returns to scale if and only if $\alpha + \beta > 1$, and decreasing returns to scale if and only if $\alpha + \beta < 1$.
	\end{myrem} 
	
	
	\section{Some technical results}\label{sec:techResults}
	First we introduce some notations and definitions that will be used throughout the rest of the supplement. For a subset $A\in \R^d$, $\overline{A}$ denotes the closure of $A$ with respect the Euclidean topology.  
	For two $\otimes_{i=1}^k n_i$-tensors $\bm A$ and $\bm B$, the Frobenius inner product of $\bm A$ and $\bm B$ is defined as:
	\begin{equation}\label{eq:Frob_def}
		\langle \bm A , \bm B\rangle_F := \sum_{i_1=1}^{n_1}\ldots\sum_{i_k=1}^{n_k} A_{i_1\ldots i_k} B_{i_1\ldots i_k}\quad\text{and}\quad \|\bm A\|_F := \sqrt{\langle \bm A , \bm A   \rangle_F}.
	\end{equation}

	This section contains some technical lemmas that will be used later.
	\begin{mylem}\label{dagconv}
		For a convex set $S \subseteq \rd$, the set $S^\dagger$ is convex.
	\end{mylem}
	
	\begin{proof}
		Take $\bm Y , \bm Z \in S^\dagger$ and $\lambda \in [0,1]$. Then, there exist $\bm W , \bm X \in S$, such that $\bm W \preccurlyeq \bm Y$ and $\bm X \preccurlyeq \bm Z$. Hence, $\lambda \bm W + (1-\lambda)\bm X \preccurlyeq \lambda \bm Y + (1-\lambda)\bm Z$. Now, convexity of the set $S$ implies that $\lambda \bm W + (1-\lambda)\bm X \in S$, and hence, $\lambda \bm Y + (1-\lambda)\bm Z \in S^\dagger$.
	\end{proof}
	
	\begin{mylem}\label{dagclosed}
		For a compact set $S \subseteq \rd$, the set $S^\dagger$ is closed.
	\end{mylem}
	
	\begin{proof}
		For two sets $S_1$ and $S_2 \subseteq \rd$, if we define $S_1 + S_2 := \{\bm s_1 + \bm s_2: \bm s_1 \in S_1, \bm s_2 \in S_2\}$, then note that $S^\dagger = S + \mathbf{0}_d^\dagger$. Now, $S$ being compact and $\mathbf{0}_d^\dagger$ being closed, the result follows.
	\end{proof}
	
	\begin{mylem}\label{simplequasi}
		If  $\;\widehat{\bt} := \arg \min_{\bt \in \Qc} \|\bm Y - \bt\|$, then $\|\widehat{\bt}\|_\infty \leq \|\bm Y\|_\infty$. 
	\end{mylem}
	\begin{proof}
		Suppose, towards a contradiction, that $\|\widehat{\bt}\|_\infty > \|\bm Y\|_\infty$. Define $\bt' = (\theta_1',\ldots,\theta_n')$ by $\theta_i' := \widehat{\theta}_i \mathbbm{1}\{|\widehat{\theta}_i| \leq \|\bm Y\|_\infty\} + \mathrm{sgn}(\widehat{\theta_i})\|\bm Y\|_\infty \mathbbm{1}\{|\widehat{\theta}_i| > \|\bm Y\|_\infty\}$. Then, we have:
		\begin{eqnarray*}
			&&\|\bm Y - \bt'\|^2\\ &=& \sum_{i\,:\,|\widehat{\theta}_i| \leq \|\bm Y\|_\infty} (Y_i- \widehat{\theta}_i)^2 + \sum_{i\,:\, \widehat{\theta}_i > \|\bm Y\|_\infty} (Y_i- \|\bm Y\|_\infty)^2 + \sum_{i\,:\, \widehat{\theta}_i < -\|\bm Y\|_\infty} (Y_i+ \|\bm Y\|_\infty)^2\\&<& \sum_{i\,:\, |\widehat{\theta}_i| \leq \|\bm Y\|_\infty} (Y_i- \widehat{\theta}_i)^2 + \sum_{i\,:\, \widehat{\theta}_i > \|\bm Y\|_\infty} (Y_i- \widehat{\theta}_i)^2 + \sum_{i\,:\, \widehat{\theta}_i < -\|\bm Y\|_\infty} (Y_i- \widehat{\theta}_i)^2 = \|\bm Y - \widehat\bt\|^2~.
		\end{eqnarray*}
		Note that the strict inequality above came from the fact that since $\|\widehat{\bt}\|_\infty > \|\bm Y\|_\infty$, i.e., there exists $i$ such that $\widehat{\theta}_i > \|\bm Y\|_\infty$ or $\widehat{\theta}_i < -\|\bm Y\|_\infty$. 
		
		We will now show that $\bt' \in \Qc$, which will yield a contradiction. To this end, suppose $i \in [n]$ and $S\subseteq [n]$ are such that $\bm X_i \in \textrm{Cv}^\dagger\left(\{\bm X_j : j \in S\}\right)$. By Lemma \ref{lem:primal}, we have $\widehat{\theta}_i \leq \max\{\widehat{\theta}_j : j \in S\}$. Now, suppose that $|\widehat{\theta}_i| \leq \|\bm Y\|_\infty$. Then, we have:
		\begin{align*}
			\theta_i' = \widehat{\theta}_i \leq& \min\left\{\max\{\widehat{\theta}_j : j \in S\}, \|\bm Y\|_\infty \right\}\\
			=& \max\left\{\min\{\widehat{\theta}_j,\|\bm Y\|_\infty\} : j \in S\right\}\\
			\leq& \max \{\theta_j': j \in S \}~. 
		\end{align*}
		If $\widehat{\theta}_i < -\|\bm Y\|_\infty$, then $\theta_i' = -\|\bm Y\|_\infty$, and since $\theta_j' \geq -\|\bm Y\|_\infty$ for all $j$, we trivially have $\theta_i' \leq \max \{\theta_j': j \in S \}$. Finally, if $\widehat{\theta}_i > \|\bm Y\|_\infty$, then $\theta_i' = \|\bm Y\|_\infty$. But since $\max\{\widehat{\theta}_j : j \in S\} \geq \widehat{\theta}_i$, there exists $j \in S$ such that $\widehat{\theta}_j> \|\bm Y\|_\infty$, so $\theta_j' = \|\bm Y\|_\infty$. Hence, in this case, $\max\{\theta_j': j \in S\} = \theta_i' = \|\bm Y\|_\infty$. The proof of Lemma \ref{simplequasi} is now complete.
	\end{proof}

	\begin{mylem}\label{infconv7}
		In the notations of Section \ref{fastalg2}, for any $\alpha \in \R$, we have:
		$$\widehat{\varphi}^{-1}((-\infty,\alpha]) = \frac{1}{K}\sum_{i=1}^K \widehat{\varphi}^{(i) ~-1}((-\infty,\alpha]),$$
	\end{mylem}
	\begin{proof}
		Note that $\boldsymbol{x} \in \widehat{\varphi}^{-1}((-\infty,\alpha])$ if and only if there exists $\boldsymbol{x}_1,\ldots,\boldsymbol{x}_K$ with average $\boldsymbol{x}$, such that $\widehat{\varphi}^{(i)}(\boldsymbol{x}_i) \le \alpha$, i.e. $\boldsymbol{x}_i \in \widehat{\varphi}^{(i)~-1}((-\infty,\alpha])$ for all $i$. The last statement, of course, is equivalent to saying that $\boldsymbol{x} \in \frac{1}{K}\sum_{i=1}^K \widehat{\varphi}^{(i) ~-1}((-\infty,\alpha])$.
	\end{proof}

	\section{Proof of Lemma \ref{lem:primal}}\label{primalproof}
	\noindent Let us define  $$\mathcal{B} := \bigcap_{(i,S) \in \mathcal{L}{(\mathcal{X})}} \left\{\bm z \in \rn: z_i \leq \max_{j\in S} z_j\right\}~.$$ We need to show that $\mathcal{B} = \Qc$. First, we show that $\Qc \subseteq \mathcal{B}$. To this end, choose $\bm z \in \Qc$,  and let $\psi \in \mathcal{C}$ satisfy $\psi(\bm X_i) = z_i$ for all $i \in [n]$. Choose any $(i,S)\in \mathcal{L}(\mathcal{X})$. Then, $\bm X_i \in \bm X^\dagger$ for some $\bm X \in \cv\left(\{\bm X_j: j \in S\}\right)$. Since $\psi$ is decreasing, $\psi(\bm X_i) \leq \psi(\bm X)$. Since $\psi$ is quasiconvex, $\psi(\bm X) \leq \max_{j\in S} \psi(\bm X_j)$. Thus, $z_i \leq \max_{j\in S} z_j$, and hence, $\bm z \in \mathcal{B}$, showing that $\Qc \subseteq \mathcal{B}$. 
	
	Showing the other inclusion is a bit more involved. Choose $\bm z \in \mathcal{B}$, and define a function $\bm z^*: [n]\to [n]$ inductively, as follows. Let $\ell_1 := \min\{z_i: i \in [n]\}$. Define $\bm z^*(1):= \min\{i \in [n]: z_i = \ell_1\}$. Now, assume that $\bm z^*(1),\ldots,\bm z^*(m-1)$ have already been defined for some $1 < m <n$. Define $\ell_m := \min \{z_i: i \in [n]\setminus \{\bm z^*(1),\ldots,\bm z^*(m-1)\} \}$ and $\bm z^*(m) := \min \{i \in [n]\setminus \{\bm z^*(1),\ldots,\bm z^*(m-1)\}: z_i = \ell_m\}$. Clearly, 
	\begin{equation}\label{eq:z_*_prop}
		\bm z^* \text{ is a bijection, and } \bm z_{z^*(i)} \leq \bm z_{z^*(j)},\quad  \textrm{for all}~ i < j.
	\end{equation}  For example, if $n = 9$ and $\bm z = (4,1,1,6,1,4,6,8,2)$, then $\bm z^*(1) = 2, \bm z^*(2) = 3, \bm z^*(3) = 5, \bm z^*(4) = 9, \bm z^*(5) = 1, \bm z^*(6) = 6, \bm z^*(7)=4, \bm z^*(8) = 7$ and $\bm z^*(9) = 8$.
	
	Now, define $\psi: \rd \to \re$ inductively, as follows. Set $ \psi (\bm X) = \bm z_{\bm z^*(1)}$ for all $\bm X \in  \bm X_{\bm z^*(1)}^\dagger.$ Suppose inductively, that $\psi$ has been defined on $\cv^\dagger(\{\bm X_{\bm z^*(1)},\ldots,\bm X_{\bm z^*(m-1)}\})$ for some $1 < m \leq n$. As the next step, for all $\bm X \in \cv^\dagger(\{\bm X_{\bm z^*(1)},\ldots,\bm X_{\bm z^*(m)}\}) \setminus \cv^\dagger(\{\bm X_{\bm z^*(1)},\ldots,\bm X_{\bm z^*(m-1)}\})$, define $$\psi (\bm X) = \bm z_{\bm z^*(m)} \hspace{.1cm} .$$ Thus, inductively, $\psi$ is defined on $\cv^\dagger(\{\bm X_{\bm z^*(1)},\ldots,\bm X_{\bm z^*(n)}\}) = \cv^\dagger\left(\{\bm X_1,\ldots,\bm X_n\}\right).$ Finally, define $\psi(\bm X) = \bm z_{\bm z^*(n)}$ for all $\bm X \notin\cv^\dagger\left(\{\bm X_1,\ldots,\bm X_n\}\right)$.
	
	Several things need to be checked in order, now. First, let us show that $\psi(\bm X_i) = \bm z_i$ for all $i \in [n]$, or equivalently, that $\psi(\bm X_{\bm z^*(m)}) = \bm z_{\bm z^*(m)}$ for all $m \in [n]$. Take an $m \in [n]$, and suppose that $\bm X_{\bm z^*(m)} \in \bm X_{\bm z^*(1)}^\dagger$. Since $\bm z \in \mathcal{B}$, we must have $\bm z_{\bm z^*(m)} \leq \bm z_{\bm z^*(1)}$. But the reverse inequality is true by~\eqref{eq:z_*_prop}. Hence, $\psi(\bm X_{\bm z^*(m)}) = \bm z_{\bm z^*(1)} = \bm z_{\bm z^*(m)}$. So, assume that $\bm X_{\bm z^*(m)} \notin \bm X_{\bm z^*(1)}^\dagger$, whence there exists $1<k\leq m$ such that $\bm X_{\bm z^*(m)} \in \cv^\dagger(\{\bm X_{\bm z^*(1)},\ldots,\bm X_{\bm z^*(k)}\}) \setminus \cv^\dagger(\{\bm X_{\bm z^*(1)},\ldots,\bm X_{\bm z^*(k-1)}\})$. Since $\bm z \in \mathcal{B}$, we must have $\bm z_{\bm z^*(m)} \leq \bm z_{\bm z^*(k)}$. But the reverse inequality is trivially true; see~\eqref{eq:z_*_prop}. Hence, $\psi(\bm X_{\bm z^*(m)}) = \bm z_{\bm z^*(k)} = \bm z_{\bm z^*(m)}$. This completes our first verification. 
	
	Next, we show that the function $\psi$ is decreasing. For this, take $\bm X \preccurlyeq \bm Y \in \rd$. We need to show that $\psi(\bm X) \geq \psi(\bm Y)$. Suppose $\bm X \notin \cv^\dagger(\{\bm X_1,\ldots,\bm X_n\})$, then $\psi(\bm X) =\bm z_{\bm z^*(n)}$. Since $\psi$ is bounded above by $\bm z_{\bm z^*(n)}$, we are done. Now, suppose that $\bm X \in \cv^\dagger(\{\bm X_1,\ldots,\bm X_n\})$. Let 
	\[\ell := \inf\{i \in [n]: \bm X \in \cv^\dagger(\{\bm X_{\bm z^*(1)},\ldots,\bm X_{\bm z^*(i)}\})\}.\] Then, $\psi(\bm X) = \bm z_{\bm z^*(\ell)}$. Since $\bm X \preccurlyeq \bm Y$, we must have $\bm Y \in \cv^\dagger(\{\bm X_{\bm z^*(1)},\ldots,\bm X_{\bm z^*(\ell)}\})$. Hence, $\psi(\bm Y) = \bm z_{\bm z^*(k)}$ for some $k \leq \ell$. Since $\bm z_{\bm z^*(k)} \le \bm z_{\bm z^*(l)}$, our second verification is complete.
	
	Finally, we claim that the function $\psi$ is quasiconvex. Towards showing this, take $\alpha \in \mathbb{R}$. We must show that $S_\alpha(\psi)$ is a convex set. If $\alpha < \bm z_{\bm z^*(1)}$, then $S_\alpha(\psi) = \emptyset$, whereas if $\alpha \geq \bm z_{\bm z^*(n)}$, then $S_\alpha(\psi) = \rd$, and in either case, we are done. So, let us assume that $\bm z_{\bm z^*(1)} \leq \alpha < \bm z_{\bm z^*(n)}$. Then, there exists $k \in [n-1]$ such that $\bm z_{\bm z^*(k)} \leq \alpha < \bm z_{\bm z^*(k+1)}$. Since $\psi(\bm X) \leq \bm z_{\bm z^*(k)}$ for all $\bm X \in \cv^\dagger(\{\bm X_{\bm z^*(1)},\ldots,\bm X_{\bm z^*(k)}\})$, it is clear that $\cv^\dagger(\{\bm X_{\bm z^*(1)},\ldots,\bm X_{\bm z^*(k)}\}) \subseteq S_\alpha(\psi)$. On the other hand, suppose that $\bm X \notin \cv^\dagger(\{\bm X_{\bm z^*(1)},\ldots,\bm X_{\bm z^*(k)}\})$. Then $\psi(\bm X) = \bm z_{\bm z^*(j)}$ for some $j \geq k+1$, and hence, $\psi(\bm X) \geq \bm z_{\bm z^*(k+1)} > \alpha$. So, $\bm X \notin S_\alpha(\psi)$, showing that $S_\alpha(\psi) \subseteq \cv^\dagger(\{\bm X_{\bm z^*(1)},\ldots,\bm X_{\bm z^*(k)}\})$. Hence, $S_\alpha(\psi) = \cv^\dagger(\{\bm X_{\bm z^*(1)},\ldots,\bm X_{\bm z^*(k)}\})$. Our final claim now follows from Lemma \ref{dagconv}. 
	
	We thus conclude that $\mathcal{B} \subseteq \Qc$, and the proof of Lemma \ref{lem:primal} is now complete. 
	
	\section{Proof of Theorem~\ref{thm:exun}} 
	\label{sec:proof_of_theorem_thm:exun}
	\textit{Existence of minimizer:} We use the primal characterization of  $\mathcal{Q}$ in Lemma~\ref{lem:primal}  to prove this. It follows from Lemma \ref{lem:primal} that the set $\Qc$ is a closed set. Let $K := \Qc \cap \overline{B}_{\|\bm Y\|+1}\left(\bm Y\right)$, where $\bm Y := (Y_1,\ldots,Y_n)$. Then, $K$ is nonempty and compact (note that $\mathbf{0}_n \in K$). Hence the continuous function $\bm z \mapsto \|\bm Y - \bm z\|$ attains minimum over $K$ at some $\bm z_0 \in K$. For any $\bm z \in \Qc \setminus \overline{B}_{\|\bm Y\|+1}\left(\bm Y\right)$, we have $\|\bm Y - \bm z\| > \|\bm Y\| + 1 > \|\bm Y - \bm z_0\|$, as $\bm z_0 \in \overline{B}_{\|\bm Y\|+1}(\bm Y)$. Hence, the function $\bm z \mapsto \|\bm Y - \bm z\|$ attains minimum over $\Qc$ at $\bm z_0\in \Qc$, proving existence of a minimizer of \eqref{opt}.
	
	\textit{Almost sure uniqueness of minimizer:} Note that the function $d: \mathbb{R}^n \to [0,\infty)$ defined by $d(\bm x, \Qc) := \inf\{\|\bm x - \bt\|: \bt \in \Qc\}$ is Lipschitz on $\mathbb{R}^n$. Thus, by Rademacher's theorem (see \cite{rademacher1,rademacher2}), $x\mapsto d(\bm x, \Qc)$ is differentiable Lebesgue almost everywhere on $\mathbb{R}^n$. Now, if $d(\bm x, \Qc)$ is differentiable at some $\bm x \in \mathbb{R}^n$, it follows from \cite{moverflow} that there exists a unique $\bt \in \Qc$ such that $d(\bm x, \Qc) = \|\bm x - \bt\|$. This shows that the set $\mathcal{K}:=\{\bm x \in \mathbb{R}^n: d(\bm x, \Qc) = \|\bm x - \bt\|~\textrm{for more than one}~\bt \in \Qc\}$ has Lebesgue measure $0$. Hence, if $Y$ has density with respect to the Lebesgue measure on $\mathbb{R}$, then so does $\bm Y$ with respect to the Lebesgue measure on $\mathbb{R}^n$, and hence, $\mathbb{P}(\bm Y \in \mathcal{K}) = 0$. A similar proof can also be found in~\citet[Proposition 6]{JMLR:v20:17-687}.

	\section{Proof of validity of Algorithm~\ref{alg1}} 
	\label{sec:proof:alg1}
	
	By Lemma \ref{lem:primal} it is clear  that if $\bm z \in \Qc$, then none of the \textbf{if} statements in Algorithm \ref{alg1} will be executed, and consequently, the algorithm will always output ``out $=1$". On the other hand, suppose that Algorithm \ref{alg1} outputs ``out $=1$". This means that none of the \textbf{if} statements was executed, which in turn, implies that for every $i \in [n]$, $\bm X_i \notin \cv^\dagger(\{\bm X_j: z_j < z_i\})$. Hence, if $\bm X_i \in \cv^\dagger\left(\{\bm X_j: j \in S\}\right)$ for some $i \in [n]$ and $S \subseteq [n]$, then $S \nsubseteq \{j \in [n]: z_j < z_i\}$, which implies that $\max_{j \in S} z_j \geq z_i$. By Lemma \ref{lem:primal}, we can then conclude that $\bm z \in \Qc$. This shows the validity of Algorithm \ref{alg1}. 
	
	\section{Proof of Lemma \ref{dual}}\label{dualproof}
	\noindent We need a preliminary lemma, to start with.
	\begin{mylem}\label{sephypd}
		Let $\bm a, \bm a_1,\ldots, \bm a_k \in \mathbb{R}^d$ be such that $\bm a \notin \cv^\dagger(\{\bm a_1,\ldots,\bm a_k\}).$ Then, there exists $\bm v \in \mathbf{0}_d^\dagger$~, such that $\bm v^\top (\bm a_i - \bm a) > 0$ for all $i \in [k].$ 
	\end{mylem}
	\begin{proof}
		Define $f : \cv^\dagger(\{\bm a_1,\ldots,\bm a_k\}) \to \mathbb{R}$ as: $$f(\bm x) = \|\bm a - \bm x\|^2 .$$ Now, since $f$ is a continuous function, it attains minimum on the compact set  $\cv^\dagger(\{\bm a_1,\ldots,\bm a_k\}) \bigcap \overline{B}_{\|\bm a - \bm a_1\|}\left(\bm a\right)$ at some point $\bm p$, where $\overline{B}_r(\bm x)$ denotes the closed $L^2$ ball of radius $r$ centered at $\bm x$. Clearly, $f(\bm p) \leq f(\bm x)$ for all $\bm x \in \cv^\dagger(\{\bm a_1,\ldots,\bm a_k\})$. Take $\bm v := \bm p - \bm a$ . We first claim that $\bm v \in \mathbf{0}_d^\dagger$. Suppose, towards a contradiction, that $p_i < a_i$ for some $i \in [d]$. Define $\tilde{\bm p} := (p_1,\ldots,p_{i-1},a_i,p_{i+1},\ldots,p_d)$. As $\bm p \preccurlyeq \tilde{\bm p}$ and $\bm p \in \cv^\dagger(\{\bm a_1,\ldots,\bm a_k\}),$ we have $\tilde{\bm p} \in \cv^\dagger(\{\bm a_1,\ldots,\bm a_k\})$. However, $$f(\tilde{\bm p})= f(\bm p) - (a_i - p_i)^2 < f(\bm p)~,$$ contradicting the minimality of $\bm p$ and proving our claim.
		Next, we show that $\bm v^\top (\bm a_i - \bm a) > 0$ for all $i \in [k].$ As $\bm p$ is the projection of $\bm a$ onto $\cv^\dagger(\{\bm a_1,\ldots,\bm a_k\})$ (a closed convex set), we have $\langle \bm p- \bm x, \bm a- \bm p\rangle \ge 0$ for all $\bm x \in \cv^\dagger(\{\bm a_1,\ldots,\bm a_k\})$.
		Therefore, for any $i \in [k]$,
		$$\bm v^\top (\bm a_i - \bm a) = \langle \bm p - \bm a_i, \bm a - \bm p\rangle + \|\bm p - \bm a\|^2 \geq \|\bm p - \bm a\|^2 > 0~.$$
		Note that the last inequality uses the fact that $\bm a \notin \cv^\dagger(\{\bm a_1,\ldots,\bm a_k\})$ and $\bm p \in \cv^\dagger(\{\bm a_1,\ldots,\bm a_k\})$, so $\bm a \neq \bm p$. The proof of Lemma \ref{sephypd} is now complete.
	\end{proof} 
	We are now ready to prove Lemma \ref{dual}. Define  $$\mathcal{J} := \bigcup\limits_{\bx_1,\ldots,\bx_n \in \mathbf{0}_d^\dagger}\hspace{0.25cm}\bigcap\limits_{(i,j)\in\mathcal{U}(\mathcal{X},\bx_1,\ldots,\bx_n)} \{\bm z \in \mathbb{R}^n: z_i \geq z_j\}~,$$ 
	where $$\mathcal{U}(\mathcal{X},\bx_1,\ldots,\bx_n) := \{(i,j) \in [n]^2: \bx_j^\top (\bm X_i - \bm X_j) \leq 0\}~.$$
	Choose $\bm z \in \mathcal{J}.$ Then, there exist $\bx_1,\ldots,\bx_n \in \mathbf{0}_d^\dagger$, such that $z_i \geq z_j$ for all $i, j \in [n]$ such that $\bx_j^\top (\bm X_i - \bm X_j) \leq 0.$ Let $j$ and $S$ be such that $j \in [n],$  $S \subseteq [n]$, and  $\bm X_j \in \cv^\dagger\left(\{\bm X_i: i \in S\}\right)$, i.e., there exists a $\bm v \in \mathbf{0}_d^\dagger$ and a nonnegative sequence $\{\lambda_i\}_{i\in S}$ satisfying $\sum_{i\in S} \lambda_i = 1$, such that $\bm X_j = \sum_{i\in S} \lambda_i \bm X_i + \bm v$.  We will now show $z_j \le \max \{z_i : i \in S\}$. Suppose, towards a contradiction, that $z_j > \max \{z_i : i \in S\}$. Then, $\bx_j^\top (\bm X_i - \bm X_j) >0$ for all $i \in S$. Hence, $\bx_j^\top (\lambda_i \bm X_i - \lambda_i \bm X_j) \geq 0$ for all $i \in S$. However, $\sum_{i\in S} \bx_j^\top (\lambda_i \bm X_i - \lambda_i \bm X_j) = -\bx_j^\top \bm v \leq 0$, and hence, $\bx_j^\top (\lambda_i \bm X_i - \lambda_i \bm X_j) = 0$ for all $i \in S$. Hence, $\lambda_i = 0$ for all $i \in S$, contradicting $\sum_{i \in S} \lambda_i = 1$. So, $z_j \leq \max \{z_i : i \in S\}$. By Theorem \ref{lem:primal}, $\bm z \in \Qc$. Hence, $\mathcal{J} \subseteq \Qc$.
	
	For showing the reverse inclusion, choose $\bm z \in \Qc$.  Fix $j \in [n]$, and first, suppose that $S_j := \{i \in [n] : z_i < z_j\} \neq \emptyset.$ By Theorem \ref{lem:primal}, $\bm X_j \notin \cv^\dagger(\{\bm X_i : i \in S_j\})$. By Lemma \ref{sephypd}, there exists $\bx_j \in \mathbf{0}_d^\dagger$, such that $\bx_j^\top(\bm X_i - \bm X_j) >0$ for all $i \in S_j$. If $S_j =\emptyset$, define $\bx_j = \mathbf{0}_d$. Thus, we have created $n$ vectors $\bx_1, \ldots, \bx_n \in \mathbf{0}_d^\dagger$, with the property that whenever $z_i < z_j$ for some $i,j \in [n]$, we have $\bx_j^\top(\bm X_i - \bm X_j) > 0$. So, $\bm z \in \mathcal{J}$, and hence, $\Qc \subseteq \mathcal{J}$, completing the proof of Theorem \ref{dual}.

	\section{Proof of Lemma~\ref{miqeqv}} 
	\label{sec:proof:miqeqv}
	
	Note that for every $M > 0$ and any feasible solution $(\bm z^\top,\bx_1^\top,\ldots,\bx_n^\top,(u_{ij})_{i\neq j})$ of the MIQO problem \eqref{miqop}, the vector $\bm z$ belongs to $\Qc$ by Lemma \ref{dual}. This direction does not need $M$ to be large and holds for any $M$. Now, suppose that $\bm z \in \Qc$. By Lemma \ref{dual}, there exist vectors $\bx_1,\ldots,\bx_n \in \mathbf{0}_d^\dagger$ such that $z_i \geq z_j$ for all $i,j$ satisfying $\bx_j^\top(\bm X_i - \bm X_j) \leq 0$. For each $i \neq j$, set
	\[   
	u_{ij} = 
	\begin{cases}
		0 &\quad\text{if}\quad\bx_j^\top(\bm X_i - \bm X_j) \leq 0,\\
		1 &\quad\text{if}\quad\bx_j^\top(\bm X_i - \bm X_j) > 0.\\
	\end{cases}
	\]                                                            
	Then, it is easy to check that $(\bm z^\top,\bx_1^\top,\ldots,\bx_n^\top,(u_{ij})_{i\neq j})$ is a feasible solution of the MIQO problem \eqref{miqop} whenever $M > \max_{i\neq j}\left\{|z_i-z_j|\vee \left|\bx_j^\top(\bm X_i - \bm X_j)\right|\right\} =: M_0$. This is thus the direction, where we need to take $M$ large. The MIQO problem \eqref{miqop} is thus indeed equivalent to the problem \eqref{opt} for all $M > M_0$. Lemma \ref{miqeqv} now follows, by observing that $\Rc_M \subseteq \Rc_{M+1}$ for all $M \geq 1$. 
	{
		\section{Proof of Theorem \ref{mainthm}}\label{sub:proof_of_theorem_mainthm}
		In the proof of Theorem \ref{mainthm}, we will use the following standard notations for a function $h: \mathbb{R}^d\times R \to \mathbb{R}$:
		$$\p_n(h) := \frac{1}{n}\sum_{i=1}^n h(\bm X_i, Y_i)\quad \textrm{and}\quad \p(h) := \int_{\mathbb{R}^d} h(\bm X,  Y) ~\mathrm{d} \p(\bm X, Y)~.$$ Finally, we use $C$ to denote a constant. By ``constant'' we will always mean a quantity that does not depend on $n$ but might depend on the various parameters 
		introduced in our assumptions.  In the rest of this paper, we make the convention that the constant $C$ is not necessarily the same on each occurrence.
		
		For any function $h: \mathbb{R}^d\to \mathbb{R}$, define
		\begin{equation}\label{eq:gamma}
			\gamma(h, (\bm X, Y)) := (Y- h(\bm X))^2.
		\end{equation}
		Recall that $\widehat{\varphi}= \argmin1_{h\in \mathcal{H}_{d, \Gamma}} \p_n \gamma(h, \cdot).$  Defining $$\bar{\varphi}_{\Gamma} := \argmin1_{h \in \Hh_{d,\Gamma}}\int_{\mathbb{R}^d}\left(h(\bm x) - \varphi(\bm x)\right)^2~\mathrm{d} \mathbb{P}(\bm x),$$
		we have the following basic inequality $\p_n \gamma(\widehat{\varphi}, \cdot) \le \p_n \gamma(\bar{\varphi}_{\Gamma}, \cdot).$ Finally it follows from $\e(\varepsilon|\bm X) = 0$, that
		\begin{align}\label{eq:basic_steps}
			\begin{split}
				P(\widehat{\varphi}- {\varphi})^2&= P(\gamma(\widehat{\varphi}, \cdot ))- P(\gamma({\varphi}, \cdot ))\\
				&= P(\gamma(\widehat{\varphi}, \cdot )) - P(\gamma(\bar{\varphi}_{\Gamma}, \cdot )) +  P(\gamma(\bar{\varphi}_{\Gamma}, \cdot ))- P(\gamma({\varphi}, \cdot ))\\
				&=P(\gamma(\widehat{\varphi}, \cdot )) - P(\gamma(\bar{\varphi}_{\Gamma}, \cdot )) +  P(\varphi-\bar{\varphi}_{\Gamma} )^2 +\p_n \gamma(\bar{\varphi}_{\Gamma}, \cdot)- \p_n \gamma(\bar{\varphi}_{\Gamma}, \cdot) \\
				&\le P(\gamma(\widehat{\varphi}, \cdot )) - P(\gamma(\bar{\varphi}_{\Gamma}, \cdot )) +  P(\varphi-\bar{\varphi}_{\Gamma} )^2  +\p_n \gamma(\bar{\varphi}_{\Gamma}, \cdot)- \p_n \gamma(\widehat{\varphi}, \cdot) \\
				&= P(\varphi-\bar{\varphi}_{\Gamma} )^2 + (\p_n -P) [\gamma(\bar{\varphi}_{\Gamma}, \cdot)- \gamma(\widehat{\varphi}, \cdot)] \\
				&\le  P(\varphi-\bar{\varphi}_{\Gamma} )^2 +  \sup_{h\in \Hh_{d,\Gamma}} \big| (\p_n -P) [\gamma(\bar{\varphi}_{\Gamma}, \cdot)- \gamma(h, \cdot)]\big|.
			\end{split}
		\end{align} 
		Taking expectation on both sides of the above display, the proof will be complete if we prove:
		\begin{align}\label{eq:Gnbound}
			&\E \bigg[ \sup_{h\in \Hh_{d,\Gamma}} \big| (\p_n-P) \big[\gamma(\bar{\varphi}_{\Gamma}, \cdot)- \gamma(h, \cdot)\big]\big|  \bigg]\nonumber\\ &\le   C \Gamma \max\{\Gamma, C_\varepsilon, C_\varphi\} \times \begin{cases}
				n^{-1/2}  &\text{ when } d = 2,\\
				n^{-1/2} \log n  &\text{ when } d = 3,\\
				n^{-2/(d+1)} &\text{ when } d\ge 4.
			\end{cases}
		\end{align} 
		\textbf{Proof of~\eqref{eq:Gnbound}:}
		Observe that 
		\begin{align}
			&\E \bigg[ \sup_{h\in \Hh_{d,\Gamma}} \big| (\p_n-P) \big[\gamma(\bar{\varphi}_{\Gamma}, \cdot)- \gamma(h, \cdot)\big]\big|  \bigg] \nonumber\\
			\le{}& \E \bigg[ \sup_{h\in \Hh_{d,\Gamma}} \Big\{\big| (\p_n-P) [ h-\bar{\varphi}_{\Gamma} ]^2 \big| + \big|2 (\p_n-P) [(h- \bar{\varphi}_{\Gamma} (\bm X)) (Y- \bar{\varphi}_{\Gamma} (\bm X))]\big| \Big\}     \bigg]\nonumber\\
			\le{}& \E \bigg[ \sup_{h, g\in \Hh_{d,\Gamma}} \big| (\p_n-P) [ h-g]^2\big|  \bigg] +2 \E \bigg[ \sup_{h\in \Hh_{d,\Gamma}}  \Big|(\p_n-P) \big[(h- \bar{\varphi}_{\Gamma}) (Y- \bar{\varphi}_{\Gamma}) \big]    \Big|  \bigg]\nonumber\\\label{eq:proof12.1}
		\end{align}
		
		Lemma~\ref{lemma7} provides an upper bound for the first term in~\eqref{eq:proof12.1}. We will now bound the second term in~\eqref{eq:proof12.1}. Observe that by symmetrization~\citet[Lemma 2.3.1]{vander}
		\begin{align}\label{eq:12.1secondterm}
			\begin{split}
				&2  \E \bigg[ \sup_{h\in \Hh_{d,\Gamma}}  \Big|(\p_n-P) \big[(h(\bm X)- \bar{\varphi}_{\Gamma}(\bm X)) (Y- \bar{\varphi}_{\Gamma}(\bm X))\big]    \Big|  \bigg]\\
				\le{}& 4  \E \bigg[ \sup_{h\in \Hh_{d,\Gamma}}  \Big|(\p_n-P) \big[ (Y- \bar{\varphi}_{\Gamma}(\bm X)) h(\bm X)\big]    \Big|  \bigg]\\
				\le{}& 4  \E \bigg[ \sup_{h\in \Hh_{d,\Gamma}}  \Big|(\p_n-P) \big[ ({\varepsilon} + \varphi(\bm X) - \bar{\varphi}_{\Gamma}(\bm X)) h(\bm X)\big]    \Big|  \bigg]\\
				\le{}& 4  \E \bigg[ \sup_{h\in \Hh_{d,\Gamma}}  \Big|(\p_n-P) \big[ {\varepsilon} h(\bm X)\big]    \Big|  \bigg]+  4  \E \bigg[ \sup_{h\in \Hh_{d,\Gamma}}  \Big|(\p_n-P) \big[ ( \varphi(\bm X) - \bar{\varphi}_{\Gamma}(\bm X)) h(\bm X)\big]    \Big|  \bigg]\\
				={}& 4  \E \bigg[ \sup_{h\in \Hh_{d,\Gamma}}  \Big|\p_n \big[ {\varepsilon} h(\bm X)\big]    \Big|  \bigg]+  4  \E \bigg[ \sup_{h\in \Hh_{d,\Gamma}}  \Big|(\p_n-P) \big[ ( \varphi(\bm X) - \bar{\varphi}_{\Gamma}(\bm X)) h(\bm X)\big]    \Big|  \bigg].
			\end{split}
		\end{align}
		Lemma~\ref{lemma6} (see~\eqref{eq:multiplier} and~\eqref{eq:multiplier_mh}) provides an upper bound for the both of the above quantities. Combining all this, we get~\eqref{eq:Gnbound}.

		%
		\section{Necessary Lemmas for Section \ref{sub:proof_of_theorem_mainthm}}\label{sec:necLemmas} 
		\begin{mylem}\label{lemma10andCor1} 
			Let $\p = f(x)dx$ be a probability measure with a continuous density function on $\R^d$ such that
			\begin{equation}\label{eq:Tail_cond_lemma}
				f(x) \le C (1+ \|x\|)^{-r} \quad \text{ for some }\quad r> (d^2+ 1)/(d-1).
			\end{equation} Let $\bm X_1,...,\bm X_n\sim \p$ be i.i.d.~samples from $\p$. Then for any $d\ge2$ and a fixed positive valued function $m$ such that $\|m\|_\infty < \infty $, 
			\begin{equation}\label{eq:10_1}
				\E \left[\sup_{K\in \Kk_d} |(\p_n-\p) [m(\cdot) \mathbf{1}(\cdot \in K)] \right]  \le C_d C_m  n^{-2/(d+1)}  \log n^{\mathbf{1}(d=3)},
			\end{equation}
			and 
			\begin{equation}\label{eq:10_1_ep}
				\E \left[\sup_{K\in \Kk_d} |(\p_n-\p) [\varepsilon \mathbf{1}(\cdot \in K)] \right] \le C_d  C_\varepsilon  n^{-2/(d+1)}  \log n^{\mathbf{1}(d=3)},
			\end{equation}
			
			for some constants  $C_m$, $
			C_d$, and  $C_\varepsilon$  that  depend on $m$, $d$, and $\varepsilon$ only, respectively.  
		\end{mylem}

		\begin{proof}
			We will apply Theorem~2.1 of~\cite{han2021set} to prove the above result. Define 
			\begin{equation}\label{eq:F_def}
				\Ff_m := \{ f: \R^d \to \R | f(\bm x) = m(\bm x) \mathbf{1}(\bm x \in K) \text{ for some } K \in \Kk_d\}.
			\end{equation}
			In Lemma~\ref{lem:entropy}, {we} show that for every $d\ge 2$,
			\begin{equation}\label{eq:entropy}
				\log N_{[]}(\delta, \Ff_m, L_1(P)) \le C_d C_m \delta^{-(d-1)/2}.
			\end{equation}
			Then choosing $\sigma^2$ in~\cite{han2021set} to be equal to $P m^2$,  Theorems 2.1-(2) and~ 2.3  and Remark~2.4-(2) of~\cite{han2021set} implies that 
			\begin{align}\label{eq:10_1_proof}
				\E \left[\sup_{K\in \Kk_d} \big|(\p_n-\p) [m(\cdot) \mathbf{1}(\cdot \in K)]\big| \right] &= \E \left[\sup_{f\in \Ff_m} |(\p_n-P) f |\right]\nonumber\\ &\le C_d C_m  n^{-2/(d+1)}  \log n^{\mathbf{1}(d=3)}.
			\end{align}
			The proof of~\eqref{eq:10_1_ep} is similar. The main difference is, now we choose $\sigma^2$ in Theorem~2.1-(2) to be $\E(\varepsilon^2)$. 
		\end{proof}

		\begin{mylem}\label{lem:entropy} Let  \begin{equation}\label{eq:F_ep}
				\Ff_\varepsilon := \{ g: \R^d \to \R | g(\bm x) =  \varepsilon \mathbf{1}(\bm x \in K) \text{ for some } K \in \Kk_d\},
			\end{equation}
			where $\|\E(|\varepsilon| \big| \bm X=\cdot)\|< \infty$ and for  some fixed function $m:\R^d\to\R$, let \begin{equation}\label{eq:F_m}
				\Ff_m := \{ g: \R^d \to \R | g(\bm x) = m(\bm x) \mathbf{1}(\bm x \in K) \text{ for some } K \in \Kk_d\}.
			\end{equation}
			If $\,P= f(\bm x)d\bm x$ satisfies the assumptions of Lemma~\ref{lemma10andCor1}, then 
			\begin{equation}\label{eq:entropy_m}
				\log N_{[],1}(\delta, \Ff_m, L_1(P)) \le C_d [\delta/\|m\|_\infty ]^{-(d-1)/2}
			\end{equation}
			\begin{equation}\label{eq:entropy_ep}
				\log N_{[],1}(\delta, \Ff_\varepsilon, L_1(P)) \le C_d [\delta/C_\varepsilon ]^{-(d-1)/2},
			\end{equation}
			where $C_\varepsilon := \big\|\E\big(|\varepsilon|\big| \bm X= \cdot)\big\|_{\infty}$.
		\end{mylem}

		\begin{proof}
			We will use the following bound on the bracketing numbers by~\cite{bronshtein1976varepsilon} (also~\citet[Eq. (33) and Lemma 3]{kur2019optimality} and~\citet[Theorem 8.4.1]{dudley2014uniform}), for every $d\ge 2$ and for every  small enough $\epsilon:$
			\begin{equation}\label{eq:bron}
				\log N_{[],1}(\epsilon,\Kk_d^{(R)}, \mathrm{vol}) \le C d^{(d+4)/2} (\mathrm{vol} B_d(0,1))^{(d-1)/2} (\epsilon/R^d)^{-(d-1)/2},
			\end{equation}
			where for any $R>0$, $\Kk_d^{(R)}$ is the set of convex bodies contained in $B_d(0, R)$, the centered {E}uclidean ball of radius $R.$
			As $\Kk_d$ is unbounded, we will partition any set $K\in \Kk_d$ via the following partition  for every $i\ge 1$, define $D_i= \{x\in \R^d: i-1 \le \|x\|\le i\}.$ Noting that $\cup_{i\ge1} D_i =\R^d$, observe that $K= \cup_{i\ge 1}K\cap D_i$. Since we can write
			\begin{align}\label{eq:f_g_part}
				\begin{split}
					g(\bm x)= m(x) \mathbf{1}(\bm x \in K) &=\sum_{i\ge 1}  m(x) \mathbf{1}(\bm x \in K\cap D_i):= \sum_{i\ge 1} g_i(\bm x).
				\end{split}
			\end{align}
			and
			\begin{equation}\label{eq:ep_part}
				h(\bm x)= \varepsilon \mathbf{1}(\bm x \in K) =\sum_{i\ge 1}  \varepsilon \mathbf{1}(\bm x \in K\cap D_i):= \sum_{i\ge 0} h_i(\bm x).
			\end{equation}
			Now for each fixed integer $R\ge 1$, let 
			\begin{equation}\label{eq:M_R}
				M_R:= \sup_{x \in D_R} f(x) = C  R^{-r}.
			\end{equation}
			Then by observing that for any two sets $A, A' \in \Kk_d^{(R)}$, we have  $\p(X\in A\Delta A') \le M_R \mathrm{vol}(A\Delta A')$, we have that 
			\begin{equation}\label{eq:entropyofSet}
				\log N_{[],1}(\epsilon,\Kk_d^{(R)}, P) \le \log N_{[],1}(\epsilon/M_R,\Kk_d^{(R)}, \mathrm{vol}) \le C_d (\epsilon/M_R R^d)^{-(d-1)/2} =: N(R, \epsilon).
			\end{equation}
			Thus there exist sets $\{(\ubar{S}_j ,\bar{S}_j) \}_{j=1}^{\exp(N(R, \epsilon))}$ that form an $L_1(P)$ bracket  for $\Kk_d^{(R)}$, i.e., $\ubar{S}_j\subset \bar{S}_j$ for every $j \le N(R, \epsilon)$ and  for any set $S\in \Kk_d^{(R)},$ there exists $k\le \exp(N(R, \epsilon))$ such that $\ubar{S}_k \subset S\subset \bar{S}_k$ and $P(\bar{S}_j\Delta\ubar{S}_j) \le \epsilon.$ This implies that 
			\begin{equation}\label{eq:an_bracket}
				\log N_{[],1}(\epsilon,\Kk_d^{(R)}\cap D_R, P) \le N(R, \epsilon),
			\end{equation}
			and $\{(\ubar{S}_j\cap D_R, \bar{S}_j\cap D_R )\}_{j=1}^{\exp(N(R, \epsilon))}$ form an $L_1(P)$ bracket (of width $\epsilon$) for $\Kk_d^{(R)}\cap D_R$, where that for any set of sets $\mathcal{A}$ and a set $B$, $\mathcal{A}\cap B:= \{A\cap B: A\in \mathcal{A}\}.$\newline

			\noindent \textbf{Proof of~\eqref{eq:entropy_m}:}
			Let $m^+(\bm x) := \max(0,m(\bm x) )$ and $m^-(\bm x) = \max(0, -m(\bm x))$ define the positive and negative part of $m$, respectively. Defining 
			\[   l_{j, R, \epsilon} := {m^+(\cdot)\mathbf{1}\big(\cdot \in \ubar{S}_j\cap D_R\big)- m^-(\cdot)\mathbf{1}\big(\cdot \in \bar{S}_j\cap D_R\big)}
			\] and 
			\[ u_{j, R,\epsilon} :={ m^+(\cdot)\mathbf{1}\big(\cdot \in \bar{S}_j\cap D_R\big)- m^-(\cdot)\mathbf{1}\big(\cdot \in \ubar{S}_j\cap D_R\big)}.\]
			We will now show that 
			\begin{equation}\label{eq:bracket_mold}
				\Big\{\Big[ {l_{j, R, \epsilon}},{u_{j, R,\epsilon}} \Big]\Big\}_{j=1}^{\exp(N(R, \epsilon))}
			\end{equation}
			forms an $L_1(P)$ bracket (of width $\|m\|_\infty \epsilon$) for
			\[\Ff_{m,R} := \{ g: \R^d \to \R | g(\bm x) = m(\bm x) \mathbf{1}(\bm x \in K) \text{ for some } K \in \Kk_d^{R}\cap D_R\}.\]
			
			Fix some $S\in \Kk_d^R\cap D_r$, then by~\eqref{eq:an_bracket}, there exists an $k\le {\exp} (N(R, \epsilon))$ such that 
			$\ubar{S}_k\cap D_R \subset S\subset \bar{S}_k\cap D_R$. Thus we have that  
			\begin{align}\label{eq:both_sides}
				\begin{split}
					m^+(\cdot)\mathbf{1}\big(\cdot \in \ubar{S}_k\cap D_R\big) &\le m^+(\cdot)\mathbf{1}\big(\cdot \in {S}\big) \le m^+(\cdot)\mathbf{1}\big(\cdot \in \bar{S}_k\cap D_R\big),\\
					-m^-(\cdot)\mathbf{1}\big(\cdot \in \bar{S}_k\cap D_R\big) &\le -m^-(\cdot)\mathbf{1}\big(\cdot \in {S}\big) \le -m^-(\cdot)\mathbf{1}\big(\cdot \in \ubar{S}_k\cap D_R\big),
				\end{split}
			\end{align}
			Combining the above two inequalities we get that $l_{k, R, \epsilon} \le m(\cdot)\mathbf{1}\big(\cdot \in {S}\big) \le u_{k, R, \epsilon}.$
			Thus~\eqref{eq:bracket_m} forms a bracket for $\Ff_{m, R}$. We will now find the width of this bracket:
			\begin{align}\label{eq:width}
				\begin{split}
					&P\big( \big|u_{k, R,\epsilon}- l_{k, R,\epsilon} \big|\big)
					\\={}&P\bigg( \bigg|m^+(\cdot)\mathbf{1}\big(\cdot \in (\bar{S}_k\Delta \ubar{S}_k)\cap D_R\big)+  m^-(\cdot)\mathbf{1}\big(\cdot \in (\bar{S}_k\Delta \ubar{S}_k)\cap D_R\big) \bigg|\bigg)\\
					={}&P\Big( |m(\cdot)|\mathbf{1}\big(\cdot \in (\bar{S}_k\Delta \ubar{S}_k)\cap D_R\big)\Big)\\
					={}&\|m\|_\infty \p\big(\bm X\in (\bar{S}_k\Delta \ubar{S}_k)\cap D_R\big) \le\|m\|_{\infty} \epsilon.
				\end{split}
			\end{align}
			Thus we have that 
			\begin{equation}\label{eq:an_bracket_final}
				\log N_{[],1}(\epsilon ,\Ff_{m, R}, P) \le N(R, \epsilon/ \|m\|_{\infty}) = C_d [\epsilon/ (\|m\|_\infty M_R R^d)]^{-(d-1)/2} := Q(m, R, \epsilon),
			\end{equation}
			where $M_R$ is defined in~\eqref{eq:M_R}.
			We will use the above entropy bound to find an $L_1(P)$ bracket for $\Ff_m$ of width $\epsilon$. We will do this by combining $\epsilon_R := \epsilon/ (C_\alpha R^\alpha)$ brackets for $\Ff_{m, R}$ where 
			\begin{equation}\label{eq:alpha}
				\alpha := 1+ (r- (d^2+ 1)/(d-1))/2,
			\end{equation} where $C_\alpha :=\sum_{R\ge 1} R^{-\alpha}$; note that $C_\alpha < {\infty}$ as $\alpha >1.$ In particular, fix $g\in \Ff_m$, recalling ~\eqref{eq:f_g_part}, $g= \sum_{R\ge 1} g_R$ {($g_R \in \mathcal{G}_{m,R}$)}. By~\eqref{eq:an_bracket_final} and~\eqref{eq:an_bracket}, we have that there exists $k_R \le \exp(Q(m, R, \epsilon_R))$ (see~\eqref{eq:an_bracket_final}) such that 
			\[ l_{k_R, R,\epsilon_R} \le g_R\le u_{k_R, R,\epsilon_R},\] 
			such that $P(|u_{j, R,\epsilon_R}- l_{j, R,\epsilon_R}|) \le\epsilon_R $, where $\epsilon_R=\epsilon/ (C_\alpha R^\alpha).$ Hence it is easy to see that 
			\begin{equation}\label{eq:stich_bracket}
				\left\{\Big[\sum_{R\ge 1}l_{k_R, R,\epsilon_R}, \sum_{R\ge 1}u_{k_R, R,\epsilon_R} \Big]\big| k_R\in [\exp(Q(m, R, \epsilon_R)] \text{ for every }  R\ge 1 \right\} ,
			\end{equation}
			forms an $\epsilon$ bracket for $\Ff_m$ with respect  to the $L_1(P)$ norm.   Thus
			\begin{align}\label{eq:final_calc_f_m1}
				\begin{split}
					\log N_{[],1}(\epsilon ,\Ff_{m}, P) &\le  \sum_{R\ge 1} Q(m, R, \epsilon_R)\\
					&\le\sum_{R\ge 1}  C_d [\epsilon_R/ (\|m\|_\infty M_R R^d)]^{-(d-1)/2}\\
					&\le C_d \sum_{R\ge 1}   [\epsilon/ (C_\alpha R^\alpha \|m\|_\infty M_R R^d)]^{-(d-1)/2}\\
					&\le C_d [\epsilon/(\|m\|_\infty C_\alpha)]^{-(d-1)/2} \sum_{R\ge 1} [  R^\alpha   C  R^{-r} R^d]^{(d-1)/2}\\
					&\le C_d [\epsilon/(\|m\|_\infty C_\alpha)]^{-(d-1)/2}\\
					&\le C_d [\epsilon/\|m\|_\infty ]^{-(d-1)/2},
				\end{split}
			\end{align}
			as by~\eqref{eq:Tail_cond_lemma} and~\eqref{eq:alpha}, we have that $(\alpha + d-r)(d-1)/2 < -1.$\newline
			
			\noindent\textbf{Proof of~\eqref{eq:entropy_ep}:} This proof will be similar to the proof of~\eqref{eq:entropy_m} above. Defining $\varepsilon^+ := \max (0, \varepsilon)$, $\varepsilon^- := \max (0, -\varepsilon)$, 
			\[   L_{j, R, \epsilon} := {\varepsilon^+(\cdot)\mathbf{1}\big(\cdot \in \ubar{S}_j\cap D_R\big)- \varepsilon^-(\cdot)\mathbf{1}\big(\cdot \in \bar{S}_j\cap D_R\big)}
			\] and 
			\[ U_{j, R,\epsilon} :={ \varepsilon^+(\cdot)\mathbf{1}\big(\cdot \in \bar{S}_j\cap D_R\big)- \varepsilon^-(\cdot)\mathbf{1}\big(\cdot \in \ubar{S}_j\cap D_R\big)}.\]
			We will now show that 
			\begin{equation}\label{eq:bracket_m}
				\Big\{\Big[ {L_{j, R, \epsilon}},{U_{j, R,\epsilon}} \Big]\Big\}_{j=1}^{\exp(N(R, \epsilon))}
			\end{equation}
			forms an $L_1(P)$ bracket of width $C_\varepsilon \epsilon$ (see Lemma~\ref{lem:entropy} for a definition) for 
			\[\Ff_{\varepsilon,R} := \{ g: \R^d \to \R | g(\bm x) = \varepsilon \mathbf{1}(\bm x \in K) \text{ for some } K \in \Kk_d^{R}\cap D_R\}.\]
			Following arguments similar to~\eqref{eq:both_sides}, we see that~\eqref{eq:bracket_m} form a valid bracket. We will now find its width (wrt $L_1(P)$ norm).
			\begin{align}\label{eq:width_ep}
				\begin{split}
					&P\big( \big|U_{k, R, \epsilon} - L_{k, R, \epsilon}  \big|\big)\\={}&P\bigg( \bigg|\varepsilon^+(\cdot)\mathbf{1}\big(\cdot \in (\bar{S}_k\Delta \ubar{S}_k)\cap D_R\big)+  \varepsilon^-(\cdot)\mathbf{1}\big(\cdot \in (\bar{S}_k\Delta \ubar{S}_k)\cap D_R\big) \bigg|\bigg)\\
					={}&P\Big( |\varepsilon|\mathbf{1}\big(\cdot \in (\bar{S}_k\Delta \ubar{S}_k)\cap D_R\big)\Big)\\
					={}&P\Big( \E\big(|\varepsilon|\big| \cdot\big)\mathbf{1}\big(\cdot \in (\bar{S}_k\Delta \ubar{S}_k)\cap D_R\big)\Big)\\
					\le{}& C_\varepsilon P\Big(\bar{S}_k\Delta \ubar{S}_k)\cap D_R\Big),
				\end{split}
			\end{align}
			where $C_\varepsilon := \big\|\E\big(|\varepsilon|\big| \bm X= \cdot)\big\|_{\infty}.$ Thus similar to~\eqref{eq:an_bracket_final},  we have that 
			\begin{equation}\label{eq:an_bracket_final_ep}
				\log N_{[],1}(\epsilon ,\Ff_{\varepsilon, R}, P) \le N(R, \epsilon/ C_\varepsilon) = C_d [\epsilon/ (C_\varepsilon M_R R^d)]^{-(d-1)/2} := Q_\varepsilon(m, R, \epsilon).
			\end{equation}
			Thus just as in~\eqref{eq:stich_bracket}, we have that 
			\begin{equation}\label{eq:stich_bracket_ep}
				\left\{\Big[\sum_{R\ge 1}L_{k_R, R,\epsilon_R}, \sum_{R\ge 1}U_{k_R, R,\epsilon_R} \Big]\big| k_R\in [\exp(Q_\varepsilon(m, R, \epsilon_R)] \text{ for every }  R\ge 1 \right\} ,
			\end{equation}
			forms an $\epsilon$ bracket for $\Ff_\varepsilon$ wrt to $L_1(P)$ norm.   Thus
			\begin{align}\label{eq:final_calc_f_m}
				\begin{split}
					\log N_{[],1}(\epsilon ,\Ff_{\varepsilon}, P) &\le  \sum_{R\ge 1} Q_\varepsilon(m, R, \epsilon_R)\\
					&\le\sum_{R\ge 1}  C_d [\epsilon_R/ (\|m\|_\infty M_R R^d)]^{-(d-1)/2}\\
					&\le C_d \sum_{R\ge 1}  C_d [\epsilon/ (C_\alpha R^\alpha C_\varepsilon M_R R^d)]^{-(d-1)/2}\\
					&\le C_d [\epsilon/(C_\varepsilon C_\alpha)]^{-(d-1)/2} \sum_{R\ge 1} [  R^\alpha   C (1+ R)^{-r} R^d]^{(d-1)/2}\\
					&\le C_d [\epsilon/(C_\varepsilon C_\alpha)]^{-(d-1)/2}\\
					&\le C_d [\epsilon/C_\varepsilon ]^{-(d-1)/2},
				\end{split}
			\end{align}
			as by~\eqref{eq:Tail_cond_lemma} and~\eqref{eq:alpha}, we have that $(\alpha + d-r)(d-1)/2 < -1.$
		\end{proof}
		
		\begin{mylem}[Lemma 8 of~\cite{kur2019optimality}]\label{lemma8}
			Let $\mathcal{H} \subseteq \{h : \mathbb{R}^d \mapsto [0,\Gamma]\}$ be a class of non-negative, bounded functions, and let $\mathcal{C}:= \{h^{-1}([0,\alpha]): h\in \mathcal{H}, \alpha \in [0,\Gamma]\}$ be the corresponding collection of lower level sets. Then for any fixed function $m(\cdot)$, 
			\begin{align}\label{eq:mh_process}
				&\e \sup_{h \in \mathcal{H}}\left|\frac{1}{n} \sum_{i=1}^n r_i m(\bm X_i) h(\bm X_i)\right|\nonumber\\ &\leq 2\Gamma \cdot \e \sup_{C \in \mathcal{C}}\big|(\p_n-P) [m(\cdot) \mathbf{1}(\cdot \in C)]  \big| + C{\Gamma (\|m\|_{\infty}+ \|m\|_{L_2(P)})}n^{-1/2}~,
			\end{align} where $r_1,\ldots, r_n$ are i.i.d. Rademacher random variables and
			\begin{equation}\label{eq:mh_centered}
				\E  \sup_{h\in \Hh}  \Big|(\p_n-P) \big[ m(\cdot) h(\cdot)\big]    \Big| \le \Gamma\, \e \sup_{C\in \mathcal{C}} \big|   (\p_n-P) [m(\cdot) \mathbf{1}(\cdot \in C)]\big|+ { \Gamma \|m\|_{L_2(P)} }n^{-1/2}.
			\end{equation}Furthermore, 
			\begin{equation}\label{eq:mult_process}
				\E  \sup_{h\in \Hh}  \Big|(\p_n-P) \big[ \varepsilon h(\bm X)\big]    \Big| \le \Gamma\, \e \sup_{C\in \mathcal{C}} \big|   (\p_n-P) [\varepsilon \mathbf{1}(  X \in C)]\big|+   \Gamma \mathrm{Var}(\varepsilon) n^{-1/2}.
			\end{equation}
		\end{mylem}

		\begin{proof} We will first prove~\eqref{eq:mh_centered} and use that to prove~\eqref{eq:mh_process}. The reduction scheme here is inspired by~\cite{carpenter2018near}; also see~\cite{han2021set} and~\cite{kur2019optimality}.\newline
			
			\noindent\textbf{Proof of~\eqref{eq:mh_centered}:} 
			Noting that {for $h \in \mathcal{H}$}
			\[
			h(\bm x) = \Gamma -\int_{0}^{\Gamma} \mathbf{1}(h(\bm x)\le t) dt,
			\]
			we have
			\begin{align}\label{eq:mh_final}
				\begin{split}
					&\e \sup_{h\in \mathcal{H}}|\p_n(mh) - P(mh)|\\\
					\le{}& \e\sup_{h\in \mathcal{H}} \left|\int_{0}^\Gamma \Big(P\big(m(\bm X) \mathbf{1}(h(\bm X) \le t)\big) - \p_n\big(m(\bm X) \mathbf{1}(h(\bm X) \le t)\big)\Big)~dt  \right|+ \Gamma\E \Big| (\p_n- P )m\Big|  \\\
					\le{}& \int_0^\Gamma \e \sup_{h\in \mathcal{H}} \big|   (\p_n-P) [m(\cdot) \mathbf{1}(h(\bm X) \le t)]\big|~dt +{ \Gamma \|m\|_{L_2(P)} }n^{-1/2}\\\
					\le{}&  \Gamma\, \e \sup_{h\in \mathcal{H}, t\in [0, \Gamma]} \big|   (\p_n-P) [m(\cdot) \mathbf{1}(h(\bm X) \le t)]\big|+ { \Gamma \|m\|_{L_2(P)} }n^{-1/2}\\
					={}&  \Gamma\, \e \sup_{C\in \mathcal{C}} \big|   (\p_n-P) [m(\cdot) \mathbf{1}(\bm X \in C)]\big|+ { \Gamma \|m\|_{L_2(P)} }n^{-1/2}.
				\end{split}
			\end{align}
			\noindent\textbf{Proof of~\eqref{eq:mh_process}:} Observe that
			\begin{align}\label{firstline}
				\begin{split}
					&\e \sup_{h \in \mathcal{H}}\left|\frac{1}{n} \sum_{i=1}^n r_i  m(\bm X_i) h(\bm X_i)\right| \\
					={}& \e \sup_{h \in \mathcal{H}}\left|\frac{1}{n} \sum_{i=1}^n r_i \left(m(\bm X_i) h(\bm X_i)-P(hm) + P(hm)\right)\right|\\
					\le{}& \e \sup_{h \in \mathcal{H}}\left|\frac{1}{n} \sum_{i=1}^n r_i \left(m(\bm X_i) h(\bm X_i)-P(mh)\right)\right| + \e\sup_{h\in \mathcal{H}}\Big|P(mh) \cdot \frac{1}{n}\sum_{i=1}^n r_i\Big|~.
				\end{split}
			\end{align} 
			Let us now bound the two terms in \eqref{firstline}. The second term in \eqref{firstline} can be bounded as follows:
			\begin{eqnarray*}\label{secondline}
				\e \sup_{h\in \mathcal{H}}\left| P(mh) \cdot \frac{1}{n}\sum_{i=1}^n r_i\right| &\le& \Gamma \|m\|_{\infty}\cdot \e\left|\frac{1}{n}\sum_{i=1}^n r_i\right|\\ &\le& \Gamma \|m\|_{\infty} \cdot \sqrt{\e \left[\left(\frac{1}{n}\sum_{i=1}^n r_i\right)^2\right]} = \frac{\Gamma \|m\|_{\infty}}{\sqrt{n}}. 
			\end{eqnarray*}
			For bounding the first term in \eqref{firstline}, observing that 
			\[
			h(\bm x) = \Gamma -\int_{0}^{\Gamma} \mathbf{1}(h(\bm x)\le t) dt~,
			\]
			and  appealing to the symmetrization lemma (see Lemma 2.3.6 in \cite{vander}), to conclude that:
			\begin{equation}\label{thirdline}
				\e \sup_{h \in \mathcal{H}}\left|\frac{1}{n} \sum_{i=1}^n r_i \left(m(\bm X_i) h(\bm X_i)-P(mh)\right)\right|\le 2\e \sup_{h\in \mathcal{H}}|\p_n(mh) - P(mh)|.
			\end{equation}
			Our result \eqref{eq:mh_process} now follows from ~\eqref{eq:mh_final}, \eqref{firstline}, \eqref{secondline} and \eqref{thirdline}.\newline

			\noindent\textbf{Proof of~\eqref{eq:mult_process}:} Observe that
			We will now prove~\eqref{eq:mult_process}. Observe that
			\begin{align}\label{eq:integral}
				\begin{split}
					&\e \sup_{h\in \Hh}  \Big|(\p_n-P) \big[ \varepsilon h(\bm X)\big]    \Big|  \\
					={}& \e\sup_{h\in \mathcal{H}} \left|\int_{0}^\Gamma \Big(P\big(\varepsilon \mathbf{1}(h(\bm X) \le t)\big) - \p_n\big(\varepsilon \mathbf{1}(h(\bm X) \le t)\big)\Big)~dt  \right|+ \Gamma\E \Big| (\p_n- P )\varepsilon\Big|\\
					\le{}& \int_0^\Gamma \e \sup_{h\in \mathcal{H}} \big|   (\p_n-P) [\varepsilon \mathbf{1}(h(\bm X) \le t)]\big|~dt+  \Gamma \mathrm{Var}(\varepsilon) n^{-1/2}\\
					\le{}&  \Gamma\, \e \sup_{h\in \mathcal{H}, t\in [0, \Gamma]} \big|   (\p_n-P) [\varepsilon \mathbf{1}(h(\bm X) \le t)]\big|+  \Gamma \mathrm{Var}(\varepsilon) n^{-1/2}\\
					={}&  \Gamma\, \e \sup_{C\in \mathcal{C}} \big|   (\p_n-P) [\varepsilon \mathbf{1}(\bm X \in C)]\big|+  \Gamma \mathrm{Var}(\varepsilon) n^{-1/2}.
				\end{split}
			\end{align}
		\end{proof}

		\begin{mylem}\label{lemma6}
			Assume that $d\geq 2$. Suppose that $\varepsilon$ has a $2+\delta$ moment bounded by $L$ for some $\delta > 0$, and also assume that $\,P= f(\bm x)d\bm x$ satisfies the assumptions of Lemma~\ref{lemma10andCor1}, then 
			\begin{equation}\label{eq:multiplier_r}
				\E \bigg[ \sup_{h\in \Hh_{d,\Gamma}}  \Big|\p_n \big[ r h(\cdot)\big]    \Big|  \bigg]\le  CC_d \Gamma \times \begin{cases}
					n^{-1/2}  &\text{ when } d = 2,\\
					n^{-1/2} \log n  &\text{ when } d = 3,\\
					n^{-2/(d+1)} &\text{ when } d\ge 4,
				\end{cases}
			\end{equation}
			\begin{equation}\label{eq:multiplier}
				\E \bigg[ \sup_{h\in \Hh_{d,\Gamma}}  \Big|\p_n \big[ \varepsilon h(\cdot)\big]    \Big|  \bigg]\le C C_\varepsilon  C_d \Gamma \times \begin{cases}
					n^{-1/2}  &\text{ when } d = 2,\\
					n^{-1/2} \log n  &\text{ when } d = 3,\\
					n^{-2/(d+1)} &\text{ when } d\ge 4,
				\end{cases}
			\end{equation}
			and 
			\begin{equation}\label{eq:multiplier_mh}
				\E \bigg[ \sup_{h\in \Hh_{d,\Gamma}}  \Big|(\p_n-P) \big[ ( \varphi(\cdot) - \bar{\varphi}_{\Gamma}(\cdot)) h(\cdot)\big]    \Big|  \bigg]\le  CC_\varphi C_d \Gamma \times \begin{cases}
					n^{-1/2}  &\text{ when } d = 2,\\
					n^{-1/2} \log n  &\text{ when } d = 3,\\
					n^{-2/(d+1)} &\text{ when } d\ge 4,
				\end{cases}
			\end{equation}
			where   $C_\varepsilon := \|\E(\varepsilon|\bm X=\cdot)\|_{\infty} + \mathrm{var}(\varepsilon)$, $C_\varphi:= \|\varphi - \bar{\varphi}_{\Gamma}\|_{L_2(P)}+ \|\varphi(\bm X_i) - \bar{\varphi}_{\Gamma}(\bm X_i)\|_{\infty}$ and $C_d$ is a constant depending on $d$ only.
		\end{mylem}

		\begin{proof}  Let $\mathcal{H}_{d,\Gamma,1}^+$ (resp. $\mathcal{H}_{d,\Gamma,2}^+$) denote the set of all non-negative quasiconvex (resp. quasiconcave) functions on $\mathbb{R}^d$, bounded by $\Gamma$. Since lower level sets of quasiconvex functions (resp. upper level sets of quasiconcave functions) are convex, by choosing $m(\cdot)\equiv 1$, it follows from~\eqref{eq:10_1} (Lemma~\ref{lemma10andCor1}) and~\eqref{eq:mh_process} (Lemma~\ref{lemma8}), that for $j \in \{1,2\}$,
			\begin{equation}\label{rademacher1}
				\e \sup_{h \in \mathcal{H}_{d,\Gamma,j}^+} \left|\frac{1}{n} \sum_{i=1}^n r_i h(\bm X_i)\right| \le C \Gamma n^{-1/2} + \Gamma C_d  n^{-2/(d+1)}  \log n^{\mathbf{1}(d=3)}.
			\end{equation}
			where $r_1,\ldots,r_n$ are i.i.d. Rademacher random variables. {Hence}, for $d \ge 2$ we have 
			\begin{equation}\label{rademacher}
				\e \sup_{h \in \mathcal{H}_{d,\Gamma,j}^+} \left|\frac{1}{n} \sum_{i=1}^n r_i h(\bm X_i)\right| \le  C_d \Gamma \times \begin{cases}
					n^{-1/2}  &\text{ when } d = 2,\\
					n^{-1/2} \log n  &\text{ when } d = 3,\\
					n^{-2/(d+1)} &\text{ when } d\ge 4.
				\end{cases}
			\end{equation}
			
			Similarly, using~\eqref{eq:10_1_ep} (Lemma~\ref{lemma10andCor1}) and~\eqref{eq:mult_process} (Lemma~\ref{lemma8}), that for $j \in \{1,2\}$, we have
			\begin{equation}\label{rademacher1_ep}
				\e \sup_{h \in \mathcal{H}_{d,\Gamma,j}^+} \left|\frac{1}{n} \sum_{i=1}^n \varepsilon_i h(\bm X_i)\right| \le \Gamma \mathrm{var}(\varepsilon) n^{-1/2} + \Gamma C_\varepsilon C_d   n^{-2/(d+1)}  \log n^{\mathbf{1}(d=3)}.
			\end{equation}
			Hence, for $d \ge 2$, we have,
			\begin{equation}\label{rademacher_ep}
				\e \sup_{h \in \mathcal{H}_{d,\Gamma,j}^+} \left|\frac{1}{n} \sum_{i=1}^n \varepsilon_i h(\bm X_i)\right| \le  C_d C_\varepsilon \Gamma \times \begin{cases}
					n^{-1/2}  &\text{ when } d = 2,\\
					n^{-1/2} \log n  &\text{ when } d = 3,\\
					n^{-2/(d+1)} &\text{ when } d\ge 4.
				\end{cases}
			\end{equation}
			
			Finally, using~\eqref{eq:10_1} (Lemma~\ref{lemma10andCor1}) and~\eqref{eq:mh_centered} (Lemma~\ref{lemma8}), that for $j \in \{1,2\}$, we have
			\begin{equation}\label{rademacher1_mh}
				\e \sup_{h \in \mathcal{H}_{d,\Gamma,j}^+} \left| (\p_n -P) [( \varphi(\cdot) - \bar{\varphi}_{\Gamma}(\cdot)) h(\cdot)]\right| \le \Gamma C C_\varphi n^{-1/2} + \Gamma C_\varphi C_d   n^{-2/(d+1)}  \log n^{\mathbf{1}(d=3)}.
			\end{equation}
			where $C_\varphi := \|\varphi - \bar{\varphi}_{\Gamma}\|_{L_2(P)}+ \|\varphi(\bm X_i) - \bar{\varphi}_{\Gamma}(\bm X_i)\|_{\infty}$. Hence, for $d \ge 2$, we have,
			\begin{equation}\label{rademacher_mh}
				\e \sup_{h \in \mathcal{H}_{d,\Gamma,j}^+} \big| (\p_n -P) [( \varphi(\cdot) - \bar{\varphi}_{\Gamma}(\cdot)) h(\cdot)]\big| \le  C_d C_\varphi\Gamma \times \begin{cases}
					n^{-1/2}  &\text{ when } d = 2,\\
					n^{-1/2} \log n  &\text{ when } d = 3,\\
					n^{-2/(d+1)} &\text{ when } d\ge 4.
				\end{cases}
			\end{equation}
			
			To complete the proof of Lemma \ref{lemma6}, we note that for any quasiconvex function $h$, the function $h^+ := \max\{h,0\}$ is non-negative, quasiconvex, and the function $h^{-} = \max\{-h,0\}$ is non-negative, quasiconcave. Hence, we have:
			\begin{align}\label{finst}
				\begin{split}
					\e\sup_{h \in \mathcal{H}_{d,\Gamma}}\left|\frac{1}{n} \sum_{i=1}^n h(\bm X_i) \varepsilon_i \right| &\le \e\sup_{h \in \mathcal{H}_{d,\Gamma}}\left|\frac{1}{n} \sum_{i=1}^n h^+(\bm X_i) \varepsilon_i \right| + \e\sup_{h \in \mathcal{H}_{d,\Gamma}}\left|\frac{1}{n} \sum_{i=1}^n h^-(\bm X_i) \varepsilon_i \right|\\
					&\le \e\sup_{h \in \mathcal{H}_{d,\Gamma,1}^+}\left|\frac{1}{n} \sum_{i=1}^n h(\bm X_i) \varepsilon_i \right| + \e\sup_{h \in \mathcal{H}_{d,\Gamma,2}^+}\left|\frac{1}{n} \sum_{i=1}^n h(\bm X_i) \varepsilon_i \right|. 
				\end{split}
			\end{align}
			The results~\eqref{eq:multiplier_r},~\eqref{eq:multiplier},~\eqref{eq:multiplier_mh} of Lemma \ref{lemma6} now follows from by combining~\eqref{finst} with \eqref{rademacher}, \eqref{rademacher_ep}, and \eqref{rademacher_mh}, respectively. 
		\end{proof}

		\begin{mylem}\label{lemma7}
			Assume that $d \geq 2$, and that the distribution $P$ satisfies the assumptions in~Lemma~\ref{lemma6}. Then,
			$$\e \sup_{f,g \in \mathcal{H}_{d,\Gamma}} \left|\p_n(f-g)^2 - P(f-g)^2\right| \le C C_d \Gamma^2 \times \begin{cases}
				n^{-1/2}  &\text{ when } d = 2,\\
				n^{-1/2} \log n  &\text{ when } d = 3,\\
				n^{-2/(d+1)} &\text{ when } d\ge 4.
			\end{cases}$$
		\end{mylem}

		\begin{proof}
			The following proof is a slight modification of Lemma 7 of ~\cite{kur2019optimality}. 
			By Theorem 2.1 in \cite{koltchinskii2011oracle}, we have:
			\begin{equation}\label{1st}
				\e \sup_{f,g \in \mathcal{H}_{d,\Gamma}} \left|\p_n(f-g)^2 - P(f-g)^2\right| \le 2 \e \sup_{f,g \in \mathcal{H}_{d,\Gamma}} \left|\frac{1}{n} \sum_{i=1}^n r_i (f(\bm X_i) - g(\bm X_i))^2 \right|~,
			\end{equation}
			where $r_1,\ldots,r_n$ are i.i.d. Rademacher. By Corollary 3.2.2 in \cite{gine}, the right hand side of \eqref{1st} can be bounded as follows:
			\begin{equation}\label{2nd}
				\e \sup_{f,g \in \mathcal{H}_{d,\Gamma}} \left|\frac{1}{n} \sum_{i=1}^n r_i (f(\bm X_i) - g(\bm X_i))^2 \right| \le 4 \Gamma \cdot \e \sup_{f,g \in \mathcal{H}_{d,\Gamma}} \left|\frac{1}{n} \sum_{i=1}^n r_i (f(\bm X_i) - g(\bm X_i)) \right|~.
			\end{equation}
			Combining \eqref{1st} and \eqref{2nd}, we have:
			\begin{align}\label{3rd}
				\e \sup_{f,g \in \mathcal{H}_{d,\Gamma}} \left|\p_n(f-g)^2 - P(f-g)^2\right|&\le 8 \Gamma \cdot \e \sup_{f,g \in \mathcal{H}_{d,\Gamma}} \left|\frac{1}{n} \sum_{i=1}^n r_i (f(\bm X_i) - g(\bm X_i)) \right|\nonumber\\ &\le 16\Gamma\cdot \e \sup_{f\in \mathcal{H}_{d,\Gamma}} \left|\frac{1}{n} \sum_{i=1}^n r_i f(\bm X_i) \right|.
			\end{align}
			The last inequality in \eqref{3rd} follows from the triangle inequality. Lemma \ref{lemma7} now follows from \eqref{3rd},~\eqref{rademacher}, and~\eqref{finst}. 
		\end{proof}
	}
	
\end{document}